\documentclass[11pt,a4paper]{article}
\usepackage{amsmath,amsfonts,amssymb,graphicx,color}

\numberwithin{equation}{section}

\newcommand{\nc}{\newcommand}
\nc{\rnc}{\renewcommand}
\rnc{\title}[1]{{\Large\bf\mbox{}\\\medskip#1\bigskip\medskip\\}}
\rnc{\author}[1]{{\large #1\smallskip\\}}
\nc{\address}[1]{{\em #1\medskip\\}}
\def\gauss#1#2{\mbox{\small $\left[#1\atop #2\right]$}}
\def\gausst#1#2{\mbox{\small $\left[#1\atop #2\rule{0pt}{9pt}\right]$}}

\newcommand{\sm}[1]{{\scriptstyle #1}}

\newcommand{\spos}[2]{\makebox(0,0)[#1]{$\sm{#2}$}}

\newcommand{\W}[5]{W\!\left(\,\begin{array}{@{}cc|@{\:}}#4&#3\\
#1&#2\end{array}\;#5\right)}
\newcommand{\B}[6]{B^{#1 #2}\!\left(#3\;\;\;\begin{array}{@{}c|@{\:}}#4\\
#4\end{array}\;#5,\,#6\right)}

\newcommand{\xilatt}{\xi_{\mbox{\scriptsize latt}}}
\newcommand{\face}[5]{\begin{picture}(1.6,1.6)
\multiput(0.3,0.3)(1,0){2}{\line(0,1){1}}
\multiput(0.3,0.3)(0,1){2}{\line(1,0){1}}
\put(0.26,0.26){\spos{tr}{#1}}\put(1.34,0.26){\spos{tl}{#2}}
\put(1.34,1.34){\spos{bl}{#3}}
\put(0.26,1.34){\spos{br}{#4}}
\put(0.8,0.8){\spos{}{#5}}\end{picture}}

\newcommand{\boundary}[3]{
\begin{picture}(2.7,2.6)(-0.5,0.1)
\put(0.5,1.5){\line(1,1){1}}\put(0.5,1.5){\line(1,-1){1}}
\put(1.5,0.5){\line(1,0){0.5}}\put(1.5,2.5){\line(1,0){0.5}}
\multiput(2,0.5)(0,0.3){7}{\line(0,1){0.2}}
\put(0.34,1.5){\makebox(0,0)[r]{$\scriptstyle #1$}}
\put(1.4,2.8){\makebox(0,0){$\scriptstyle #2$}}
\put(1.4,0.2){\makebox(0,0){$\scriptstyle #2$}}
\put(1.4,1.5){\makebox(0,0){#3}}
\end{picture}}

\newcommand{\inside}[2]{\makebox(0,0){$\begin{array}{l} \sm{#1} \\ \sm{#2}
\end{array}$}}

\newcommand{\re}{\mbox{Re}}
\newcommand{\im}{\mbox{Im}}
\newcommand{\ir}{^{\mbox{\scriptsize IR}}}
\newcommand{\uv}{^{\mbox{\scriptsize UV}}}
\newcommand{\irt}{^{\mbox{\tiny IR}}}
\newcommand{\uvt}{^{\mbox{\tiny UV}}}

\makeatletter
\renewcommand{\@makecaption}[2]{
   \vskip\abovecaptionskip
   \sbox\@tempboxa{#1. #2}%
   \ifdim \wd\@tempboxa >\hsize
     #1. #2\par
   \else
     \global \@minipagefalse
     \hb@xt@\hsize{\hfil\box\@tempboxa\hfil}%
   \fi
   \vskip\belowcaptionskip}
\makeatother

\begin{document}

\begin{center}

\title{Exact $\varphi_{1,3}$ boundary flows in the tricritical Ising model}

\author{Giovanni Feverati\footnote{feverati@ms.unimelb.edu.au},
Paul A. Pearce\footnote{P.Pearce@ms.unimelb.edu.au}}

\address{Department of Mathematics and Statistics\\
University of Melbourne, Parkville, Victoria 3010, Australia}

\author{Francesco Ravanini\footnote{ravanini@bologna.infn.it}}

\address{INFN Sezione di Bologna\\
Via Irnerio 46, 40126 Bologna, Italy}

\end{center}

\setcounter{footnote}{0}

\begin{abstract}
We consider the tricritical Ising model on a strip or cylinder under 
the integrable perturbation by the thermal $\varphi_{1,3}$ boundary 
field. This perturbation induces five distinct renormalization group 
(RG)  flows between Cardy type boundary conditions labelled by the 
Kac labels $(r,s)$. We study these boundary RG flows in detail for 
all excitations. Exact Thermodynamic Bethe Ansatz (TBA) equations are 
derived using the lattice approach by considering the continuum 
scaling limit of the $A_4$ lattice model with integrable boundary 
conditions. Fixing the bulk weights to their critical values, the 
integrable boundary weights admit a thermodynamic boundary field 
$\xi$ which induces the flow and, in the continuum scaling limit, 
plays the role of the perturbing boundary field $\varphi_{1,3}$. The 
excitations are completely classified, in terms of string content, 
by $(m,n)$ systems and quantum numbers but the string content changes 
by either two or three well-defined mechanisms along the flow. We 
identify these mechanisms and obtain the induced maps between the 
relevant finitized Virasoro characters. We also solve the TBA 
equations numerically to determine the boundary flows for the leading 
excitations.
\end{abstract}

\section{Introduction}
Quantum Field Theories with a boundary have received
a lot of attention recently due to their applications in Condensed Matter,
Solid State Physics and String Theory (D-branes). A problem of great
interest is the Renormalization Group (RG) flow between different boundary
fixed points of a Conformal Field Theory (CFT) that remains conformal 
in the bulk. Many interesting
results have been achieved, and flows have been studied for minimal
models and for $c=1$ CFT (see e.g. \cite{GRW}
and references therein). Numerical scaling functions for the flow
of states interpolating two different boundary conditions can be systematically
explored by use of the approximate Truncated Conformal Space Approach 
(TCSA) \cite{TCSA,bdyFlows,GRW2}.

A beta function can be defined for the boundary deformations, much
the same as for the bulk perturbations of conformal field theories
\cite{gthm}. The conformal boundary conditions play the
role of ultraviolet (UV) and infrared (IR) points of the flow. One
flows away from the UV fixed point by perturbing with a relevant 
boundary operator
and flows into an IR fixed point attracted by irrelevant boundary operators.

Among the possible boundary perturbations of a CFT there are some
  \emph{integrable} perturbations that preserve an infinite number of 
conservation laws.
In this case the flows are amenable to investigation by exact methods such as
commuting transfer matrices and Bethe ansatz techniques.
One of the most celebrated of these methods is the Thermodynamic Bethe
Ansatz (TBA) \cite{TBA} giving a set of non-linear coupled
integral equations governing the scaling functions along the RG flow.
Boundary TBA equations were first obtained with the usual scattering
approach \cite{btba, bdyFlows} for the groundstate, a few excited states
and the boundary entropy in the
minimal models. In particular, the groundstate and  boundary entopies 
of the Tricritical Ising Model have been studied~\cite{NA} within 
this approach.

In this paper we use a lattice approach to obtain exact TBA equations
for all the excitations of the integrable boundary flows of the 
Tricritical Ising Model (TIM) with
central charge $c=\frac{7}{10}$. This model has interesting applications
in Solid State Physics and Statistical Physics. Its Kac table of 
conformal weights $h_{r,s}$ is
\vspace{0.3cm}
\begin{center}
\setlength{\unitlength}{0.35mm}
\begin{picture}(170,100)(20,-60)
\put(0,0)
\mbox{\hspace{.5in} \Large
\begin{tabular}{r|c|c|c|l}
\multicolumn{5}{l}{$s$}\\
\cline{2-4}
2&$\ {1\over 10}\rule[-12pt]{0pt}{32pt}\ $&$\ {3\over 80}\ $&
$\ {3\over 5}\ $&\\[6pt]
\cline{2-4}
1&0&${7\over 16}\rule[-12pt]{0pt}{32pt}$&${3\over 2}$&\\[6pt]
\cline{2-4}
\end{tabular}}
\put(-154,-56){\Large
\begin{tabular}{rcccl}
&$\qquad ~~\, 1\ $&$\ \ 2\ $&$\ \ 3\ $&$\ r\ $
\end{tabular}}
\put(-92,-1){\vector(0,-1){10}}
\put(-58,-1){\vector(0,-1){10}}
\put(-26,-1){\vector(0,-1){10}}
\put(-81.5,1){\vector(1,-1){15}}
\put(-34.4,1){\vector(-1,-1){15}}
\multiput(-68.8,-0.8)(-2,-2){7}{$\cdot$}
\multiput(-49.9,-0.8)(2,-2){7}{$\cdot$}
\put(-80.5,-11){\vector(-1,-1){3}}
\put(-35.4,-11){\vector(1,-1){3}}
\end{picture}
\end{center}
\vspace{0.3cm}
We have drawn only half of the table, by taking into account the well
known $\mathbb{Z}_{2}$ Kac table symmetry $(r,s)\equiv (4-r,5-s)$.
The number of independent chiral primary fields is thus six.

Cardy-type conformal boundary conditions for minimal models with 
diagonal modular
invariant partition functions of type $(A_{p},A_{p'})$ are in 
one-to-one correspondence
with the chiral primary fields in the Kac table.
So the TIM $(A_{3},A_{4})$ admits six different conformal boundary 
conditions which
we denote by $B_{(r,s)}$, $r=1,2,3$ and $s=1,2$.

Let us consider the TIM defined on a strip of width $L$
in the two-dimensional $(x,t)$ plane so that $0\leq x\leq L$ and
$-\infty <t<+\infty $. Let us impose boundary conditions
  $B_{(r,s)}$  and $B_{(r',s')}$ on the left ($x=0$) and right ($x=L$) 
edge respectively.
We denote this situation by $B_{(r,s)|(r',s')}$. In the sequel
we are interested in boundaries of type $(r,s)|(1,1)$ which we denote
by $\mathcal{B}_{(r,s)}=B_{(r,s)|(1,1)}\equiv B_{(r,1)|(1,s)}$. For 
these boundary conditions the partition function reduces
to a single character \cite{Cardy}
\begin{equation}
Z_{\mathcal{B}_{(r,s)}}(q)=Z_{(r,s)|(1,1)}(q)=\chi _{r,s}(q)
\end{equation}
where $\chi _{r,s}(q)$ denotes the character of the irreducible
Virasoro representation labelled by $(r,s)$ at central charge $c=\frac{7}{10}$.

For each given boundary condition $\mathcal{B}_{(r,s)}$ there is
a set of boundary operators $\varphi_{u,v}$ that live on the edge.
If we want to keep the boundary condition unchanged all along the 
edge, these operators
must be restricted to the conformal families appearing in the OPE
fusion of the Virasoro family $(r,s)$ with itself: $(u,v)\in 
(r,s)\times (r,s)$.
They are distinguished in terms of their conformal dimensions $h_{r,s}$
as relevant ($h_{r,s}<1$) and irrelevant ($h_{r,s}>1$). Of course
only relevant perturbations break scale invariance at the boundary
in such a way to get out of the fixed boundary point of a specific
boundary condition and flow to another one. So if we want to consider
possible relevant boundary perturbations of the TIM, i.e. QFTs described
by the action
\begin{equation}
S=S_{(r,s)}+\lambda \int _{-\infty }^{+\infty }dt\, \varphi_{u,v}(x=0,t)
\end{equation}
where $S_{(r,s)}$ denotes the action of TIM with boundary condition
$\mathcal{B}_{(r,s)}$, we have to restrict to the possibilities
as shown in Table~\ref{t_bc}.

\begin{table}\caption{\label{t_bc}}\begin{center}
\begin{tabular}{|c|c|c|}
\hline
\multicolumn{2}{|c|}{boundary condition}&
boundary perturbations\\
\hline
\hline
$\mathcal{B}_{(1,1)}\equiv \mathcal{B}_{(3,4)}$&
$-$&
none\\
\hline
$\mathcal{B}_{(2,1)}\equiv \mathcal{B}_{(2,4)}$&
$0$&
none\\
\hline
$\mathcal{B}_{(3,1)}\equiv \mathcal{B}_{(1,4)}$&
$+$&
none\\
\hline
$\mathcal{B}_{(1,2)}\equiv \mathcal{B}_{(3,3)}$&
$-0$&
$\varphi _{1,3}$\\
\hline
$\mathcal{B}_{(2,2)}\equiv \mathcal{B}_{(2,3)}$&
$d$&
$\varphi _{1,2},\, \varphi _{1,3}$\\
\hline
$\mathcal{B}_{(3,2)}\equiv \mathcal{B}_{(1,3)}$&
$0+$&
$\varphi _{1,3}$\\
\hline
\end{tabular}\end{center}\end{table}

Actually there are two physically different flows for each {}``pure''
perturbation (i.e. containing only one operator $\varphi _{1,2}$ or
$\varphi _{1,3}$), flowing to two possibly different IR destinies.
This is achieved by taking different signs in the coupling constant
in the case of $\varphi _{1,3}$ flows, and real or purely imaginary
coupling constant in case of $\varphi _{1,2}$ ones. The boundary condition
$\mathcal{B}_{(2,2)}$ can be perturbed by any linear combination
of the fields $\varphi _{1,2}$ and $\varphi _{1,3}$. The symbols 
$+,-,0+,-0,0,d$
represent other ways~\cite{Chim,affleck} to denote the TIM conformal 
boundary conditions
and we give them only to make contact with the existing literature.

\begin{figure} \begin{center}
\includegraphics[width=0.7\linewidth]{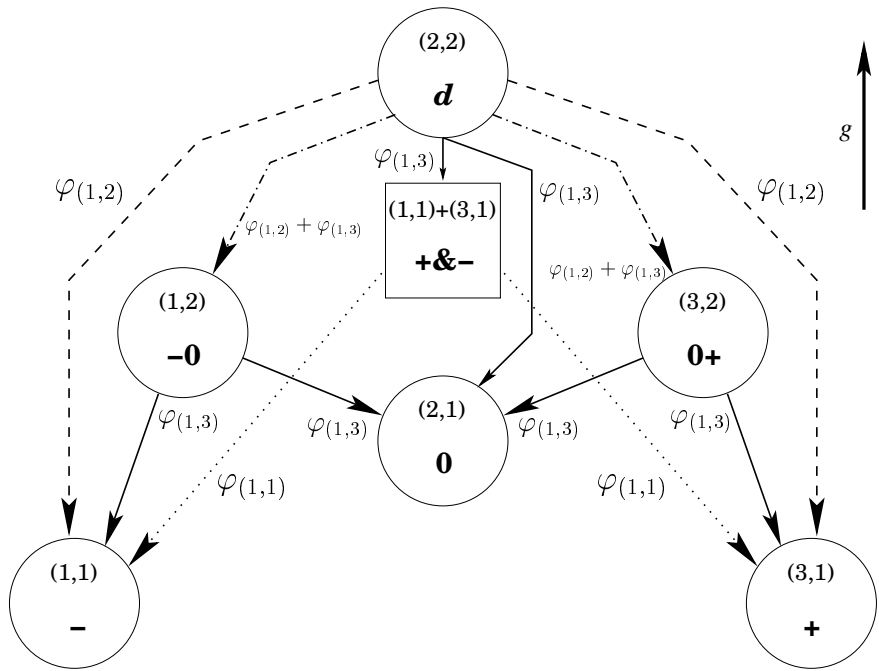}
\caption{\label{fig-flows}\small The boundary flows between TIM 
conformal boundary conditions. Pure
$\varphi _{1,3}$ and $\varphi _{1,2}$ flows are integrable. More information
can be found in \cite{affleck}. The boundary entropy associated with 
the $+\&-$ boundary condition is linear in the Cardy type boundary 
entropies: $g_{(+\&-)}=g_{(1,1)}+g_{(3,1)}$.}
\end{center}
\end{figure}

Along the physical flow from UV to IR, the boundary entropy associated
to each $\mathcal{B}_{(r,s)}$
\begin{equation}
g_{(r,s)}=\left(\frac{2}{5}\right)^{1/4}\frac{\sin \frac{\pi 
r}{4}\sin \frac{\pi s}{5}}{\sqrt{\sin \frac{\pi }{4}\sin \frac{\pi 
}{5}}}
\end{equation}
decreases \cite{gthm}, so we only expect flows between
boundary conditions where the initial boundary entropy
is larger than the final boundary entropy.
The possible conformal boundary conditions have been studied
by Chim \cite{Chim} and the flows connecting them by Affleck \cite{affleck}.
The picture is summarized in Figure~\ref{fig-flows}. Integrability
can be investigated in a manner similar to the bulk perturbations
and it turns out that the flows generated by pure $\varphi _{1,3}$
and $\varphi _{1,2}$ perturbations are integrable. In contrast, the flow
starting at $B_{(2,2)}$ as a perturbation which is a linear combination
of $\varphi _{1,2}$ and $\varphi _{1,3}$ is strongly suspected to
be non-integrable.
Notice the $\mathbb{Z}_{2}$ symmetry of Figure~\ref{fig-flows}, which
is related to the supersymmetry of the TIM. Investigation
of the supersymmetric aspects of the TIM boundary flows is, however,
beyond the scope of the present paper.

In this paper we derive exact TBA equations for all excitations of 
the integrable boundary flows of the TIM from the functional 
relations for the transfer matrix of the $A_4$ lattice RSOS 
model~\cite{HardSq} which is a member of the $A_L$ 
Andrews-Baxter-Forrester models~\cite{ABF}. Using general 
techniques~\cite{KlumpP,PearceN} of the lattice approach for turning 
functional equations into non-linear integral equations, we derive 
the TBA equations for all excitations and solve them numerically to 
determine the flows for the leading excitations. Our preliminary 
results for the flow $\chi_{1,2} \mapsto \chi_{2,1}$ were announced 
in \cite{FPR}.
Although we only consider the TIM here, our approach is quite general 
and should apply,
for example, to the integrable boundary flows of all minimal models. 
A similar lattice approach has been successfully applied~\cite{PCA} 
to the massive and massless bulk thermal RG flows of the TIM.
We stress that we are not addressing here the interesting question 
of predicting the allowed pattern of flows between fixed points. Rather, 
given a particular flow, the thrust of the lattice approach is to calculate 
the flow of all excitations and to predict, excitation by excitation, where 
these levels flow.

The layout of this paper is as follows.
In Section~2, we discuss the TIM as the continuum scaling limit of 
the $A_4$ lattice model. We describe the integrable boundary 
conditions~\cite{BP} that give rise, in the continuum scaling limit, 
to the conformal boundary conditions labelled by the Kac labels 
$(r,s)$, the associated double row transfer matrices~\cite{BPO} and 
their functional equations. We summarize the results of O'Brien, 
Pearce and Warnaar~\cite{OPW} which apply at the conformal  fixed 
point of the TIM before the boundary perturbation $\xi=\im(\xilatt)$ 
is turned on. In particular, we review the classification of the 
allowed patterns of zeros of the transfer matrix eigenvalues in terms 
of $(m,n)$ systems and quantum numbers and summarize the TBA 
equations and their solutions for the finite-size energy spectra in 
terms of finitized characters. In Section~3, we consider the three 
boundary flows with constant $r$, namely, $\chi_{1,2} \mapsto 
\chi_{1,1}$, $\chi_{3,2} \mapsto \chi_{3,1}$ and $\chi_{2,2} \mapsto 
\chi_{2,1}$. We describe two mechanisms A, B under which the patterns 
of zeros change during these flows and in each case we show that this 
is consistent with an explicit map between finitized characters. In 
each case, we also solve the functional equations to obtain the 
boundary TBA equations describing the flow. In Section~4, we repeat 
these considerations for the two boundary flows with variable $r$, 
namely, $\chi_{1,3} \mapsto \chi_{2,1}$ and $\chi_{1,2} \mapsto 
\chi_{2,1}$. In these cases we find three mechanisms A, B, C. In 
Section~5, we discuss the numerical solution of the boundary TBA 
equations for all five thermal boundary flows. Finally, we finish 
with some conclusions in Section~6.

\section{The tricritical Ising model}
The conformal unitary minimal model with central charge $c=7/10$ describes the
critical behaviour of the tricritical Ising model. A lattice realization
of this universality class is given by the critical
$A_4$ ABF model on the critical line separating regimes III and IV.
Since we are interested in working in the presence of a boundary we define, following \cite{OPW}, a
double row transfer matrix with $(r_1,a_1)$ and $(r_2,a_2)$ boundary
conditions by the following diagrammatic representation
\setlength{\unitlength}{10mm}
\begin{equation}\label{DRTMdef}
\mathbf{D}^{N}_{r_1\,a_1|r_2\,a_2}(u,\xi_1,\xi_2)_{\sigma,\sigma'}
=\sum_{\tau_{0},\dots,\tau_{N}}
\raisebox{-1.4\unitlength}[1.3\unitlength][1.1\unitlength]
{\begin{picture}(8,2.4)(-0.7,0.1)
\put(0,0.5){\line(1,1){1}}\put(0,2.5){\line(1,-1){1}}
\multiput(-0.5,0.5)(0,2){2}{\line(1,0){0.5}}
\multiput(-0.5,0.5)(0,0.3){7}{\line(0,1){0.2}}
\put(6,1.5){\line(1,1){1}}\put(6,1.5){\line(1,-1){1}}
\multiput(7,0.5)(0,2){2}{\line(1,0){0.5}}
\multiput(7.5,0.5)(0,0.3){7}{\line(0,1){0.2}}
\multiput(1,0.5)(1,0){3}{\line(0,1){2}}
\multiput(5,0.5)(1,0){2}{\line(0,1){2}}
\multiput(1,0.5)(0,1){3}{\line(1,0){5}}
\put(0,0.45){\spos{t}{r_1}}\put(1,0.45){\spos{t}{r_1}}
\put(2,0.45){\spos{t}{\sigma_1}}\put(3,0.45){\spos{t}{\sigma_2}}
\put(5,0.45){\spos{t}{\sigma_{N-1}}}\put(6,0.45){\spos{t}{r_2}}
\put(7,0.45){\spos{t}{r_2}}
\put(0,2.6){\spos{b}{r_1}}\put(1,2.6){\spos{b}{r_1}}
\put(2,2.6){\spos{b}{\sigma'_{1}}}\put(3,2.6){\spos{b}{\sigma'_2}}
\put(5,2.6){\spos{b}{\sigma'_{N-1}}}\put(6,2.6){\spos{b}{r_2}}
\put(7,2.6){\spos{b}{r_2}}
\put(1.05,1.45){\spos{tl}{\tau_0}}\put(2.05,1.45){\spos{tl}{\tau_1}}
\put(3.05,1.45){\spos{tl}{\tau_2}}\put(4.99,1.45){\spos{tr}{\tau_{N-1}}}
\put(5.99,1.45){\spos{tr}{\tau_{N}}}
\multiput(1.5,1)(1,0){2}{\spos{}{u}}\put(5.5,1){\spos{}{u}}
\multiput(1.5,2)(1,0){2}{\spos{}{\lambda\!-\!u}}
\put(5.5,2){\spos{}{\lambda\!-\!u}}
\put(0.12,1.5){\makebox(0,0){\scriptsize $ \begin{matrix}r_1 ,\, a_1 \\
\lambda \!-\!u ,\,\xi_1 \end{matrix} $}}
\put(6.9,1.5){\makebox(0,0){\scriptsize $ \begin{matrix}r_2 ,\, a_2 \\
u ,\,\xi_2 \end{matrix} $}}
\multiput(0.0,0.5)(0,2){2}{\makebox(1,0){\dotfill}}
\multiput(6,0.5)(0,2){2}{\makebox(1,0){\dotfill}}
\end{picture}}
\end{equation}
The bulk weights are fixed to their critical values
\begin{equation}\label{bulkw}
\setlength{\unitlength}{8mm}
\W{a}{b}{c}{d}{u}=
\raisebox{-0.7\unitlength}[0.8\unitlength][0.7\unitlength]
{\ \face{a}{b}{c}{d}{u}}
\ = {\sin(\lambda-u)\over \sin\lambda}\,\delta_{a,c}+
{\sin u\over\sin\lambda}\sqrt{S_aS_c\over S_bS_d}\,\delta_{b,d}
\end{equation}
where the heights $a,b,c,d\in\{1,2,3,4\}$ on any edge must respect the
$A_4$ adjacency rule.
The physical range of the spectral parameter is $0<u<\lambda$ where
$\lambda={\pi\over 5}$ is the crossing parameter. The crossing factors
are $S_a=\sin a\lambda / \sin \lambda$.

The most general integrable boundaries for critical bulk weights
are given by (3.28) in \cite{BP}. They are labelled by a pair $(r,a)$ 
and depend upon boundary spins $b$, $c$, $d$, 
the spectral parameter $u$ and a boundary thermodynamic field $\xi$
(interaction)
\begin{equation}
B^{r a}\!\left(c\;\;\;\begin{array}{@{}c|@{\:}}b\\
d\end{array}\;u,\,\xi\right)
\end{equation}
We only need to consider
diagonal boundary weights in the sense that $b=d$. Explicitly, the relevant 
boundary weights for our purposes are
\begin{equation} \label{bweight} \begin{array}{l}
\setlength{\unitlength}{6mm}  \B{r}{a}{c}{c\pm 1}{u}{\xi} =
\raisebox{-1.25\unitlength}{\boundary{c}{c\pm 1}
{\scriptsize $\begin{matrix} r \; a \\  u \; \xi \end{matrix}$}} \\[8mm]
\hspace{25mm} = \displaystyle
\frac{ \left[ \begin{matrix} S_{(r\mp c+a)/2} \, S_{(c\pm a\mp r)/2}
\sin(\xi+u) \, \sin(\lambda r +\xi -u) \hspace{23mm} \\[2mm] \hspace{20mm} +
S_{(r\pm c+a)/2} \, S_{(c\mp a\pm r)/2}\sin(\xi-u)\,\sin(\lambda r +\xi +u)
\end{matrix} \right] }
{\displaystyle \sin \! \lambda ~ S_r \,(S_c\, S_{c\pm 1})^{1/2}\,
\cosh2\im(\xi)}
\end{array}\end{equation}
where $r=1,\ldots,3$ and $a=s=1,\ldots,4$. Here $a$ and $b=d=c\pm 1$ must be 
adjacent at fusion level $r$, that is, $a-b=\pm 1$ for $r=1$, $a+b=4,6$ for $r=2$ and $a+b=5$ for $r=3$. 
The normalization factor chosen here is different from that used in \cite{BP}.
It does not effect any result but simply keeps the boundary weights finite
in the limit $\im(\xi)\rightarrow \pm\infty$.
An obvious and important symmetry is given by
$\xi\rightarrow -\xi-r\lambda$
\begin{equation}
\B{r}{a}{c}{c\pm 1}{u}{\xi}=\B{r}{a}{c}{c\pm 1}{u}{-\xi-r\lambda}.
\end{equation}
A special case is obtained setting $a=1$
\begin{equation} \label{bweight_1}
\setlength{\unitlength}{6mm}
\B{r}{1}{r\pm 1}{r}{u}{\xi}=\raisebox{-1.25\unitlength}[0.8\unitlength]
{\boundary{r\pm 1}{r}{\scriptsize $ u, \; \xi $}}
=\frac{ S_{r\pm 1}^{1/2} \, \sin(\xi\pm u) \, \sin(\lambda r +\xi \mp u) }
{\displaystyle \sin \! \lambda ~ S_r^{1/2} \, \cosh2\im(\xi)}
\end{equation}
In this case we can use the equivalent form of (3.32) in \cite{BP} 
arising from its construction
\begin{equation} \label{bweightalter}
\begin{array}{c}
\B{r}{1}{r\pm 1}{r}{u}{\xi}=- \displaystyle \frac{\sin \lambda}{\cosh2\im(\xi)}
\; \prod_{k=1}^{r-2} \frac{\sin ^{2} \lambda}
{\sin(u-(k+1)\lambda-\xi) \sin(u+k \lambda +\xi)} \\[5mm]
\times \;\displaystyle{\sum_{\{e_i\},\{f_i\}}} \; \displaystyle 
\frac{\displaystyle\sqrt{2 \cos \lambda} \, 
\prod_{m=2}^{r}S_m}{\sqrt{\displaystyle\prod_{m=1}^{r-2}
S_{e_m} S_{f_m} }} ~
\setlength{\unitlength}{13mm}
\raisebox{-1.3\unitlength}
{\hspace{0.2\unitlength}\begin{picture}(4,2.8)
\multiput(0.3,0.3)(1,0){3}{\line(0,1){2}}
\multiput(0.3,0.3)(0,1){3}{\line(1,0){5}}
\multiput(4.3,0.3)(1,0){2}{\line(0,1){2}}
\put(0.26,0.1){$\sm{r}$} \put(-0.17,1.24){$\sm{r \pm 1}$}
\put(1.26,0.1){$\sm{e_1}$}
\put(1.26,2.45){$\sm{f_1}$}
\put(0.26,2.45){$\sm{r}$}
\put(2.26,0.1){$\sm{e_2}$}
\put(2.26,2.45){$\sm{f_2}$}
\put(4.26,0.1){$\sm{e_{r-2}}$}
\put(4.26,2.45){$\sm{f_{r-2}}$}
\put(5.26,0.1){$\sm{1}$} \put(5.4,1.24){$\sm {2}$}
\put(5.26,2.45){$\sm{1}$}
\put(0.8,0.8){\inside{u-\xi}{-(r-1)\lambda}}
\put(0.8,1.8){\inside{-u-\xi}{-(r-2)\lambda}}
\put(1.8,0.8){\inside{u-\xi}{-(r-2)\lambda}}
\put(1.8,1.8){\inside{-u-\xi}{-(r-3)\lambda}}
\put(4.8,1.8){\spos{}{-u-\xi}}
\put(4.8,0.8){\spos{}{u-\xi-\lambda}}
\end{picture}}
\end{array}
\end{equation}
As in \cite{FPR}, we will vary the imaginary part $\im(\xi)$
of the field $\xi$ to interpolate
between the conformal fixed points.

The given bulk and boundary weights
satisfy the Yang-Baxter and Boundary Yang-Baxter equations and lead 
to commuting
double row transfer matrices and an integrable model.
The double row transfer matrices are also periodic, crossing symmetric and
transpose symmetric
\begin{eqnarray}
\mathbf{D}^{N}_{r_1\,a_1|r_2\,a_2}(u,\xi_1,\xi_2) &=&
\mathbf{D}^{N}_{r_1\,a_1|r_2\,a_2}(u+\pi,\xi_1,\xi_2) , \\
\mathbf{D}^{N}_{r_1\,a_1|r_2\,a_2}(u,\xi_1,\xi_2) &=&
\mathbf{D}^{N}_{r_1\,a_1|r_2\,a_2}(\lambda-u,\xi_1,\xi_2) \\
\mathbf{D}^{N}_{r_1\,a_1|r_2\,a_2}(u,\xi_1,\xi_2) &=&
(\mathbf{D}^{N}_{r_1\,a_1|r_2\,a_2}(u,\xi_1,\xi_2))^{t}  \label{transpose}
\end{eqnarray}
for arbitrary complex values of the spectral and boundary parameters.

Moreover, for $a_1=a_2=1$, the normalized transfer matrices
\begin{eqnarray}
\label{normal}
\mathbf{t}(u) & = &
\mathbf{D}^{N}_{r_1\,1|r_2\,1}(u,\xi_1,\xi_2) \,
S_{r_1}(u,\xi_1)\, S_{r_2}(u,\xi_2) \, S(u)\,
\left[ \frac{\sin (u+2\lambda )\sin \lambda }
{\sin (u+3\lambda )\sin (u+\lambda )}\right] ^{2N}
\end{eqnarray}
with
\begin{eqnarray}
S(u)&=&\frac{\sin ^{2}(2u-\lambda )}{\sin (2u+\lambda )\sin (2u-3\lambda)}
\label{normal_s} \\[2mm]
S_{r}(u,\xi) & = &h_{r}(u-\xi)\, h_{-r}(u+\xi) \:
\frac{\cosh2\im(\xi)}{\sin \lambda}
\label{normal_sr} \\[2mm]
h_{r}(u)&=&\frac{\sin \lambda \sin (u+(3-r)\lambda )\sin (u+(1-r)\lambda )}
{\sin u\sin (u-\lambda )\sin (u+2\lambda )}  \label{normal_h}
\end{eqnarray}
satisfy the functional equation
\begin{equation} \label{functional}
\mathbf{t}(u) \: \mathbf{t}(u+\lambda)=
1+ \mathbf{t}(u+3\lambda)
\end{equation}
where we have suppressed the $\xi$ dependence.
Periodicity, crossing symmetry and transpose symmetry extend to this
operator and its factors $S(u)$, $S_{r}(u,\xi)$.
The transfer matrix $\mathbf{D}$ is an entire function of $u$
whereas the normalized transfer matrix $\mathbf{t}$  is a meromorphic 
function with poles
arising from the normalization factor.
We need to determine appropriate
choices of the boundaries to interpolate between distinct conformal 
boundary conditions of types $(r,s)$ by varying the imaginary part of 
$\xi$. We then need to solve the double row transfer matrix 
functional equation in the continuum scaling limit for the induced 
flow of the resulting eigenvalue (energy) spectra.

In the scaling limit, the large $N$ corrections to the
eigenvalues\footnote{We use $D,~t$ to indicate eigenvalues of
$\mathbf{D}, ~\mathbf{t}$.} of
the double row transfer matrix are related to the excitation energies of
the associated perturbed conformal field theory by
\begin{eqnarray} \label{leading}
-\frac{1}{2} \log D^{N}_{r_1\,a_1|r_2\,a_2}(u) &=&
N f_b(u) +f_{r_1\,a_1|r_2\,a_2}(u)+\frac{2\pi \sin 5 u}{N}
\: E_{r_1\,a_1|r_2\,a_2}(\xi) +o(\frac{1}{N})
\end{eqnarray}
where $f_b$ is the bulk free energy, $f_{r_1\,a_1|r_2\,a_2}$ is the surface
(i.e. boundary dependent) free energy and
$E_{r_1\,a_1|r_2\,a_2}(\xi)$ is a scaling function that, at the boundary
critical points, reduces to
\begin{equation}
\left. E_{r_1\,a_1|r_2\,a_2}\right|_{\mbox{\scriptsize crit}}=
-\frac{c}{24}+h +n, \qquad n \in \mathbb{N}
\end{equation}
where $h$ is one of the conformal weights allowed\footnote{If the 
lattice is wrapped on a cylinder, its conformal partition function is 
given by a sum of characters determined by a ``fusion'' of the
boundaries $(r_1\,a_1|r_2\,a_2)$.} by the boundaries
$(r_1\,a_1|r_2\,a_2)$.
Observe that $\mathbf{D}^{N}_{r_1\,a_1|r_2\,a_2}(u,\xi_1,\xi_2)$
in (\ref{leading}) can be a function of up to two boundary
parameters but we find that only one boundary parameter $\xi$ needs 
to flow to reproduce the $\varphi_{1,3}$ thermal boundary flows.

For our computations, we need certain information about the analyticity
of the double row transfer matrices encoded in the zeros and poles of 
their eigenvalues. Because of integrability,
$\mathbf{D}^{N}_{r_1\,a_1|r_2\,a_2}(u,\xi_1,\xi_2)$ at different spectral
parameters $u,v$ forms a commuting family of operators so that its
eigenstates can be chosen independent of the spectral parameter $u$ 
and only the eigenvalues depend on $u$. These eigenvalues are Laurent 
polynomials in the variable $z=\exp(iu)$.
We have performed numerical diagonalization for small sizes (up to
$N=16$ faces in a row). For these sizes, the zeros of the eigenvalues 
are extracted numerically and we extrapolate their pattern to the 
limit of large $N$.
We will refer to  numerical observations obtained in this way
as ``numerics on $D$''.

\subsection{Critical point: classification of the zeros and TBA\label{ss_crit}}
In the scaling limit at the isotropic point $u=\lambda/2$,
$B^{ra}$ is critical if the boundary parameter is
$\xi=\lambda/2$ or $-\lambda/2$. These two choices yield the fixed 
and semi-fixed
boundaries introduced in \cite{SB}. Actually, the parameter can be
chosen in the regions $\xi \in [\lambda/2,(5-r-1/2)\lambda]$,
$\xi \in [(1/2-r)\lambda,-\lambda/2]$ respectively. Indeed, these are the
maximal intervals where each single boundary contribution remains 
nonnegative at the isotropic point.
The ``numerics on $D$'' shows that the scaling properties (in particular
the classification of zeros that will be introduced shortly) do not depend
on the actual value of $\xi$ inside the intervals \cite{OPW}, so in this sense
$\re(\xi)$ is an irrelevant variable close to the critical point.

In general, a cylinder partition function is a superposition of characters. 
However, a single character is obtained if  the double row transfer 
matrix is built~\cite{OPW} with one $r$-type and one $s$-type 
boundary.
The $r$-type boundary is obtained from $B^{r1}$ by
choosing $\xi \in [\lambda/2,(5-r-1/2)\lambda]$ or from $B^{r+1,1}$
by choosing $\xi \in [-(1/2+r)\lambda,-\lambda/2]$.
At the isotropic point
$u=\lambda /2$ and with $\xi=\lambda/2$ we have explicitly (fixed 
boundary, ``$-$" case)
\begin{equation} \label{rtype}
\begin{array}{l}
\B{r}{1}{r-1}{r}{\frac{\lambda}{2}}{\frac{\lambda}{2}}  =
\setlength{\unitlength}{6mm}
\raisebox{-1.25\unitlength}{\boundary{r-1}{r}
{$\scriptstyle \frac{\lambda}{2},\:\frac{\lambda}{2}$}}
= 0  \\[12mm]
\B{r}{1}{r+1}{r}{\frac{\lambda}{2}}{\frac{\lambda}{2}}  =
\setlength{\unitlength}{6mm}
\raisebox{-1.25\unitlength}{\boundary{r+1}{r}
{$\scriptstyle \frac{\lambda}{2},\:\frac{\lambda}{2}$}} =
\sin \lambda \sqrt{S_{r+1} \: S_r}
\end{array} \end{equation}
Similarly, at $\xi=-\lambda/2$ in $B^{r+1,1}$ we have explicitly 
(fixed boundary, ``$+$'' case)
\begin{equation} \label{rtype2}
\begin{array}{l}
\B{r+1,}{1}{r}{r+1}{\frac{\lambda}{2}}{-\frac{\lambda}{2}}  =
\setlength{\unitlength}{6mm}
\raisebox{-1.25\unitlength}{\hspace{-3mm}\boundary{r}{r+1}
{$\scriptstyle \frac{\lambda}{2}\!,-\frac{\lambda}{2}$}}
= \sin \lambda \sqrt{S_{r+1} \: S_r}   \\[12mm]
\B{r+1,}{1}{r+2}{r+1}{\frac{\lambda}{2}}{-\frac{\lambda}{2}}  =
\setlength{\unitlength}{6mm}
\raisebox{-1.25\unitlength}{\boundary{r+2}{r+1}
{$\scriptstyle \frac{\lambda}{2}\!,-\frac{\lambda}{2}$}} = 0
\end{array} \end{equation}
In both the cases we are left with an alternating chain of $r,~r+1$ sites
on the boundary.

The $s$-type boundary weights are obtained as the braid limit of the $r$-type boundary weights with $r=s$
\begin{equation} \label{stype} \begin{array}{r}
\displaystyle \lim_{\mbox{\scriptsize Im} (\xi) \rightarrow - \infty}
\B{s}{1}{s\pm 1}{s}{u}{\xi} =
\setlength{\unitlength}{6mm}
\raisebox{-1.25\unitlength}{\boundary{s\pm 1}{s}{}}
= \sqrt{\frac{S_{s\pm1}}{S_s}} 
~\frac{e^{i(2\mbox{\scriptsize Re}(\xi)+s\lambda)}}
{-2 \sin \! \lambda} \\[7mm]
=\B{1}{s}{s\pm 1}{s}{u}{\xi} \displaystyle \frac{\cosh2\im(\xi)}
{\sin (\xi+u) \sin(\lambda+\xi-u)}
~\frac{e^{i(2\mbox{\scriptsize Re}(\xi)+s\lambda)}}{-2 \sin \! \lambda}
\end{array}
\end{equation}
A similar expression holds for the limit $\im(\xi)\rightarrow +\infty$.
The $u$ and $\im(\xi)$ dependence in the last term cancels out explicitly. 
The overall factor is immaterial for our purposes because it 
corresponds to a trivial factor multiplying the transfer matrix.
Among the $r$- and $s$-type boundaries there is the special common 
boundary condition with $r=s=1$ corresponding to the vacuum.

In the case of one $r$-type and one $s$-type boundary condition which 
leads to the single Virasoro character $\chi_{r,s}(q)$, a detailed 
analysis of the
structure of zeros of the eigenvalues of the double row transfer 
matrices has been performed in \cite{OPW}.
In the large $N$ limit, each
eigenvalue of the double row tranfer matrix is characterized by a specific
pattern of zeros, organized as 1-strings or 2-strings in two 
analyticity strips \cite{OPW}
\begin{equation}
\mbox{Strip 1: \quad} -\frac{\lambda}{2}<\re(u)<\frac{3\lambda}{2}, 
\qquad \qquad
\mbox{Strip 2: \quad} 2\lambda<\re(u)<4 \lambda
\end{equation}
The 1-strings are single zeros
appearing in the center of each strip
\begin{equation}
\re(u)=\frac{\lambda}{2} \mbox{~ or ~} 3\lambda
\end{equation}
while the 2-strings are pairs of zeros $(u,u')$ appearing on the edges of a
strip, with the same imaginary part
\begin{equation}
(\re(u),\re(u'))= \left\{
\begin{array}{lc} (-\lambda/2,3\lambda/2), ~~~& \mbox{strip 1,}\\[2mm]
(2\lambda,4\lambda), & \mbox{strip 2.}
\end{array} \right.
\end{equation}
We use $m_1$, $m_2$ to denote the number of 1-strings in each strip
and $n_1$, $n_2$ to denote the number of 2-strings in each strip.
For each $(r,s)$ boundary condition these numbers satisfy particular 
$(\boldsymbol{m},\boldsymbol{n})$ systems.
The complete classification of the allowed patterns of 1- and 
2-strings in terms of $(\boldsymbol{m},\boldsymbol{n})$ systems and 
quantum numbers is summarized in Table~\ref{critical_TBA}.
\begin{table}[thb]
\caption{\label{critical_TBA}\small
Classification, for all $(r,s)$ boundary conditions of the TIM, of 
the allowed patterns of 1- and 2-strings by 
$(\boldsymbol{m},\boldsymbol{n})$ system and quantum numbers.
The parity $\sigma=\pm 1$ occurs when there are frozen zeros. The 
parities $s_1,s_2=\pm 1$ occur in the TBA equations. The expressions 
for $n_1$,
are only used on a finite lattice because in
the scaling limit $n_1 \sim N/2 \rightarrow \infty$. The number of 
faces in a row is
even or odd according to $N=(r-s)~\mbox{mod}~2 $.}
$$
\begin{array}{|c|l|l|l|}
\hline
\chi_{r,s}(q)&\mbox{$(\boldsymbol{m},\boldsymbol{n})$ 
system}&\mbox{parities}&\mbox{quantum numbers}\\
\hline
  \chi_{1,1}(q) &
\begin{array}{l}  m_{1},\, m_{2} \mbox{~even} \\ n_2=m_1/2-m_2 \\
n_1=(N+m_2)/2-m_1 \end{array}&
\begin{array}{l}  s_{1}=1 \\ s_{2}=1  \end{array} &
\begin{array}{l}
  n^{(1)}_{k}=2(I_{k}^{(1)}+m_{1}-k)+1-m_{2} \\[2mm]
  n^{(2)}_{k}=2(I_{k}^{(2)}+m_{2}-k)+1-m_{1}
\end{array}\\
\hline
\chi_{1,2}(q) &
\begin{array}{l}
m_{1} \mbox{~odd}, ~m_{2}  \mbox{~even} \\n_2=(m_1-\sigma)/2-m_2 \\
n_1=(N+m_2+\sigma)/2-m_1 \end{array} &
\begin{array}{l}  s_{1}=1  \\  s_{2}=-1  \end{array} &
\begin{array}{c}
n^{(1)}_{k}=2(I_{k}^{(1)}+m_{1}-k)+1-m_{2}-\sigma  \\[2mm]
n^{(2)}_{k}=2(I_{k}^{(2)}+m_{2}-k)+1-m_{1}+\sigma
\end{array}\\
\hline
\chi_{2,1}(q) &
\begin{array}{l}  m_{1},\, m_{2} \mbox{~odd} \\ n_2=(m_1+1)/2-m_2 \\
n_1=(N+m_2)/2-m_1 \end{array}&
\begin{array}{l}  s_{1}=-1 \\ s_{2}=-1 \end{array}&
\begin{array}{c}
n^{(1)}_{k}=2(I_{k}^{(1)}+m_{1}-k)+1-m_{2} \\[2mm]
n^{(2)}_{k}=2(I_{k}^{(2)}+m_{2}-k)+1-m_{1}
\end{array}\\
\hline
\chi_{2,2}(q) &
\begin{array}{l} m_{1} \mbox{~even}, ~m_{2} \mbox{~odd} \\
n_2=(m_1-\sigma+1)/2-m_2 \\ n_1=(N+m_2+\sigma)/2-m_1 \end{array}&
\begin{array}{l} s_{1}=-1 \\ s_{2}=1 \end{array}&
\begin{array}{c}
n^{(1)}_{k}=2(I_{k}^{(1)}+m_{1}-k)+1-m_{2}-\sigma  \\[2mm]
n^{(2)}_{k}=2(I_{k}^{(2)}+m_{2}-k)+1-m_{1}+\sigma
\end{array}\\
\hline
\chi_{3,1}(q) &
\begin{array}{l} m_{1} \geqslant 2 \mbox{~even}, ~m_{2} \mbox{~odd} \\
n_2=m_1/2-m_2 \\ n_1=(N+m_2+1)/2-m_1 \end{array} &
\begin{array}{l}  s_{1}=1  \\  s_{2}=-1 \end{array}&
\begin{array}{c}
n^{(1)}_{k}=2(I_{k}^{(1)}+m_{1}-k)+1-m_{2} \\[2mm]
n^{(2)}_{k}=2(I_{k}^{(2)}+m_{2}-k)+1-m_{1}
\end{array}\\
\hline
\chi_{3,2}(q) &
\begin{array}{l} m_{1},\, m_{2} \mbox{~odd} \\ n_2=(m_1-\sigma)/2-m_2 \\
n_1=(N+m_2+\sigma+1)/2-m_1 \end{array} &
\begin{array}{l} s_{1}=1 \\ s_{2}=1 \end{array}&
\begin{array}{c}
n^{(1)}_{k}=2(I_{k}^{(1)}+m_{1}-k)+1-m_{2}-\sigma  \\[2mm]
n^{(2)}_{k}=2(I_{k}^{(2)}+m_{2}-k)+1-m_{1}+\sigma
\end{array}\\
\hline
\end{array} $$
\end{table}

For a generic function $h(u)$, for example the normalized transfer 
matrix eigenvalues $t(u)$, it is natural to introduce two functions 
$h_1(x)$, $h_2(x)$ with reference to the center lines of the two 
analyticity strips
\begin{eqnarray} \label{h1}
h_1(x)=h(\frac{\lambda}{2}+i\frac{x}{5}) & &\mbox{for }|\im(x)|\leqslant
\pi\\[2mm]
\label{h2}
h_2(x)=h(3\lambda+i\frac{x}{5}) & & \mbox{for } |\im(x)| \leqslant \pi
\end{eqnarray}
At the critical point of the TIM, the TBA equations derived in 
\cite{OPW} for the six $(r,s)$ conformal boundary conditions in 
Table~\ref{critical_TBA} are
\begin{eqnarray}
\epsilon _{1}(x) & = & -\sum _{k=1}^{m_{1}}\log \tanh 
(\frac{y_{k}^{(1)}-x}{2})-K*\log (1+s_{2}e^{-\epsilon _{2}(x)})\label{OPW1}\\
\epsilon _{2}(x) & = & 4e^{-x}-\sum _{k=1}^{m_{2}}\log \tanh 
(\frac{y_{k}^{(2)}-x}{2})-K*\log (1+s_{1}e^{-\epsilon _{1}(x)})\label{OPW2}
\end{eqnarray}
where  $y_k^{(j)}$ are the (scaled) locations of the 1-strings in the
strip $j=1,2$, that is to say, they are zeros of the normalized (and scaled)
transfer matrices $\hat{t}_i(x)=s_i e^{-\epsilon_i (x)}$.
The parities $ s_{i} $ are given in Table~\ref{critical_TBA}.
The locations of the zeros are determined by a set of {\em 
non-degenerate quantum numbers}
$n_k^{(i)}\in \mathbb{Z}$ by the quantization conditions
\begin{eqnarray}
\epsilon _{1}(y_k^{(2)}-i\frac{\pi}{2}) & = & i\pi n_k^{(2)},\qquad 
k=1,2,\ldots,m_2  \\
\epsilon _{2}(y_k^{(1)}-i\frac{\pi}{2}) & = & i\pi n_k^{(1)},\qquad 
k=1,2,\ldots,m_1
\end{eqnarray}
The {\em non-negative quantum numbers} $\{I_k^{(j)}\}$ have
the topological meaning that
for a given 1-string $y_k^{(j)}$, $I_k^{(j)}$ is the number of 2-strings
with larger coordinate $z_l^{(j)}>y_k^{(j)}$.
The conventional order of the zeros is
$y_1^{(j)}<y_2^{(j)}<\ldots <y_{m_j}^{(j)}$.
The notation
$(I^{(1)}_1,\ldots,I^{(1)}_{m_1}|I^{(2)}_1,\ldots,I^{(2)}_{m_2})_\sigma$
uniquely labels states.

The corresponding expression for the scaling energy (\ref{leading}) 
is given by \cite{OPW}
\begin{eqnarray} \label{energy}
E_{rs}&=&\frac{2}{\pi} \sum_{k=1}^{m_1} e^{-y_k^{(1)}}
- \int _{-\infty} ^{+\infty} \frac{dy}{\pi^2} \, e^{-y} \,
\log(1+s_2 e^{-\epsilon_2(y)} )\\ \nonumber
&=&- \frac{c}{24}+h_{rs}+\frac{(r-s)(s-r-1)}{4}
+\frac{\boldsymbol{m}^{t}C\boldsymbol{m}}{4}
-\frac{\boldsymbol{A}\cdot\boldsymbol{m}}{2}+\sum_{j=1}^{2}
\sum_{k=1}^{m_{j}} I_k^{(j)}
\end{eqnarray}
where $C$ is the $A_2$ Cartan matrix,
$\boldsymbol{A}$ is a vector
\begin{equation}
C=\begin{pmatrix} 2 & -1 \\ -1 & 2 \end{pmatrix},  \qquad
\boldsymbol{A}= (1,-1)\,\sigma
\end{equation}
and $\sigma=\pm 1$ according to Table~\ref{critical_TBA} if $s=2,3$
and $\sigma=0$ otherwise.
The conformal partition function obtained from these
energies yields a
finitized fermionic character
\begin{equation}
Z_N(q)=\sum _{\mbox{\small all states}} q^{E_{rs}}=
\chi_{r,s}^{(N)}(q)
\mathop{\longrightarrow}_{\scriptstyle N \rightarrow \infty} \chi_{r,s}(q)
\end{equation}
where the sum is over all states allowed by the finite 
$(\boldsymbol{m},\boldsymbol{n})$ system and
\begin{equation}
\chi_{r,s}^{(N)} (q)=q^{-\frac{c}{24}+h_{rs}+\frac{1}{4}(r-s)(s-r-1)}
\sum_{m_1, m_2,\sigma} q^{\frac{1}{4}\boldsymbol{m}^{t}C\boldsymbol{m}
-\frac{1}{2}\boldsymbol{A}\cdot\boldsymbol{m}}
\left[ {m_1+n_1-\delta_{\sigma,1} \atop m_1-\delta_{\sigma,1}} \right]
\left[ {m_2+n_2 \atop m_2} \right]
\end{equation}
with the previous convention on $\sigma$.

\subsection{Boundary flows\label{ss_bound}}
The TIM admits 6 Cardy-type boundary
conditions each corresponding to a single Virasoro character.
Interpolating between them, as described in \cite{affleck} and shown 
in Figure~\ref{fig-flows},
there are 7 integrable renormalization group flows, 5 generated by the
perturbing operator $\varphi_{1,3}$ and 2 by $\varphi_{1,2}$.
In this paper we consider only those described by the $A_4$ RSOS
lattice model of \cite{ABF}, that is to say, the $\varphi_{1,3}$ flows.
We expect that the $\varphi_{1,2}$ boundary flows will be given by 
consideration of the dilute $A$ RSOS lattice realization of the TIM.
There is an important reason why $A_4$ cannot describe both the perturbations,
namely, the existence of two independent perturbing parameters would
suggest the presence of an integrable surface instead of just the
integrable line that is known to exist.

 From the $A_4$ lattice description of the TIM we observe the presence 
of similarities
between the $\varphi_{1,3}$ flows related by inversion of the spin direction
in the Affleck \cite{affleck} description. Moreover, from the conformal
description, we also notice that three flows share the same ``active''
boundary, namely an $s=2$ type boundary becomes an $s=1$ type boundary,
while the $r$-type boundary is just a spectator. We therefore divide the 
flows according
to whether $r$ remains constant or not
$$
\begin{array}{cc@{\hspace{10mm}}cc}
\mbox{constant $r$:} &
\begin{array}{c@{~ \mapsto ~}c}
\chi_{1,2} \equiv  (0-)  &  \chi_{1,1} \equiv (-)  \\
\chi_{3,2} \equiv  (0+) &  \chi_{3,1} \equiv (+) \\
\chi_{2,2} \equiv  (d) &  \chi_{2,1} \equiv (0)
\end{array} &
\mbox{variable $r$:} &
\begin{array}{c@{~ \mapsto ~}c}
\chi_{1,2} \equiv  (0-) &  \chi_{2,1} \equiv (0) \\
\chi_{3,2} \equiv  (0+) &  \chi_{2,1} \equiv (0)
\end{array}
\end{array}
$$
We will see that these similarities also arise in our TBA description
of the flows.

\section{Boundary flows with constant $r$}
The common feature of this family of flows is the change from an $s=2$ to
an $s=1$ label in the conformal character $\chi_{r,s}(q)$, keeping $r$ fixed.
Clearly, on the lattice this means that the boundary of type $r$
is a spectator and the dynamics is given by the $s$ type boundary.

For convenience, from now on the complex boundary thermodynamic field
on the lattice will be denoted $\xilatt$. From (\ref{stype}) we know that
$B^{2,1}(u,\,\xilatt)$ in
the limit $\im(\xilatt)\rightarrow \pm \infty$
(and for all $\re(\xilatt) \in\mathbb{R}$)
is an $s=2$ type boundary.
We can take advantage of the fact that $\re(\xilatt)$ is irrelevant
at criticality
to fix it to a convenient value. Using (\ref{bweight}) and (\ref{bweightalter})
we have
\begin{equation} \label{boundary21}
\B{2,}{1}{2\pm 1}{2}{u}{\xilatt}=
\setlength{\unitlength}{10mm}
\raisebox{-1.3\unitlength}{\hspace{-3mm}\boundary{2 \pm 1}{2}{$u,\,\xilatt$}}
= \setlength{\unitlength}{10mm}
\raisebox{-1.3\unitlength}{\hspace{3mm}\begin{picture}(3.2,2.7)(0,0.1)
\multiput(0.5,0.5)(1,0){2}{\line(0,1){2}}
\multiput(0.5,0.5)(0,1){3}{\line(1,0){1}}
\put(1.5,1.5){\line(1,1){1}}\put(1.5,1.5){\line(1,-1){1}}
\multiput(2.5,0.5)(0,2){2}{\line(1,0){0.5}}
\multiput(3,0.5)(0,0.3){7}{\line(0,1){0.2}}
\multiput(1.5,0.5)(0,2){2}{\makebox(1,0){\dotfill}}
\put(0.5,0.1){\spos{}{2}} \put(0.1,1.5){\spos{}{2 \pm 1}}
\put(1.5,0.1){\spos{}{1}}
\put(1.5,2.8){\spos{}{1}}
\put(0.5,2.8){\spos{}{2}}
\put(1.03,1){\makebox(0,0){\scriptsize$\begin{array}{l} u-\lambda \\
-\xilatt \end{array}$}}
\put(1.03,2){\makebox(0,0){\scriptsize$\begin{array}{l} -u \\ -\xilatt
\end{array}$}}
\put(1.8,1.5){\spos{}{2}}
\put(2.5,2.8){\spos{}{1}}
\put(2.5,0.1){\spos{}{1}}
\end{picture}} 
\frac{-\sin^2\lambda}{\sin(\xilatt+u)\sin(\lambda+\xilatt-u)}
\end{equation}
and we see that if $\re(\xilatt)=-\lambda$, $\im(\xilatt)=0$,
this boundary weight can be interpreted as the addition of a column to the
lattice in presence of a $B^{1,1}$ boundary that is of
$s=1$ type. So, on the lattice, the desired flow is driven by $\im(\xilatt)$
and can be summarized as
\begin{equation} \label{boundary21-lim}
B^{2,1}(u,-\lambda+i\im(\xilatt))= \left\{ \begin{array}{c@{,\hspace{5mm}}lc}
s=2 & \im(\xilatt) =\pm\infty & \\[3mm]
s=1 & \im(\xilatt) =0, & 1\mbox{ column added}
\end{array} \right.
\end{equation}
We form a double row transfer matrix
$\mathbf{D}_{r1|2,1}^N (u,-\lambda+i\im(\xilatt))$
by coupling $B^{2,1}$ with an $r$-type boundary on the left having no free
boundary parameters. We are left with just one boundary parameter $\xilatt$.

By (\ref{boundary21-lim}), if $\im(\xilatt)=\pm \infty$, on the right
we have an $s=2$-type boundary coupled with an $r$ type on the left,
leading to the character $\chi_{r,2}$.
If $\im(\xilatt)=0$, the $s=1$ boundary on the right is coupled with an
$r$-type on the left, corresponding to $\chi_{r,1}$.
We now introduce a simplified notation $\xilatt=-\lambda+i\xi/5$ with 
$\xi$ real and
\begin{equation} \label{Dflow}
\mathbf{D}(u,\xi)\equiv\mathbf{D}_{r,1|2,1}^N (u,-\lambda+i\xi/5)
\end{equation}
If we explicitly compute $B^{2,1}(u,-\lambda+i\xi/5)$ we see that it is real
for arbitrary $\xi$ and real $u$. The boundary weight on the
left and the bulk weights are also real so each single entry of the
transfer matrix is real for real $u$. This is nothing but the
real analyticity property\footnote{We use $~^*$ to indicate complex 
conjugation.}
\begin{equation}
\mathbf{D}_{r,1|2,1}^N (u,-\lambda+i\xi/5)=
(\mathbf{D}_{r,1|2,1}^N (u^{*},-\lambda+i\xi/5)) ^{*}, \qquad \mbox{for all}
~\xi.
\end{equation}
The same property holds for the normalized transfer matrix.
Combining real analyticity, periodicity and crossing symmetry
and using the notation introduced in (\ref{h1}) and (\ref{h2}) we find
the following reality condition for $x\in \mathbb{R}$
\begin{eqnarray}
\mathbf{D}(\frac{\lambda}{2}+i\frac{x}{5},\xi) &=&
(\mathbf{D}(\frac{\lambda}{2}+i\frac{x}{5},\xi))^{*} , \\
\mathbf{D}(3\lambda+i\frac{x}{5},\xi)&=&
(\mathbf{D}(3\lambda+i\frac{x}{5},\xi))^{*}, \\
\mathbf{t}_j(x,\xi)&=&\mathbf{t}_j(x,\xi)^{*}. \label{t_real}
\end{eqnarray}
so that the transfer matrices are real along the center line of each strip.
They also are transpose symmetric (\ref{transpose}); this implies that they
are real symmetric and thus they have real eigenvalues leading to 
real energies. The previous equations are also true for the 
eigenvalues.
Of course, if $\re(\xilatt)\neq -\lambda$, the boundary weights are no
longer real, the reality condition is lost, the transfer matrices
become complex and so too do the scaling energies. Again, this is
confirmed by the ``numerics on $D$''.


We now proceed with the flow $\chi_{1,2} \mapsto \chi_{1,1}$ which is 
the prototype among
the flows in this family and so we describe it in detail.

\subsection{RG flow $\chi_{1,2} \mapsto \chi_{1,1}$\label{ss_1211}}
 From the previous discussion, the relevant double row transfer matrix is
$\mathbf{D}(u,\xi)\equiv\mathbf{D}_{1,1|2,1}^N (u,-\lambda+i\xi/5)$
containing one boundary thermodynamic field $\xi$.
It is convenient to restrict to $\xi \leqslant 0$ so that
$\xi=-\infty$ corresponds to the
ultraviolet (UV) fixed point ($\chi_{1,2}$) and $\xi =0$ to the 
infrared (IR) fixed point ($\chi_{1,1}$).
The adjacency rules force $N$ to be odd.
The addition of a column implemented by the boundary interaction
(\ref{boundary21}) is consistent with the different parity of the number of
faces in $\chi_{1,2}$ and $\chi_{1,1}$, as  required by the adjacency rules
\begin{equation}\label{parity12}
N\uv, \mbox{~odd} \longmapsto N\ir=N\uv+1, \mbox{~even}
\end{equation}

\subsubsection{Two mechanisms for changing the string content\label{mech1211}}
As in \cite{FPR}, it is sometimes convenient mathematically to consider
  flows which are the reverse of the physical flows. In this section 
the actual flow we consider is
$\mbox{IR}=\chi_{1,1} \mapsto \chi_{1,2}=\mbox{UV}$.

\begin{figure}[t]
\begin{center}
\includegraphics[width=0.31\linewidth]{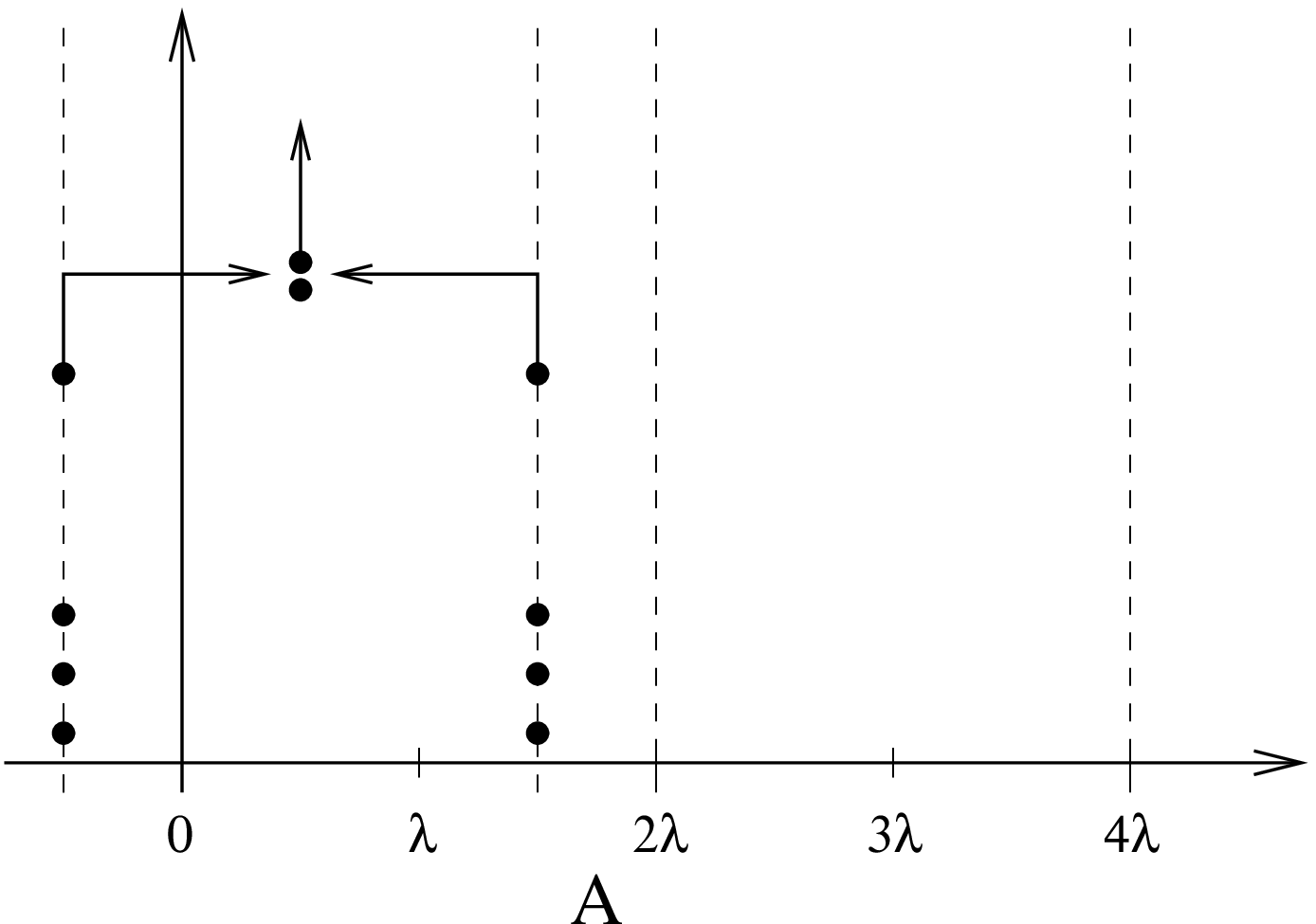}
\hspace{15mm}\includegraphics[width=0.31\linewidth]{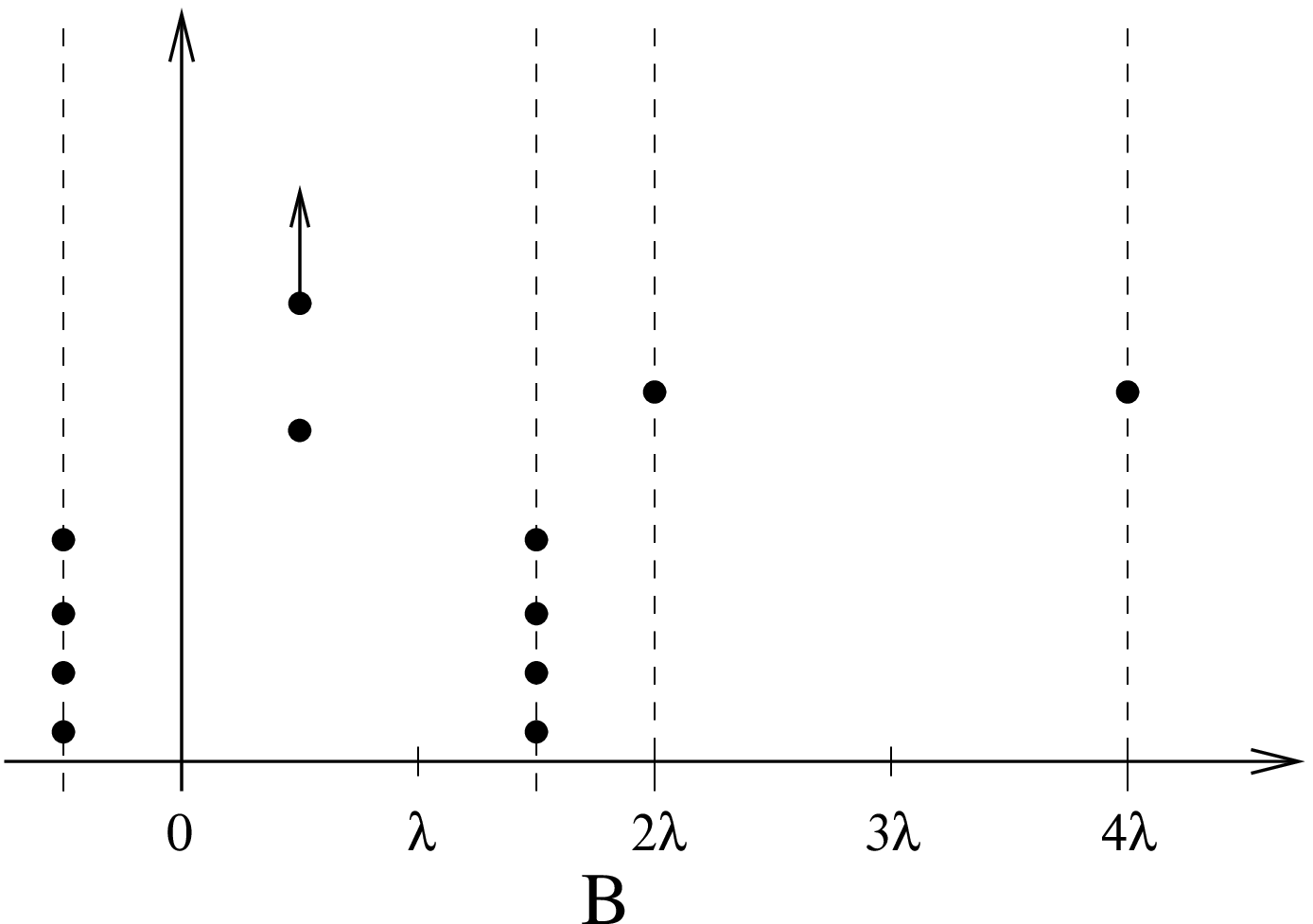}
\end{center}
\caption{\small \label{ABmech} The two mechanisms A, B respectively
that change string content during the flow
$\chi_{1,1} \mapsto \chi_{1,2}$.
Note that this is the reverse of the physical flow.
These mechanisms are illustrated in the upper-half $u$-plane for the states:
A, $(~)\,\mapsto\,(0)_{+}$;
B, $(0\,0)\,\mapsto\,(0)_{-}$.}
\end{figure}

The pattern of zeros of the eigenvalues $D(u,\xi)$ change
along the flow interpolating the extreme configurations described in
Table~\ref{critical_TBA}.
In particular, $m_1$ must change its parity from even to odd.
Using (\ref{bulkw}), a simple counting of powers of the variable $\exp(i u)$
shows that $D(u,\xi)$
is a polynomial of order $2(N+1)$  for $\xi \leqslant 0$ ($\chi_{1,1}$
is included in this range) and becomes a polynomial of order $2N$ in the
limit  $\xi \rightarrow -\infty$.
So, inside the periodicity strip
$-\lambda <\re(u) \leqslant 4\lambda$, precisely two zeros must ``escape to
infinity'' during the flow, one in the upper- and one in the 
lower-half $u$-plane.
Indeed, from the ``numerics on $D$'' we find just two
mechanisms for changing the string content during the flow.
As shown in Figure~\ref{ABmech},
these involve the objects (1 or 2-string) which are furthest from the 
real axis in strip 1 at the IR point.
Strip~2 is not directly involved.
\begin{list}{}{}
\item[A.]
When the furthest object in strip 1 is a 2-string
it moves away from the real axis, heading to $+\infty$; then it collapses
into a pair of 1-strings; one of them disappears moving to $+\infty$ 
(it is actually reached only at the UV point), the other remains in 
the scaling
region and becomes the top 1-string in strip 1.
Of course, it cannot have a 2-string above it, so it is frozen and in the
UV $\sigma=1$. This pair of 1-strings, called {\em correlated 
1-strings}, has special features
that will be explained in the numerical section.
The disappearance of the top 2-string leads to a decrease
by one unit in the quantum numbers in strip 1.
\item[B.]
If the furthest object in strip 1 is a 1-string it moves to $+\infty$.
The remaining 1-strings can be excited and are therefore not frozen, 
so in the UV we have
  $\sigma=-1$.
\end{list}
This qualitative description of the mechanisms implies the following 
changes in the values of the quantum numbers
\begin{eqnarray}
\mbox{A: }~~ {I^{(1)}_{m_1\irt}}\ir \geqslant 1:&&\qquad
\begin{array}{rcl}
m_1\ir & \mapsto  & m_1\uv=m_1\ir+1, ~~~ \sigma=+1 \\[2mm]
{I^{(1)}_k}\ir & \mapsto & {I^{(1)}_k}\uv={I^{(1)}_k}\ir-1, ~~~
{I^{(1)}_{m_1\uvt}}\uv=0  \label{mechanismA} \\[2mm]
n_1\ir & \mapsto  & n_1\uv=n_1\ir-1
\end{array}\\[3mm]
\mbox{B: }~~ {I^{(1)}_{m_1\irt}}\ir=0:&&\qquad\quad
m_1\ir \  \mapsto  \
m_1\uv=m_1\ir-1, ~~~ \sigma=-1  \label{mechanismB}
\end{eqnarray}
Observe that the change of parity of $m_1$ is consistent with the
known parities at the two endpoints of the flow, as in 
Table~\ref{critical_TBA}.
The consistency and completeness of these two mechanisms is demonstrated by
the explicit mapping between the finitized characters associated with the UV
and IR fixed points, as shown in the next section. Moreover,
we observe these same mechanisms when solving the TBA equations.

The application of mechanisms A and B to the first few 
states is summarized in Table~\ref{t_s1211}.

\begin{table}[htbp]
\caption{\label{t_s1211}
\small
Flow $\chi_{1,1} \mapsto \chi_{1,2}$ (reverse of the physical flow).
We present the explicit mapping of states from IR to UV up to the UV 
level 6. Here
$n\ir,\:n\uv$ are the excitation levels
above the ground states, respectively $h=0$ and $h=1/10$.}
\begin{center}
\begin{tabular}{|c|r@{$\:\:\mapsto\:\:$}l|c|c||c|r
@{$\:\:\mapsto\:\:$}l|c|c|}
\hline
$n\ir$\rule[-4mm]{0mm}{10mm}& \multicolumn{3}{c|}{Mapping of states --
mechanism} & $n\uv$ &
   $n\ir$ & \multicolumn{3}{c|}{Mapping of states -- mechanism} & $n\uv$\\
\hline
0\rule[-1mm]{0mm}{6mm} & $(\,)$ & $(0)_{+}$ & A & 0 &
           6 & $(4\,0)$ & $(4)_{-}$ & B & 5 \\[1mm]
2 & $(0\,0)$ & $ (0)_{-}$ & B & 1 &
           6 & $(3\,1)$ & $(2\,0\,0)_{+}$ & A & 5 \\[1mm]
3 & $(1\,0)$ & $(1)_{-}$ & B & 2 &
           7 & $(1\,0\,0\,0|0\,0)$ & $(1\,0\,0|0\,0)_{-}$ & B &5 \\[1mm]
4 & $(1\,1)$ & $(0\,0\,0)_{+}$ & A & 3 &
           7 & $(3\,2)$ & $(2\,1\,0)_{+}$ & A & 6 \\[1mm]
4 & $(2\,0)$ & $(2)_{-}$ & B & 3 &
           7 & $(4\,1)$ & $(3\,0\,0)_{+}$ & A & 6 \\[1mm]
5 & $(2\,1)$ & $(1\,0\,0)_{+}$ & A & 4 &
           7 & $(5\,0) $ & $(5)_{-}$ & B & 6 \\[1mm]
5 & $(3\,0)$ & $(3)_{-}$ & B & 4 &
           8 & $(0\,0\,0\,0)$ & $(0\,0\,0)_{-}$ & B & 6 \\[1mm]
6 & $(0\,0\,0\,0|0\,0)$ & $(0\,0\,0|0\,0)_{-}$ & B & 4 &
           8 & $(1\,1\,0\,0|0\,0)$ & $(1\,1\,0|0\,0)_{-}$ & B & 6 \\[1mm]
6 & $(2\,2)$ & $(1\,1\,0)_{+}$ & A & 5 &
           8 & $(2\,0\,0\,0|0\,0)$ & $(2\,0\,0|0\,0)_{-}$ & B & 6 \\[1mm]
\hline
\end{tabular}
\end{center}
\end{table}

\subsubsection{RG mapping between finitized characters\label{sss_map1211}}
In this section we show that the two mechanisms A, B are compatible with the
counting of states at the two endpoints of the flow, given by the finitized
characters.

We observe in (\ref{mechanismA}) and (\ref{mechanismB}) that, 
starting in the IR, each
state changes unambiguously by precisely one of the
two possible mechanisms so that the counting of states is complete.
Moreover, the IR finitized character naturally splits into two terms 
corresponding precisely to
the two mechanisms A, B (see (\ref{recourrence}))
\begin{eqnarray}
\chi_{1,1}^{(N\irt)}(q) & = &
q^{-\frac{c}{24}} \sum_{m_1\irt,\,m_2}
q^{\frac{1}{4}{\boldsymbol{m}\irt}C\boldsymbol{m}\irt}
\gausst{m_1\irt+n_1\irt}{m_1\irt} \gausst{m_2+n_2}{m_2} \nonumber \\
& = &
q^{-\frac{c}{24}}\sum_{\text{A}}
q^{\frac{1}{4}{\boldsymbol{m}\irt}C\boldsymbol{m}\irt}
q^{m_1\irt} \gausst{m_1\irt+n_1\irt-1}{m_1\irt}
\gausst{m_2+n_2}{m_2} \label{char11} \\
& + & q^{-\frac{c}{24}}\sum_{\text{B}}
q^{\frac{1}{4}{\boldsymbol{m}\irt}C\boldsymbol{m}\irt}
\gausst{m_1\irt+n_1\irt-1}{m_1\irt-1} \gausst{m_2+n_2}{m_2}
\nonumber
\end{eqnarray}
where we attach the label IR or UV only to the variables that change 
under the flow. The labels A and B  on the sums
indicate that the sums on $m_1\irt$, $m_2$ are
restricted by the constraints imposed by the two mechanisms
(\ref{mechanismA}) and (\ref{mechanismB}).
This can be understood by using (\ref{Ipos}) for the first strip in the
second line of (\ref{char11}) so that the sum on $m_1\ir \geqslant 1$
required by the mechanism A is manifest.
Similarly, the factor related to the first strip in the last line of
(\ref{char11}) can be rewritten using (\ref{Izero}) as a sum with the
constraint $I_{m_1} = 0$ required by mechanism B.

The single IR energy level at the base of a tower of states fixed by
the string content $(m_1,m_2)$ maps to a UV energy level according to
\begin{equation} \label{energyAB}
\begin{array}{l@{\hspace{7mm}}r@{~~\mapsto~~}c}
\mbox{A:} & q^{-\frac{c}{24}} \,
q^{\frac{1}{4}\boldsymbol{m}\irt C\boldsymbol{m}\irt} \,q^{m_1\irt} \,
& q^{-\frac{c}{24} + \frac{1}{10}} \,
q^{\frac{1}{4}\boldsymbol{m}\uvt C\boldsymbol{m}\uvt} \,
q^{-\frac{1}{2}(m_1\uvt-m_2)}  \
\\[2mm]
\mbox{B:} & q^{-\frac{c}{24}}  \,
q^{\frac{1}{4}\boldsymbol{m}\irt C\boldsymbol{m}\irt} \,
& q^{-\frac{c}{24} + \frac{1}{10}} \,
q^{\frac{1}{4}\boldsymbol{m}\uvt C\boldsymbol{m}\uvt} \,
q^{+\frac{1}{2}(m_1\uvt-m_2)}
\end{array}
\end{equation}
(these mappings of energies are fixed by the known expression for the
energies at the IR and UV critical points).
Using (\ref{mechanismA}) and (\ref{mechanismB}), the mapping of the 
$q$-binomials (counting
polynomials) is given by
\begin{equation} \label{countingAB}
\begin{array}{l@{\hspace{7mm}}l}
\mbox{A:} &
\gausst{m_1\irt+n_1\irt-1}{m_1\irt}  =
\gausst{m_1\uvt+n_1\uvt-1}{m_1\uvt-1}= \gausst{m_1\uvt+n_1\uvt-
\delta_{1,\sigma}}{m_1\uvt-\delta_{1,\sigma}}
\\[5mm]
\mbox{B:} &
\gausst{m_1\irt+n_1\irt-1}{m_1\irt-1} =
\gausst{m_1\uvt+n_1\uvt}{m_1\uvt}= \gausst{m_1\uvt+n_1\uvt-
\delta_{1,\sigma}}{m_1\uvt-\delta_{1,\sigma}}
\end{array}
\end{equation}
Combining
(\ref{char11}) to (\ref{countingAB})
we obtain the finitized UV character
\begin{eqnarray}
\chi_{1,1}^{(N\irt)}(q) & \mapsto &
q^{-\frac{c}{24} + \frac{1}{10}}
\sum_{\sigma,m_1\uvt,m_2}
q^{\frac{1}{4}\boldsymbol{m}\uvt C\boldsymbol{m}\uvt}
q^{-\frac{1}{2}(m_1\uvt-m_2)\sigma}
\gausst{m_1\uvt+n_1\uvt-\delta_{1,\sigma}}{m_1\uvt-\delta_{1,\sigma}}
\gausst{m_2+n_2}{m_2} \nonumber \\
& = & \chi_{1,2}^{(N\uvt)}(q)
\end{eqnarray}
This shows that the completeness of the IR counting of states, together with
the mechanisms A and B, implies the consistency and
completeness of the UV counting.

\subsubsection{Solution of the functional equations\label{sss_func}}
To solve the functional equation for the eigenvalues of the double 
row transfer matrix  $\mathbf{D}(u,\xi)\equiv\mathbf{D}_{1,1|2,1}^N 
(u,-\lambda+i\xi/5)$, we follow steps similar to those in \cite{OPW} 
but taking particular care in
managing the terms containing $\xi$.
The functional equation can be rewritten using (\ref{h1}) and 
(\ref{h2}) as a coupled
system between the two analyticity strips
\begin{eqnarray}
\label{t1system}
t_1(x+i\frac{\pi}{2}) \: t_1(x-i\frac{\pi}{2}) &=& 1+t_2(x) \\[2mm]
\label{t2system}
t_2(x+i\frac{\pi}{2}) \: t_2(x-i\frac{\pi}{2}) &=& 1+t_1(x)
\end{eqnarray}
\begin{equation}
\label{interval}
\im(x) \in (-\frac{\pi}{2},\frac{\pi}{2})
\end{equation}
These equations can be solved by taking the Fourier transform
of the logarithmic derivative of the equations, taking care to remove 
all of the
zeros and poles that can generate a singularity in $\log t_j(x)$.
It is convenient to consider separately the order $N$, order 1 and 
order $1/N$ contributions to $t_j(x)$.

The analyticity of the order $N$ contribution is contained in the 
last factor of
(\ref{normal}), leading to a zero of order $2N$ at $u=3\lambda$
and to two poles of the same order at $u=2\lambda, ~4 \lambda$.
These are at the edges of the second strip so
they do not effect the RHS terms $1+t_1$ and $1+t_2$
within (\ref{interval}).
The order $N$ contribution $f(u)$ is thus given by the inversion relations
\begin{eqnarray}\label{f1_def}
f_1(x+i\frac{\pi}{2}) \: f_1(x-i\frac{\pi}{2}) & = & 1 \\[2mm]
\label{f2_def}
f_2(x+i\frac{\pi}{2}) \: f_2(x-i\frac{\pi}{2}) & = & 1
\end{eqnarray}
The unique solution~\cite{KlumpP} of these equations with the required analyticity properties is
\begin{eqnarray}
f_1(x) & = &  1 \\
f_2(x) & = & \tanh^{2N} \frac{x}{2}
\end{eqnarray}
If we divide (\ref{t1system}), (\ref{t2system}) by (\ref{f1_def}), 
(\ref{f2_def}),
we obtain a new system of equations containing the function
$t_i(x)/f_i(x)$ on the LHS and free of order $N$ zeros and poles in
(\ref{interval})
\begin{eqnarray}
\label{t1f1system}
\frac{t_1}{f_1}(x+i\frac{\pi}{2}) \: \frac{t_1}{f_1}(x-i\frac{\pi}{2})
&=& 1+t_2(x) \\[2mm]
\label{t2f2system}
\frac{t_2}{f_2}(x+i\frac{\pi}{2}) \: \frac{t_2}{f_2}(x-i\frac{\pi}{2})
&=& 1+t_1(x)
\end{eqnarray}

\begin{figure}[t]
\begin{center}
\includegraphics[width=0.6\linewidth]{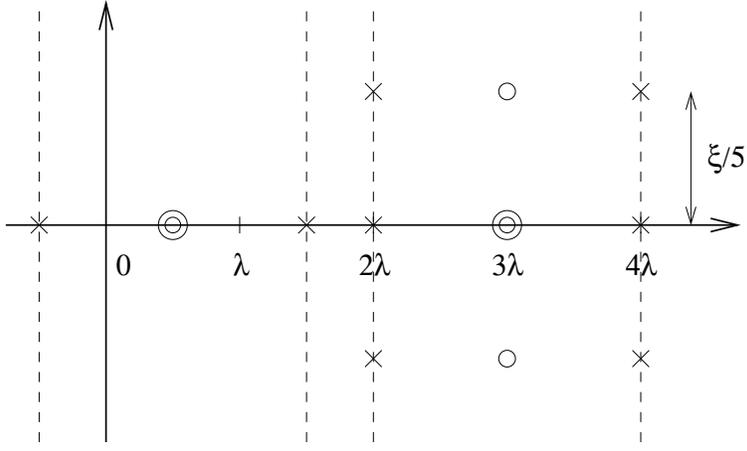}
\end{center}
\caption{\small Analyticity, in the complex $u$-plane, of the order 1 
contribution to the normalized transfer matrix eigenvalues $t$ for the 
flow $\chi_{1,2} \mapsto \chi_{1,1}$.
Circles are zeros and crosses are poles. Double circles are order 2 zeros.
\label{order1anal}}
\end{figure}

The analyticity of the order 1 contribution in (\ref{t1f1system}), 
(\ref{t2f2system}), represented in Figure \ref{order1anal}, is 
specific to each boundary condition, being
contained in the factors $S(u)$,
$S_{1}(u)$ and $S_{2}(u,\xilatt)$  of (\ref{normal}).
$S(u)$ leads $t_1(x)$ to have a double zero at $x=0$ and two poles at
$x=\pm i\pi$ and leads $t_2(x)$ to have a double zero at $x=0$ and two poles
at $x=\pm i\pi$, that do not depend on the boundary parameter.
$S_{1}(u)$ simplifies with an analogous term in
the transfer matrix generated by $B^{1,1}$ yielding a constant term to the
normalized transfer matrices $t_j(x)$.
$S_{2}(u,\xilatt)$ is crucial and specific for this family
of boundary flows (in general it will be $S_{r}(u,\xilatt)$ with
$\re(\xilatt) \neq -\lambda $)
\begin{equation} \label{s2}
S_{2}(u,\xilatt)=\frac{\sin \!\lambda \, \sin(\xilatt-u-\lambda)\,
\sin(\xilatt+u+3\lambda)\,\cosh 
2\im(\xilatt)}{\sin(\xilatt-u)\,\sin(\xilatt-u-2\lambda)
\,\sin(\xilatt+u-\lambda)\,\sin(\xilatt+u+2\lambda)}
\end{equation}
We see that $S_{2}(u,-\lambda+i\xi/5)$ has zeros and poles in the second
strip only:
two single zeros at $u=3\lambda \pm i\xi/5$ ($x=\pm \xi$) and four single
poles at $u=2\lambda \pm i\xi/5$, $u=4\lambda \pm i\xi/5$
($x=\pm i\pi \pm \xi$). The factor $\cosh(2\im(\xilatt))$ is compensated by
the corresponding normalization assumed in (\ref{bweight}) and always
disappears from the equations.
Observe that the given poles of order 1 are on the edges of the two 
analyticity strips
so $1+t_2(x)$ and $1+t_1(x)$ in (\ref{t1system}) and (\ref{t2system})
are free of poles inside (\ref{interval}).
They are also non-zero in (\ref{interval}). To see this it is enough to
observe that all factors $t_j/f_j(x\pm i\pi/2)$ have no order one zeros in
$|\im(x)|<\pi/2$.
So we conclude that the order 1 contribution satisfies
\begin{eqnarray}
\label{g1equation} g_1(x+i\frac{\pi}{2}) \: g_1(x-i\frac{\pi}{2}) &= &1 \\
\label{g2equation} g_2(x+i\frac{\pi}{2})\: g_2(x-i\frac{\pi}{2}) &= &1
\end{eqnarray}
The unique solution with the analyticity determined by the structure of 
zeros and poles just discussed is
\begin{eqnarray}
\label{g1} g_1(x,\xi) &= &\tanh ^2 \frac{x}{2} \\
\label{g2} g_2(x,\xi) &= &\tanh ^2 \frac{x}{2} \tanh \frac{x-\xi}{2}
\tanh \frac{x+\xi}{2}
\end{eqnarray}

Clearly, the functions $l_j(u)=t_j(u)/(f_j(u) g_j(u,\xi))$ giving the 
finite-size corrections satisfy the equations
\begin{eqnarray}
\label{l1} l_1(x+i\frac{\pi}{2}) \: l_1(x-i\frac{\pi}{2}) &= &1+t_2(x) \\
\label{l2} l_2(x+i\frac{\pi}{2}) \: l_2(x-i\frac{\pi}{2}) &= &1+t_1(x)
\end{eqnarray}
and do not possess the zeros and poles of order $N$ and 1 that we 
have already removed. They
only have other zeros, given by the 1-string zeros of
$\mathbf{D}(u, \xi)$.
Observe that these do not effect the functions $1+t_1(x)$ and $1+t_2(x)$, that
again are non-zero on the real axis.
Assuming $N$ large, the position of the 1-string zeros $v_k^{(j)}$ is
real\footnote{Observe that the classification of zeros in
terms of 1- and 2-strings is only strictly true for $N$ sufficiently large.}
and we can remove their contribution by introducing the functions
\begin{equation}  \label{pj_zeros}
p_j(x) = \prod_{k=1}^{m_j} \tanh [\frac{1}{2}(5v_k^{(j)} -x)]
\tanh [\frac{1}{2}(5v_k^{(j)} + x)] ,
\end{equation}
in each strip.
It is simple to prove that these functions satisfy
\begin{equation}
\label{pj} p_j(x+i\frac{\pi}{2}) \: p_j(x-i\frac{\pi}{2}) = 1.
\end{equation}
So, as before, we introduce the functions
\begin{equation} \label{ltilde}
\tilde{l}_j(x)=\frac{t_j(x)}{f_j(x) g_j(x,\xi) p_j(x)}
\end{equation}
which are completely free of poles and zeros inside
(\ref{interval}). We can then solve the remaining equations
\begin{eqnarray}
\label{tildel1}  \tilde{l}_1 (x+i\frac{\pi}{2}) \:
\tilde{l}_1(x-i\frac{\pi}{2}) & = & 1+t_2(x) \\
\label{tildel2} \tilde{l}_2(x+i\frac{\pi}{2}) \: \tilde{l}_2(x-i\frac{\pi}{2})
&= & 1+t_1(x)
\end{eqnarray}
by taking the Fourier transform of the logarithmic derivative.
The solution is
\begin{eqnarray}
\label{l1_sol} \log \tilde{l}_1 (x) & = & K * \log (1+t_2) + C_1, \\
\label{l2_sol} \log \tilde{l}_2 (x) & = & K * \log (1+t_1) + C_2,
\end{eqnarray}
where $C_j$ are integration constants. The kernel of the integration and
the convolution are given by
\begin{eqnarray}
K(x) =\frac{1}{2 \pi \cosh x},\qquad\quad
(f*g)(x) = \int_{-\infty}^{+\infty} dy f(x-y) g(y)
\end{eqnarray}
Using (\ref{ltilde}) we recombine all the contributions
\begin{eqnarray}
\log t_1 (x) & = & \log g_1(x,\xi) +
\sum_{k=1}^{m_1} \log\big[ \tanh [\frac{1}{2}(5v_k^{(1)} -x)]
\tanh [\frac{1}{2}(5v_k^{(1)} + x)]\big]  \nonumber\\
& & + K * \log (1+t_2) + C_1, \label{lattice-tba1} \\
\log t_2 (x) & = & \log f_2(x) + \log g_2(x,\xi) +
\sum_{k=1}^{m_2} \log \big[\tanh [\frac{1}{2}(5v_k^{(2)} -x)]
\tanh [\frac{1}{2}(5v_k^{(2)} + x)]\big ]  \nonumber\\
& & +K * \log (1+t_1) + C_2  \label{lattice-tba2}
\end{eqnarray}
These are the lattice TBA equations.
The boundary thermodynamic field $\xi$ enters the TBA equations only 
through the boundary factors
$g_1$, $g_2$.

There are two important simplifications of the previous equations.
It will be clear in the sequel that, in all the cases of
interest,
\begin{equation}
\lim _{x\rightarrow \pm \infty} g_j(x,\xi)=1 \mbox{ or } -1.
\end{equation}
This expression and (\ref{tlimit}) from Appendix B
can be used in the TBA equations (\ref{lattice-tba1}), (\ref{lattice-tba2})
to compute the limit $x\rightarrow + \infty$.
This forces the integration constants to be an integer multiple of $i\pi$
\begin{equation}
C_j \propto i \pi
\end{equation}
so that they can be removed from the equations and replaced by a sign,
$s_j=\pm 1$, inside the LHS logarithm: $\log s_j t_j(x)$.
This amounts to fixing the branch of the logarithm.
 From (\ref{t_real}) we know that $t_j(x)$ is real for real $x$. From the
analysis of the analyticity, we also know that $1+t_j(x)\neq 0$ for real $x$
so we find that it never changes its sign. The limit in (\ref{tlimit})
forces it to be positive, $1+t_j(x) > 0$, so that
the convolution term in the TBA equations is always real.

\subsubsection{Scaling limit and TBA equations\label{sss_TBA}}
The order~$1/N$ term in (\ref{leading}), giving the scaling behaviour,
is obtained in the scaling limit.
Guided by the function
$f_2$, that has the well defined behaviour
\begin{equation}
\lim_{N \rightarrow \infty} f_2(x+\log N)=\exp(-4e^{-x}),
\end{equation}
we assume the general scaling forms
\begin{equation}
\hat{h}_j(x)=\lim_{N \rightarrow \infty} h_j(x+\log N).
\end{equation}
This is consistent with the behaviour observed for the zeros of small
transfer matrices. Increasing $N$ corresponds to adding 2-strings in the first
strip, close to the real $u$-axis. This has the effect of pushing the original
zeros $v^{(j)}_k, ~w^{(j)}_l$ away with a logarithmic growth in their 
locations so that
\begin{equation}
y^{(j)}_k=\lim_{N \rightarrow \infty}(5v^{(j)}_k-\log N), \qquad
z^{(j)}_k=\lim_{N \rightarrow \infty}(5w^{(j)}_k-\log N)
\end{equation}
are well defined. Applying these results to the lattice TBA system
(\ref{lattice-tba1}), (\ref{lattice-tba2}) we
obtain the scaled-continuum TBA equations, after introducing
the pseudoenergy functions
\begin{equation}
\epsilon_j(x) \equiv -\log s_j \hat{t}_j (x) , \qquad j=1,2
\end{equation}
The properties of $t_j(x)$ discussed at the end of the previous section imply
\begin{equation}
e^{-\epsilon_j(x)} \in \mathbb{R} , \qquad
1+\hat{t}_j(x) > 0, \qquad x\in \mathbb{R}.
\end{equation}
so that (for real $x$) the TBA equations are
\begin{eqnarray}
\label{tba1} \epsilon_1(x) &=& -\log \hat{g}_1(x,\xi)
-\sum_{k=1}^{m_1} \log (\tanh \frac{y_k^{(1)} -x}{2})
- K * \log |1+s_2 e^{-\epsilon_2(x)}|, \\
\label{tba2} \epsilon_2(x) &=& 4 e^{-x}
- \log \hat{g}_2(x,\xi) - \sum_{k=1}^{m_2} \log (\tanh \frac{y_k^{(2)} -x}{2})
-K * \log |1+s_1e^{-\epsilon_1(x)}|.
\end{eqnarray}
This means that any complex contribution to $\epsilon_j$ comes only from
the source terms in the TBA equations. Note that these perturbed TBA equations 
differ from the critical TBA equations (\ref{OPW1}) and (\ref{OPW2}) only through the
appearance of the boundary perturbation terms $\log \hat{g}_1(x,\xi)$ and 
$\log \hat{g}_2(x,\xi)$.

The boundary thermodynamic field is also scaled so that it
survives in the continuum limit
\begin{equation}
\hat{g}_j(x,\xi)=\lim_{N \rightarrow \infty} g_j(x+\log N,\xi-\log N)
\end{equation}
In the present case, we find from (\ref{g1}), (\ref{g2}) that
\begin{eqnarray}
\hat{g}_1(x,\xi) & = & 1, \label{hatg1} \\
\hat{g}_2(x,\xi) & = & \tanh \frac{x+\xi}{2} \label{hatg2}
\end{eqnarray}
so that only $\hat{g}_2$ remains in the equations.
The sign used in the scaling of $g_j$, for the boundary field,
$\xi-\log N$, corresponds to the value
$\xi=-\infty$ at the UV conformal fixed point ($\chi_{1,2}(q)$) and
$\xi=\infty$ at the IR conformal fixed point ($\chi_{1,1}(q)$).
To be more general, we will keep track of $\hat{g}_1$ in many of the 
equations that follow, because in the analysis of other flows this 
term will be non-trivial.

In the TBA literature, it is standard to define
\begin{equation} \label{elle}
L_j(x)= \log |1+\hat{t}_j(x)| =\log|1+s_j e^{-\epsilon_j(x)}| \in \mathbb{R}.
\end{equation}
The zeros of $\hat{t}_j$ are then the zeros of $L_j$ and vice versa, as is
seen if we write the TBA equations in the more physical form
\begin{eqnarray}
\label{tba1_L}
L_1(x) &=& \log \left| 1+ s_1\,\hat{g}_1(x,\xi)
\prod_{k=1}^{m_1} \tanh \frac{y_k^{(1)} -x}{2} \:
e^{ K * L_2(x)} \right|, \\[3mm]
\label{tba2_L}
L_2(x) &=& \log \left| 1+e^{-4 e^{-x}} s_2\,
\hat{g}_2(x,\xi) \prod_{k=1}^{m_2} \tanh \frac{y_k^{(2)} -x}{2} \:
e^{K * L_1(x)} \right|.
\end{eqnarray}

Again following \cite{OPW}, we can use (\ref{l1_sol}), (\ref{l2_sol}) to
determine the scaling part of the eigenvalues of  $\mathbf{D}(u,\xi)$ at the
isotropic point $u=\lambda/2$
\begin{equation}\label{scalingenergy}
E(\xi)=
\displaystyle \frac{2}{\pi} \sum ^{m_{1}}_{k=1}e^{-y^{(1)}_{k}}-
\int ^{\infty }_{-\infty }\! \! \frac{dx}{\pi^2}\,e^{-x} L_2(x)
\end{equation}

\subsubsection{Zeros and quantum numbers\label{sss_qn}}
In the scaling limit, the zeros of the double
row transfer matrix satisfy
\begin{eqnarray}
\hat{t}_j(y_k^{(j)})=0 & & \mbox{for the 1-strings} \\
\hat{t}_j(z_k^{(j)}\mp i\pi)=0 & & \mbox{for the 2-strings}
\end{eqnarray}
The normalized transfer matrix itself can have other order~1 zeros from the
factors in (\ref{normal}).
In particular,
the point $x=-\xi$ being a zero for $g_2(x,\xi)$ is also a zero
for $\hat{t}_2(x,\xi)$, i.e. a zero in the center of the second strip.
The string-like zeros can be obtained by a single equation; to see this,
we perform the scaling limit and the shift $x\mapsto x -i\pi/2$ in
(\ref{t1system}), (\ref{t2system})
\begin{eqnarray}
\hat{t}_1(x) \: \hat{t}_1(x-i\pi) &=& 1+\hat{t}_2(x-i\frac{\pi}{2}) \\[2mm]
\hat{t}_2(x) \: \hat{t}_2(x-i\pi) &=& 1+\hat{t}_1(x-i\frac{\pi}{2})
\end{eqnarray}
and find that each real solution of
\begin{equation}\label{zeros}
\hat{t}_j(x-i\frac{\pi}{2})=-1
\end{equation}
is either a 1-string or a 2-string in the strip $3\!-\!j$.
The zero $x=-\xi$ is not contained in the previous equation because
it is coupled with a pole, as in Figure \ref{ABmech}.
The 1-string or 2-string cases are mutually exclusive,
as the numerical analysis on small double rows transfer matrices confirm:
the zeros generally repel one another and only single zeros are observed.
Double zeros can appear, however, at exceptional values of $\xi$,
for example at the point
where the 2-string in mechanism A transforms into the correlated 1-strings,
as in Figure \ref{ABmech}.

The locations of the zeros are fixed by (\ref{zeros}) so we can define
the counting functions
\begin{equation} \psi_j(x) = -i\,\epsilon_j(x-i\frac{\pi}{2})
=-i\log s_j \hat{t}_j(x-i\frac{\pi}{2})
\end{equation}
that are a multiple of $\pi$ on a 1 or 2-string
and introduce non-degenerate quantum numbers $n_k^{(j)} \in \mathbb{Z}$
by the quantization conditions
\begin{eqnarray}\label{quant1}
\psi_2(y_k^{(1)}) \equiv -i\,\epsilon_2(y_k^{(1)}-i\frac{\pi}{2}) =
n_k^{(1)} \pi,  & & (n_k^{(1)}+ s_2)  =0\mbox{~mod 2},  \\[2mm]
\label{quant2}
\psi_1(y_k^{(2)}) \equiv -i\,\epsilon_1(y_k^{(2)}-i\frac{\pi}{2}) =
n_k^{(2)} \pi, & & (n_k^{(2)}+ s_1)  =0\mbox{~mod 2}.
\end{eqnarray}
Note the inversion of the indices: $\psi_1$ is for strip 2 and 
$\psi_2$ for strip 1.
The same equations hold for the 2-string locations $z_l^{(j)}$ so, each
time $\psi_j(x)= n \pi$ is satisfied for an integer $n$ with the appropriate
parity, $x$ is a 1-string or a 2-string in the strip $3\!-\!j$.

The explicit expressions for $\psi_j$ are given by:
\begin{eqnarray}
\psi_1(x) &=& i\log \hat{g}_1(x-i\frac{\pi}{2},\xi)
+i\sum_{k=1}^{m_1} \log \tanh (\frac{y_k^{(1)} -x}{2}+i\frac{\pi}{4})
\nonumber \\
& & - 
\mbox{~}_{{PV}}\!\!\!\int_{-\infty}^{+\infty}\frac{dy}{2\pi}\frac{L_2(y)}{\sinh(x-y)},
\label{psi1} \\
\psi_2(x) &=& 4 e^{-x}+i\log \hat{g}_2(x-i\frac{\pi}{2},\xi)
+i\sum_{k=1}^{m_2} \log \tanh (\frac{y_k^{(2)} -x}{2}+i\frac{\pi}{4})
\nonumber \\
& & - \mbox{~}_{{PV}}\!\!\!\int_{-\infty}^{+\infty} \frac{dy}{2\pi} 
\frac{L_1(y)}{\sinh(x-y)}
\label{psi2}
\end{eqnarray}
where the integral around the singularity $x=y$ must be understood as a
principal value.
We take the fundamental
branch for each logarithm so that in general
$\log a + \log b$ cannot be identified as $\log (a b)$.
This prescription is convenient in view of the numerical computations
and is used also in the TBA equations (\ref{tba1}), (\ref{tba2}).

With this in mind, we can easily compute the limits of $\psi_j$ for
$x\rightarrow \pm\infty$.
For $x\rightarrow -\infty$, the dominant term in the expression of $L_2$ and
$\psi_2$ is the exponential, so we have
\begin{eqnarray}\label{L2_asymp}
L_2(x) &\sim& e^{-4e^{-x}}
\to 0,\qquad\mbox{~as~} x\to -\infty, \\[3mm]
\psi_2(x) &\sim& \;4e^{-x} \;
\to \infty,\qquad\mbox{~as~} x\to -\infty
\label{psi2_asymp}
\end{eqnarray}
These reflect the emergence of an infinite number of 2-strings in
strip 1 as the size of the double row transfer matrix goes to
infinity, as in Section~\ref{sss_TBA}. This dense filling of 2-strings
forces the 1-strings in the first strip only to be far apart from the
$-\infty$ region, $-\infty<y^{(1)}_1<\ldots<y_K^{(1)}$ so we can compute
in full generality the limits
\begin{eqnarray}
\lim_{x\rightarrow -\infty} L_1(x) &= & \log(1+s_1 \hat{g}_1(-\infty,\xi))
=\left\{ \begin{matrix} \log 2 \\ -\infty \end{matrix} \right.
\label{L1_asymp} \\[3mm]
\lim_{x\rightarrow -\infty} \psi_1(x) &= & i\log
\hat{g}_1(-\infty-i\frac{\pi}{2},\xi) .
\label{psi1_asymp}
\end{eqnarray}
In the present case, we have
\begin{equation}
\lim_{x\rightarrow -\infty} \psi_1(x) = 0, \qquad \mbox{for} ~~\xi> -\infty.
\label{psi1_asymp2}
\end{equation}
The limit $x\rightarrow +\infty$ requires some care.
We assume that $L_j(+\infty)$ is finite,
as is confirmed by numerical computations, so that the
contribution from the convolution disappears due to the principal value.
In general, there are two possibilities
\begin{equation} \label{lim_psi+}
\psi_j(+\infty) = \left\{ \begin{array}{l@{\hspace{9mm}\mbox{if~~}}c}
i\log \hat{g}_j(+\infty-i\frac{\pi}{2})- \pi m_j  &
y^{(j)}_{m_j}<+\infty,  \\[3mm]
i\log \hat{g}_j(+\infty-i\frac{\pi}{2})- \pi (m_j-1)  &
y^{(j)}_{m_j} = +\infty. \end{array} \right.
\end{equation}
In the present case, from the TBA computations at the IR critical point
($\chi_{1,1}(q)$), we know that $y^{(j)}_{m_j}<+\infty$ in both
strips so, away from the UV point, the previous equations reduce to
\begin{equation} \label{lim_psi+2}
\psi_j(+\infty) = - \pi m_j, \qquad \xi>-\infty.
\end{equation}

Our main conclusion is that generally the functions $\psi_j$ are globally
decreasing, with $m_1$,  $m_2$ nonnegative and possibly large and
again this is confirmed by numerics.
There are a few exceptions that give rise to interesting behaviours
that will be discussed later.
The monotonicity is the ingredient we need to relate the two families
$\{ n_k^{(j)} \}$ and $\{I_k^{(j)}\}$ of non-degenerate and nonnegative
quantum numbers. As in the critical case in Section~\ref{ss_crit}
and in \cite{OPW}, we start from the limit $\psi_j(+\infty)$. The
1 and 2-strings occur at multiples of $\pi$ with parity dictated by
(\ref{quant1}) and (\ref{quant2}), so decreasing from one object to the next,
the function $\psi_j$ increases by $2\pi$.
The ordering of the zeros is $y_1^{(j)}<y_2^{(j)}<\ldots <y_{m_j}^{(j)}$
so above the zero $y_k^{(j)}$ there are $I_k^{(j)}$ 2-strings (by definition)
and $m_j\!-\!k$ 1-strings.

An expression for the parities $s_j$ is still missing.
If we assume the IR critical ($\chi_{1,1}(q)$) TBA equations can be 
perturbed by
the ``small'' function
$\log \hat{g}_2=\log \tanh((x+\xi)/2)$ with large $\xi$,
we are led to maintain the even parity of $m_1$, $m_2$ and
the parities  $s_1=s_2=1$, as in
Table~\ref{critical_TBA}. This forces
\begin{equation}
n_k^{(1)}, \, n_k^{(2)} \mbox{ ~odd}
\end{equation}
We observed that $\psi_j(+\infty)$ cannot be a zero. At least close to
the IR region, we have $y_k^{(j)}<+\infty$ in both strips,
consistent with the limits (\ref{lim_psi+2}) where $m_j$ are both even.
The first available position for a zero is
$\psi_j(+\infty)+\pi=\pi(1-m_j)$,
yielding the following expressions for the quantum numbers
\begin{eqnarray}
n_k^{(1)} &=& 2(I^{(1)}_k+m_1-k)+1-m_2, \label{qn1} \\
n_k^{(2)} &=& 2(I^{(2)}_k+m_2-k)+1-m_1.  \label{qn2}
\end{eqnarray}
Observe that (\ref{lim_psi+}) is not predictive because we need to know
in advance if a zero is at $+\infty$ to get the proper solution.

A final comment on the zeros is useful. On the lattice,
the 1- and 2-strings are zeros of the transfer matrix $D(u,\xi)$.
In addition to that, the normalized transfer matrix $t(u,\xi)$ can have other
zeros and poles.
After the scaling limit, only the functions $\hat{t}_j(x,\xi)$ remain 
and, from the
previous derivation, we see that $\hat{t}_2(x,\xi)$ has a zero at $x=-\xi$.
Restricting $x$ to the real axis, the zeros of $\hat{t}_j(x,\xi)$ and
of $L_j(x)$ coincide (\ref{elle}) and can be either 1-strings or the boundary
dependent zero $x=-\xi$.
In terms of the TBA functions $L_j$ (or $\epsilon_j$) and $\psi_j$,
the characterization of the lattice quantities is:
the 1-strings in strip $j$ are the real zeros in common between $L_j(x)$ and
$\psi_{3-j}(x)-2\pi n$; the 2-strings in strip $j$ are the real zeros of
$\psi_{3-j}(x)-2\pi n$ and not of $L_j(x)$; the boundary zeros are the real
zeros of $L_2(x)$ and not of $\psi_1-2\pi n$.

As a complement to the description of the mechanisms and to the TBA equations,
we summarize the main results concerning this flow starting from the IR point:
\\[3mm] \indent
\begin{tabular}{l}
{\bf mechanism A before the collapse and mechanism B:}  \\[2mm]
\hspace*{10mm} $ \begin{array}{l}
m_1, ~m_2 \mbox{ ~even}, \qquad n_2=\frac{m_1}{2}-m_2 \geqslant 0 ,\qquad
s_1=s_2=1,  \\[2mm]
n_k^{(1)} = 2(I^{(1)}_k+m_1-k)+1-m_2, \\[2mm]
n_k^{(2)} = 2(I^{(2)}_k+m_2-k)+1-m_1;  \end{array} $ \\[12mm]
{\bf mechanism A after the collapse point:} \\[2mm]
\hspace*{10mm} $ \begin{array}{l}
m_1^A=m_1+2, \qquad n^{(1)}_{m_1^A-1}=n^{(1)}_{m_1^A}=1-m_2 , \\[2mm]
\mbox{the values of $~m_2, ~n_2, ~n_k^{(2)}, ~n_k^{(1)}$ with $~k<m_1^{A}-1$,
$~s_j$, ~remain unchanged.}
\end{array} $
\end{tabular}

\subsection{RG flow $\chi_{3,2} \mapsto \chi_{3,1}$\label{ss_3231}}
As pointed out in Section~\ref{ss_bound}, this flow is
dual to $\chi_{1,2} \mapsto \chi_{1,1}$ and many of the considerations of
Section~\ref{ss_1211} apply to it.
For this flow we consider  the double row transfer matrix
$\mathbf{D}(u,\xi)\equiv\mathbf{D}_{3,1|2,1}^N (u,\xilatt)$
with an odd number of faces $N$ in accord with the adjacency rules.
We find the same change of parity in the number of columns (\ref{parity12})
and the same two mechanisms A and B (\ref{mechanismA}), 
(\ref{mechanismB}) changing the string content during the flow.
The actual parities of $N$, $m_i$ are different but
the changes are still fixed by the mechanisms A and B. The application of
mechanisms A and B to the first few states is given in
Table~\ref{t_s3231}.
\begin{table}
\caption{\label{t_s3231}
\small
Flow $\chi_{3,1} \mapsto \chi_{3,2}$ (reverse of the physical flow).
We present the explicit mapping of states from IR to UV up to the UV 
level 6. Here
$n\ir,\:n\uv$ are the excitation levels
above the ground states, respectively $h=3/2$ and $h=3/5$.}
\begin{center}
\begin{tabular}{|c|r@{$\:\:\mapsto\:\:$}l|c|c||c|r
@{$\:\:\mapsto\:\:$}l|c|c|}
\hline
$n\ir$\rule[-4mm]{0mm}{10mm}& \multicolumn{3}{c|}{Mapping of states --
mechanism} & $n\uv$ &
   $n\ir$ & \multicolumn{3}{c|}{Mapping of states -- mechanism} & $n\uv$\\
\hline
0\rule[-1mm]{0mm}{6mm} & $(0\,0|0)$ & $(0|0)_{-}$ & B & 0 &
           5 & $(4\,1|0)$ & $(3\,0\,0|0)_{+}$ & A & 5 \\[1mm]
1 & $(1\,0|0)$ & $ (1|0)_{-}$ & B & 1 &
           5 & $(3\,2|0)$ & $(2\,1\,0|0)_{+}$ & A & 5 \\[1mm]
2 & $(2\,0|0)$ & $(2|0)_{-}$ & B & 2 &
           6 & $(0\,0\,0\,0|1)$ & $(0\,0\,0|1)_{-}$ & B & 5 \\[1mm]
2 & $(1\,1|0)$ & $(0\,0\,0|0)_{+}$ & A & 2 &
           6 & $(1\,0\,0\,0|0) $ & $(1\,0\,0|0)_{-}$ & B & 5 \\[1mm]
3 & $(3\,0|0)$ & $(3|0)_{-}$ & B & 3 &
           6 & $(6\,0|0)$ & $(6|0)_{-}$ & B & 6 \\[1mm]
3 & $(2\,1|0)$ & $(1\,0\,0|0)_{+}$ & A & 3 &
           6 & $(5\,1|0)$ & $(4\,0\,0|0)_{+}$ & A & 6 \\[1mm]
4 & $(4\,0|0)$ & $(4|0)_{-}$ & B & 4 &
           6 & $(4\,2|0)$ & $(3\,1\,0|0)_{+}$ & A & 6 \\[1mm]
4 & $(3\,1|0)$ & $(2\,0\,0|0)_{+}$ & A & 4 &
           6 & $(3\,3|0)$ & $(2\,2\,0|0)_{+}$ & A & 6 \\[1mm]
4 & $(2\,2|0)$ & $(1\,1\,0|0)_{+}$ & A & 4 &
           7 & $(1\,0\,0\,0|1)$ & $(1\,0\,0|1)_{-}$ & B & 6 \\[1mm]
5 & $(0\,0\,0\,0|0)$ & $(0\,0\,0|0)_{-}$ & B & 4 &
           7 & $(2\,0\,0\,0|0)$ & $(2\,0\,0|0)_{-}$ & B & 6 \\[1mm]
5 & $(5\,0|0)$ & $(5|0)_{-}$ & B & 5 &
           7 & $(1\,1\,0\,0|0)$ & $(1\,1\,0|0)_{-}$ & B & 6 \\[1mm]
\hline
\end{tabular}
\end{center}
\end{table}
Again, we find a mapping between finitized characters
\begin{eqnarray}
\chi_{3,1}^{(N\irt)}(q) & \mapsto & \chi_{3,2}^{(N\uvt)}(q) \\[1mm]
N\ir  & \mapsto & N\uv=N\ir-1 \nonumber
\end{eqnarray}
showing the consistency of the mechanisms with the IR and UV counting 
as in Section~\ref{sss_map1211}.

In solving the functional equations, as noticed in Section~\ref{sss_func},
the order~1 behaviour is specific to each boundary condition.
Here we need to consider $S(u)$, $S_3(u,\lambda/2)$, $S_2(u,\xilatt)$
from (\ref{normal}) to (\ref{normal_sr}).
The only difference with Section~\ref{sss_func} is the term
$S_3(u,\lambda/2)$ that replaces $S_1(u)$
\begin{equation}
S_3(u,\lambda/2)=-\frac{\sin ^2 \lambda}{\sin (u+\frac{\lambda}{2})
\sin (u-\frac{3}{2}\lambda) }.
\end{equation}
It simply adds to Figure~\ref{order1anal} poles at
$u=-\lambda/2$,
$u=3\lambda/2$, that is, $x=\mp i \pi$. Being outside (\ref{interval})
they are not relevant, so we obtain the same
order~1 system (\ref{g1equation}), (\ref{g2equation}), the same solution
(\ref{g1}), (\ref{g2}) and the same scaling  terms (\ref{hatg1}), 
(\ref{hatg2}):
\begin{eqnarray}
\hat{g}_1(x) &= & 1 \\
\hat{g}_2(x) &= & \tanh  \frac{x+\xi}{2}
\end{eqnarray}
The order~$1/N$ analyticity is also specific to each boundary condition
in the sense that the parity of $m_j$ in (\ref{pj_zeros}) is boundary
dependent, but the way to proceed is exactly the same as before.
In conclusion, the equations obtained in Section~\ref{sss_TBA}
hold true also for this case.

We summarize the description of this flow starting from the IR point:
\\[3mm] \indent
\begin{tabular}{l}
{\bf mechanism A before the collapse and mechanism B:}  \\[2mm]
\hspace*{10mm} $ \begin{array}{l}
m_1 \geqslant 2 \mbox{ ~even}, ~m_2 \mbox{ ~odd}, \qquad
n_2=\frac{m_1}{2}-m_2 \geqslant 0 ,\qquad
s_1=1, ~s_2=-1,  \\[2mm]
n_k^{(1)} = 2(I^{(1)}_k+m_1-k)+1-m_2, \\[2mm]
n_k^{(2)} = 2(I^{(2)}_k+m_2-k)+1-m_1;  \end{array} $ \\[12mm]
{\bf mechanism A after the collapse:} \\[2mm]
\hspace*{10mm} $ \begin{array}{l}
m_1^A=m_1+2, \qquad n^{(1)}_{m_1^A-1}=n^{(1)}_{m_1^A}=1-m_2 , \\[2mm]
\mbox{the values of $~m_2, ~n_2, ~n_k^{(2)}, ~n_k^{(1)}$ with $~k<m_1^{A}-1$,
$~s_j$, ~remain unchanged.}
\end{array} $
\end{tabular}

\subsection{RG flow $\chi_{2,2} \mapsto \chi_{2,1}$\label{ss_2221}}
\begin{table}
\caption{\label{t_s2221}
\small
Flow $\chi_{2,1} \mapsto \chi_{2,2}$ (reverse of the physical flow).
We present the explicit mapping of states from IR to UV up to the UV 
level 6. Here
$n\ir,\:n\uv$ are the excitation levels
above the ground states, respectively $h=7/16$ and $h=3/80$.}

\begin{center}
\begin{tabular}{|c|r@{$\:\:\mapsto\:\:$}l|c|c||c|r
@{$\:\:\mapsto\:\:$}l|c|c|}
\hline
$n\ir$\rule[-4mm]{0mm}{10mm}& \multicolumn{3}{c|}{Mapping of states --
mechanism} & $n\uv$ &
   $n\ir$ & \multicolumn{3}{c|}{Mapping of states -- mechanism} & $n\uv$\\
\hline
0\rule[-1mm]{0mm}{6mm} & $(0|0)$ & $(|0)_{-}$ & B & 0 &
           5 & $(1\,1\,0|0)$ & $(1\,1|0)_{-}$ & B & 4 \\[1mm]
1 & $(1|0)$ & $ (0\,0|0)_{+}$ & A & 1 &
           5 & $(1\,0\,0|1)$ & $(1\,0|1)_{-}$ & B & 4 \\[1mm]
2 & $(2|0)$ & $(1\,0|0)_{+}$ & A & 2 &
           5 & $(5|0)$ & $(4\,0|0)_{+}$ & A &5 \\[1mm]
3 & $(0\,0\,0|0)$ & $(0\,0|0)_{-}$ & B & 2 &
           6 & $(3\,0\,0|0)$ & $(3\,0|0)_{-}$ & B & 5 \\[1mm]
3 & $(3|0)$ & $(2\,0|0)_{+}$ & A & 3 &
           6 & $(2\,1\,0|0)$ & $(2\,1|0)_{-}$ & B & 5 \\[1mm]
4 & $(1\,0\,0|0)$ & $(1\,0|0)_{-}$ & B & 3 &
           6 & $(2\,0\,0|1) $ & $(2\,0|1)_{-}$ & B & 5 \\[1mm]
4 & $(0\,0\,0|1)$ & $(0\,0|1)_{-}$ & B & 3 &
           6 & $(1\,1\,0|1)$ & $(1\,1|1)_{-}$ & B & 5 \\[1mm]
4 & $(4|0)$ & $(3\,0|0)_{+}$ & A & 4 &
           6 & $(1\,1\,1|0)$ & $(0\,0\,0\,0|0)_{+}$ & A & 5 \\[1mm]
5 & $(2\,0\,0|0)$ & $(2\,0|0)_{-}$ & B & 4 &
           6 & $(6|0)$ & $(5\,0|0)_{+}$ & A & 6 \\[1mm]
\hline
\end{tabular}
\end{center}
\end{table}

The treatment of the flow $\chi_{2,2} \mapsto \chi_{2,1}$ is very 
similar to the previous flows.
The relevant transfer matrix is now
$\mathbf{D}(u,\xi)\equiv\mathbf{D}_{2,1|2,1}^N (u,\xilatt)$
which has an even number of faces $N$.
Again, we find the same change of parity in the number of columns
(\ref{parity12}) and the same two mechanisms A and B (\ref{mechanismA}),
(\ref{mechanismB}), as illustrated in Table~\ref{t_s3231}.
We also find a mapping between finitized characters
\begin{eqnarray}
\chi_{2,1}^{(N\irt)}(q) & \mapsto & \chi_{2,2}^{(N\uvt)}(q) \\[1mm]
N\ir  & \mapsto & N\uv=N\ir-1 \nonumber
\end{eqnarray}
showing the consistency of the mechanisms with the IR and UV counting as in
Section~\ref{sss_map1211}.

In solving the functional equations, we need to take care of the
order 1 analyticity. With reference to Section~\ref{sss_func},
we need to consider the term
\begin{equation}
S_{2}(u,\frac{\lambda}{2})=\frac{\sin^2 \lambda \sin(u+\frac{1}{2}\lambda)
\sin(u+\frac{7}{2}\lambda)}{\sin^2 (u-\frac{1}{2}\lambda)
\sin(u+\frac{3}{2}\lambda)\sin(u+\frac{5}{2}\lambda)}.
\end{equation}
All of its zeros and poles are independent of $\xi$ and disappear in the
scaling limit. Many of them disappear also at the lattice level: the factors
$\sin(u+\frac{3}{2}\lambda)\sin(u+\frac{5}{2}\lambda)$
in the denominator are cancelled by equal terms in
$\mathbf{D}(u,\xi)$. Indeed, $\sin(u+\frac{3}{2}\lambda)$
explicitly appears in $B^{2,1}(u,\lambda/2)$;
$\sin(u+\frac{5}{2}\lambda)$ does not explicitly appear but we can prove its
existence by crossing symmetry:
$\mathbf{D}(u,\xi)=\mathbf{D} (\lambda-u,\xi)$
implies that if $u=-3\lambda/2$ is a zero,
then $\lambda-u=5\lambda/2 \equiv -5\lambda/2 $ must also be a zero
(the last equality is true because of periodicity).
The remaining factors contain zeros at $u=-\lambda/2$, $u=3\lambda/2$,
which are irrelevant because they are outside of (\ref{interval}), and
a double pole at $u=\lambda/2$ that compensates a double zero
from $S(u)$. So, already at the lattice level, $g_1(x)=1$. For $g_2$,
all the considerations in Section~\ref{sss_func} can be repeated here.

We conclude that the same TBA equations, quantization
conditions and energy expression as in Section~\ref{sss_TBA} hold for
this flow.

We summarize the description of this case starting from the IR point:
\\[3mm] \indent
\begin{tabular}{l}
{\bf mechanism A before the collapse and mechanism B:}  \\[2mm]
\hspace*{10mm} $ \begin{array}{l}
m_1, ~m_2 \mbox{ ~odd}, \qquad n_2=\frac{m_1+1}{2}-m_2 \geqslant 0 ,\qquad
s_1=s_2=-1,  \\[2mm]
n_k^{(1)} = 2(I^{(1)}_k+m_1-k)+1-m_2, \\[2mm]
n_k^{(2)} = 2(I^{(2)}_k+m_2-k)+1-m_1;  \end{array} $ \\[12mm]
{\bf mechanism A after the collapse:} \\[2mm]
\hspace*{10mm} $ \begin{array}{l}
m_1^A=m_1+2, \qquad n^{(1)}_{m_1^A-1}=n^{(1)}_{m_1^A}=1-m_2 , \\[2mm]
\mbox{the values of $~m_2, ~n_2, ~n_k^{(2)}, ~n_k^{(1)}$ with $~k<m_1^{A}-1$,
$~s_j$, ~remain unchanged.}
\end{array} $
\end{tabular}

\section{Boundary flows with variable $r$}

\subsection{RG flow $\chi_{3,2} \mapsto \chi_{2,1}$\label{ss_1321}}
For the flow $\chi_{3,2}\equiv \chi_{1,3} \mapsto \chi_{2,1}$ we 
consider the  double row transfer matrix
$\mathbf{D}_{1,1|3,1}^N (u,\xilatt)$ with the trivial $B^{1,1}$
boundary on the left, with no free parameter, and $B^{3,1}(u,\xilatt)$
on the right with the parameter $\xilatt \in \mathbb{C}$.
The number of faces $N$ is even.
The limit $\im(\xilatt) \rightarrow \pm \infty$ on $B^{3,1}(u,\xilatt)$
reproduces the $s=3$ boundary (\ref{stype}) and, for $\im(\xilatt)=0$,
$B^{3,1}(u,\re(\xilatt))$ yields an $r=2$ type boundary (\ref{rtype}) if
$\re(\xilatt) \in [-5/2\lambda,-\lambda/2]$.
If we choose $\xilatt=-3\lambda/2 +i\xi/5$ with $\xi$ real,
\begin{equation}
\B{3,}{1}{3\pm 1}{3}{u}{-\frac{3}{2}\lambda+i\frac{\xi}{5}} =
\sqrt{\frac{S_{3\pm1}}{S_3}}\;\frac{-\sin(\frac{3}{2}\lambda\pm u
+i\frac{\xi}{5}) \, \sin(\frac{3}{2}\lambda\pm u-i\frac{\xi}{5})}
{\sin \lambda \, \cosh2\im(\xi)}
\end{equation}
is real analytic in $u$.
Notice that any choice other than $\re(\xilatt) \neq -3\lambda/2$ would
lead to the loss of real analyticity, that is a key property
to get real flows. From the previous equation, we also
see that each single entry of the transfer matrix is real for real $u$.
This leads to the real analyticity for the whole matrix
\begin{equation}
\mathbf{D}(u,\xi) \equiv
\mathbf{D}_{1,1|3,1}^N (u,-\frac{3}{2}\lambda+i\frac{\xi}{5})=
(\mathbf{D}_{1,1|3,1}^N (u^{*},-\frac{3}{2}\lambda+i\frac{\xi}{5})) ^{*},
\qquad \xi\in \mathbb{R}.
\end{equation}
The same property holds for the normalized transfer matrix.
Combining real analyticity, periodicity and crossing symmetry
and using the notation introduced in (\ref{h1}, \ref{h2}) we obtain
the reality conditions for $x\in \mathbb{R}$
\begin{eqnarray}
\mathbf{D}(\frac{\lambda}{2}+i\frac{x}{5},\xi) &=&
(\mathbf{D}(\frac{\lambda}{2}+i\frac{x}{5},\xi))^{*} , \nonumber \\
\mathbf{D}(3\lambda+i\frac{x}{5},\xi)&=&
(\mathbf{D}(3\lambda+i\frac{x}{5},\xi))^{*}, \label{reality} \\
\mathbf{t}_j(x,\xi)&=&\mathbf{t}_j(x,\xi)^{*}, \qquad j=1,2. \nonumber
\end{eqnarray}
The transfer matrix is transpose symmetric (\ref{transpose}); with 
the previous equations, this implies that it is real symmetric so 
that its eigenvalues
and the scaling energies are also real. The previous equations are 
also true for the corresponding eigenvalues.
The integrable flow
\begin{equation} \label{1321scaling}
\mathbf{D}_{1,1|3,1}^N (u,-\frac{3}{2}\lambda+i\frac{\xi}{5})
\longrightarrow \left\{ \begin{array}{c@{\hspace{4mm}}l}
\chi_{1,3} & \mbox{if ~} \xi \rightarrow \pm \infty \\[3mm]
\chi_{2,1} & \mbox{if ~} \xi =0  \end{array}
\right.
\end{equation}
thus provides a lattice description of the renormalization group flow
$\chi_{1,3}\rightarrow \chi_{2,1}$.
The two ranges $\xi \geqslant 0$ and $\xi \leqslant 0$ actually describe
the same flow. For later convenience, we choose $\xi \leqslant 0$.

If we compare the transfer matrix under consideration with those of
Sections~\ref{ss_3231} and \ref{ss_2221}, we see that we are using
inequivalent lattice realizations of the same conformal characters. Indeed,
the parities and boundary spins of the corresponding chains are different
\begin{equation}\begin{array}{l@{\hspace{6mm}}r@{~\neq~}l}
\chi_{1,3}\equiv\chi_{3,2}: & \mathbf{D}_{1,1|3,1}^N (u,\pm i \infty) &
\mathbf{D}_{3,1|2,1}^{N'} (u,\pm i \infty)  \\[3mm]
\chi_{2,1}: &
\mathbf{D}_{1,1|3,1}^N (u,-\frac{3}{2}\lambda)&
\mathbf{D}_{2,1|1,1}^{N'} (u,\frac{\lambda}{2})
\end{array}
\end{equation}
$N$ being even, $N'$ odd, the left transfer matrices are used in the present
section and the right ones were used in the constant $r$ flows.
This means that we cannot use the rules given in Table~\ref{critical_TBA}.

\begin{table}[t] \caption{\small
Classification, for alternative realizations of $(1,3)$ and $(2,1)$ 
boundary conditions of the TIM not shown in Table~\ref{critical_TBA}, 
of the allowed patterns of 1- and 2-strings by 
$(\boldsymbol{m},\boldsymbol{n})$ system and quantum numbers.
The parity $\sigma=\pm 1$ occurs when there are frozen zeros.
The parities $s_1,s_2=\pm 1$ occur in the TBA equations. The 
expressions for $n_1$,
are only used on a finite lattice because in
the scaling limit $n_1 \sim N/2 \rightarrow \infty$. The number of 
faces in a row is
even or odd according to $N=(r-s)~\mbox{mod}~2 $.
The energy expression (\ref{energy}) still holds in these cases.
\label{new_critTBA}}
$$
\begin{array}{|c|l|l|l|}
\hline
\chi_{r,s}(q)&\mbox{$(\boldsymbol{m},\boldsymbol{n})$ 
system}&\mbox{parities}&\mbox{quantum numbers}\\
\hline
\begin{matrix} \chi_{1,3}^{(N)}(q) \\[2mm] N \in \mbox{even} \end{matrix} &
\begin{array}{l}  m_{1},\, m_{2} \mbox{~odd} \\ n_2=(m_1-\sigma)/2-m_2 \\
n_1=(N+m_2+\sigma)/2-m_1 \end{array}&
\begin{array}{l}  s_{1}=1 \\ s_{2}=1  \end{array} &
\begin{array}{l}
n^{(1)}_{k}=2(I_{k}^{(1)}+m_{1}-k)+1-m_{2}-\sigma  \\[2mm]
n^{(2)}_{k}=2(I_{k}^{(2)}+m_{2}-k)+1-m_{1}+\sigma
\end{array}\\
\hline
\begin{matrix} \chi_{2,1}^{(N)}(q) \\[2mm] N \in \mbox{even} \end{matrix} &
\begin{array}{l}
m_{1} \mbox{~odd}, ~m_{2}  \mbox{~even} \\n_2=(m_1+1)/2-m_2 \\
n_1=(N+m_2)/2-m_1 \end{array} &
\begin{array}{l}  s_{1}=-1  \\  s_{2}=1  \end{array} &
\begin{array}{c}
n^{(1)}_{k}=2(I_{k}^{(1)}+m_{1}-k)+1-m_{2} \\[2mm]
n^{(2)}_{k}=2(I_{k}^{(2)}+m_{2}-k)+1-m_{1}
\end{array}\\\hline
\end{array} $$
\end{table}

We used ``numerics on $D$'' to confirm the scaling behaviour in
(\ref{1321scaling}) and to classify the pattern of zeros appropriate for these
lattice realizations. The results are shown in Table~\ref{new_critTBA}.
We notice that for $\chi_{1,3}(q)$ we obtain the same
$(\boldsymbol{m},\boldsymbol{n})$ system as described in
Table~\ref{critical_TBA}, after a shift of the parity of the chain
$N_{1,3}=N_{3,2}+1$. So, we have two different transfer matrices that
give rise to the same finitization of the conformal character.
In contrast, for $\chi_{2,1}(q)$ we obtain a new finitized
character, built from a different $(\boldsymbol{m},\boldsymbol{n})$ system.
We argue that this must be the case from TBA considerations.
Indeed, we expect two flows related by duality to enter
$\chi_{2,1}$. Because of duality, we expect the same boundary terms in
the TBA equations, so the only way to generate different flows is to start
from different and inequivalent patterns of zeros.

\subsubsection{Three mechanisms for changing the string content}
It is again convenient to consider the reversed (unphysical) flow, 
from $\mbox{IR}=\chi_{2,1}$ to
$\mbox{UV}=\chi_{1,3}$.

\begin{figure}[t]
\includegraphics[width=0.31\linewidth]{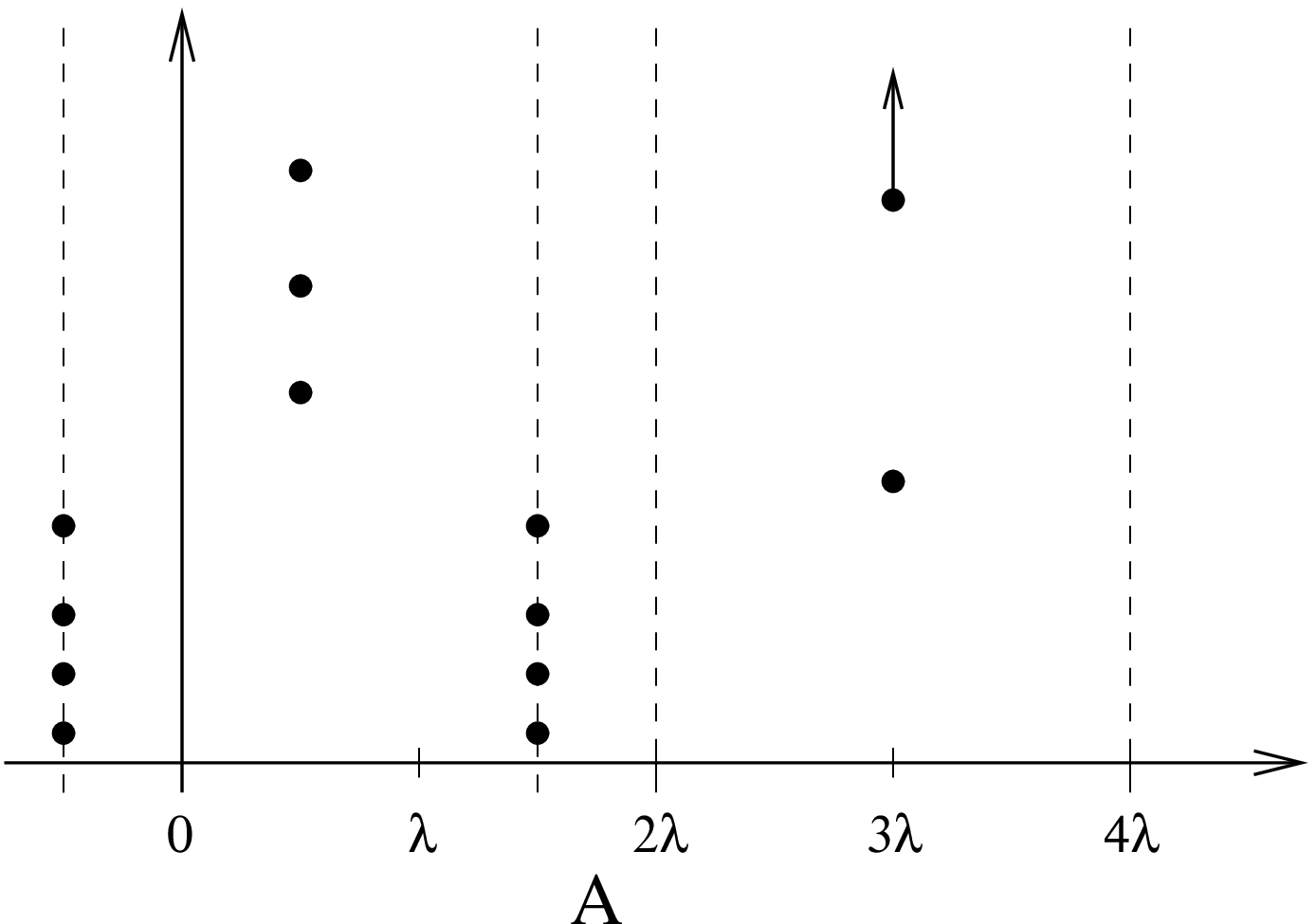}
\hfill\includegraphics[width=0.31\linewidth]{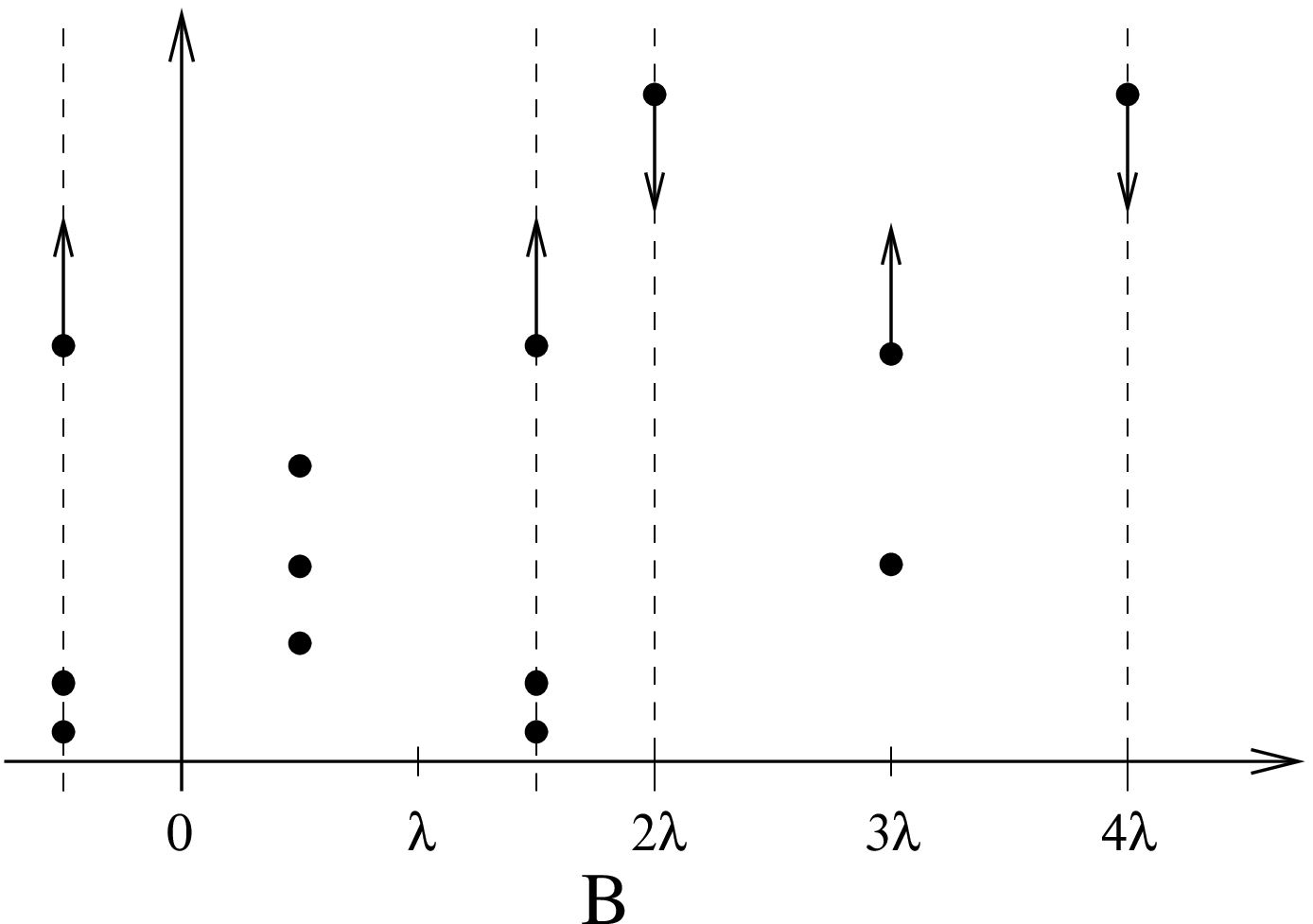}
\hfill\includegraphics[width=0.31\linewidth]{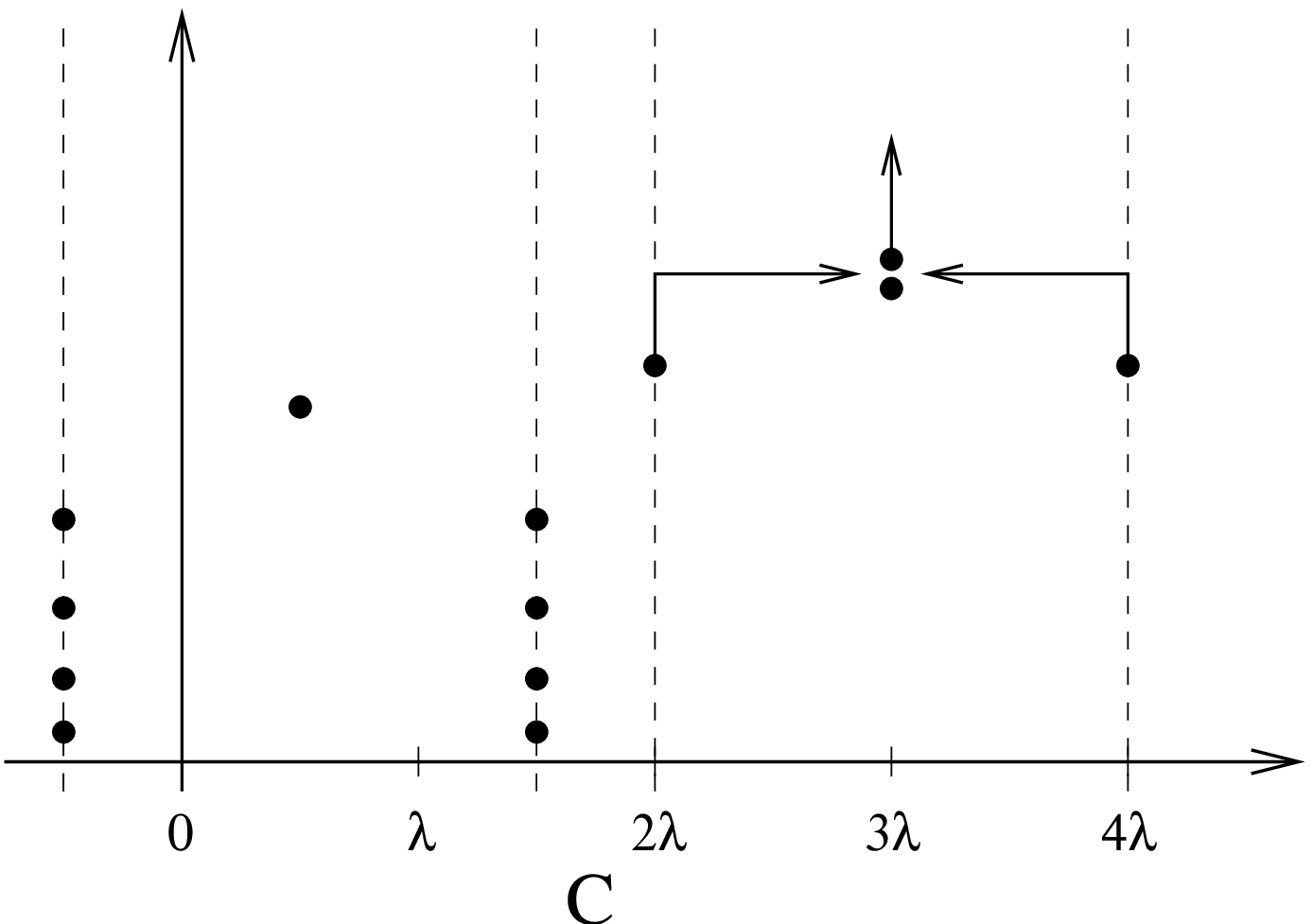}
\caption{\small \label{ABCmech} The three mechanisms A, B, C respectively
that change string content during the flow
$\chi_{2,1} \mapsto \chi_{1,3}$
which is the reverse of the physical flow.
These mechanisms are illustrated for the states:
A, $(0\,0\,0|0\,0)\,\mapsto\,(0\,0\,0|0)_{+}$;
B, $(1\,1\,1|0\,0)\,\mapsto\,(0\,0\,0|1)_{-}$;
C, $(0)\,\mapsto\,(0|0)_{-}$.}
\end{figure}

The pattern of zeros of $D(u,\xi)$ changes along the flow: $m_1$ remains
odd while $m_2$ changes from even to odd, according to Table~\ref{new_critTBA}.
$D(u,\xi)$ is a polynomial of order $2N+2$ for $\xi>-\infty$ in the variable
$\exp (iu)$ inside the periodicity strip
$-\lambda < \re(u)\leqslant 4\lambda$ and
becomes of order $2N$ at the UV point itself. We therefore expect 
precisely two zeros to
``escape to infinity'' along the flow, one for each complex half plane
($\im(u)<0$ or $\im(u)>0$).
The ``numerics on $D$'' shows three active mechanisms that, for the 
most part, only involve
the furthest zeros from the real axis in the second strip. They are distinct
from the mechanisms described for the
constant $r$ flows but show many similarities. They were described in
\cite{FPR} for the flow $\chi_{1,2}$ to $\chi_{2,1}$, dual to the
present case. More detailed ``numerics on $D$'' allows us to
be more precise in describing the 2-string movements, as given in
Figure~\ref{ABCmech} and Table~\ref{t_s1321}.
\begin{list}{}{}
\item[A.] If the top object in both strips is a 1-string, the top 1-string
in strip~2 moves to $+\infty$, decoupling from the system exactly at the UV
point where it produces a frozen state with $\sigma=1$.
\item[B.]  If there is a top 2-string in strip~1 and a top 1-string in
strip~2, they move away from the real axis towards $+\infty$.
A 2-string comes in strip~2 becoming the top 2-string.
Consequently, the quantum numbers in strip 1 decrease by 1
and those in strip 2 increase by $1$.
\item[C.] If there is a top 2-string in strip~2, it moves
away from the real axis and collapses into a pair of
{\em correlated 1-strings}. The furthest of these 1-strings from the 
real axis moves to $+\infty$,
reaching it precisely at the UV point. The 1-string closest to the 
real axis remains in the scaling region and
becomes the top 1-string at the UV point. The quantum numbers in strip~2
thus decrease by 1.
\end{list}
These mechanisms are initially observed by the ``numerics on $D$''
then verified solving the TBA equations.
For the mechanism B, the description of the movement of the 2-strings
is still incomplete. We can imagine a first scenario where the relevant
2-strings remain exactly located on the border of the corresponding strip so
that they need to reach $+\infty$ to be able to move from one strip to
the other. If this is the case, a quite unusual exchange between the 1 and
2 strings in the second strip must occur. It can be avoided in
a second scenario where the 1-string only reaches $+\infty$; the zeros
forming the 2-string leave the first strip by the two sides pointing toward
the second strip where they become the top 2-string. If this is the case,
to keep track of the
movement of the 2-string we need to solve a much more complicated system of
functional equations involving
\begin{equation}
\hat{t}_1(x) \, \hat{t}_2(x+i \frac{3}{2}\pi) =1+\hat{t}_2(x-i \frac{\pi}{2}) ,
\qquad -\pi<\im(x)<0.
\end{equation}
A 2-string located between the first and second strip can be fixed by
$\hat{t}_2(x+i\frac{3}{2}\pi)=0$ leading to
\begin{equation}
1+\hat{t}_2(x-i \frac{\pi}{2})=0, \qquad  -\frac{\pi}{2}<\im(x)<0
\end{equation}
where both the real and imaginary parts of $x$ must be determined.
This is equivalent to analytically continue $\hat{t}_1,~\hat{t}_2$ out of the
fundamental interval (\ref{interval}). Actually, the TBA equations are
completely independent of the behaviour of the 2-strings so we can 
still compute scaling functions even if we miss the complete 
description of the movement of the 2-strings.

The description of the mechanisms A,B,C forces the following changes for
the various parameters (if no label IR or UV is used the corresponding
parameter does not change during the flow):
\begin{list}{}{\itemindent 12mm}
\item[A. $I^{(1)}_{m_1}={I^{(2)}_{m_2\irt}}\ir=0$:]
\begin{eqnarray} \label{mech2A}
m_2\ir &\mapsto& m_2\uv=m_2\ir-1, \qquad \sigma=1.
\end{eqnarray}
\item[B. ${I^{(1)}_{m_1}}\ir> 0, \quad {I^{(2)}_{m_2\irt}}\ir=0$:]
\begin{eqnarray} \label{mech2B}
m_2\ir &\mapsto& m_2\uv=m_2\ir-1, \qquad \sigma=-1, \nonumber \\[2mm]
{I^{(1)}_k}\ir &\mapsto & {I^{(1)}_{k}}\uv={I^{(1)}_{k}}\ir-1,
\qquad k=1,\ldots,m_1, \\[2mm]
{I^{(2)}_k}\ir &\mapsto & {I^{(2)}_k}\uv={I^{(2)}_k}\ir+1,
\qquad k=1,\ldots,m_2\ir-1. \nonumber
\end{eqnarray}
\item[C. ${I^{(2)}_{m_2\irt}}\ir>0$:]
\begin{eqnarray} \label{mech2C}
m_2\ir &\mapsto& m_2\uv=m_2\ir+1, \qquad \sigma=-1, \qquad
{I^{(2)}_{m_2\uvt}}\uv=0, \nonumber \\[2mm]
{I^{(2)}_k}\ir &\mapsto & {I^{(2)}_k}\uv={I^{(2)}_k}\ir-1,
\qquad k=1,\ldots,m_2\ir.
\end{eqnarray}
\end{list}
The correlated 1-strings introduced here are in strip~2 while
those introduced in Section~\ref{mech1211} were in strip~1
(and connected to the frozen case).

\begin{table}[htb]
\caption{\label{t_s1321}
\small
Flow $\chi_{2,1} \mapsto \chi_{1,3}$ (reverse of the physical flow).
We present the explicit mapping of states from IR to UV up to the UV 
level 6. Here
$n\ir,\:n\uv$ are the excitation levels
above the ground states, respectively $h=7/16$ and $h=3/5$.}
\begin{center}
\begin{tabular}{|c|r@{$\:\:\mapsto\:\:$}l|c|c||c|r@{$\:\:\mapsto\:\:$}l|c|c|}
\hline
$n\ir$\rule[-4mm]{0mm}{10mm}& \multicolumn{3}{c|}{Mapping of states --
mechanism} & $n\uv$ &
   $n\ir$ & \multicolumn{3}{c|}{Mapping of states -- mechanism} & $n\uv$\\
\hline
0\rule[-1mm]{0mm}{6mm} & $(0)$ & $(0|0)_{-}$ & C & 0 &
                  5 & $(1\,1\,0|0\,0)$ & $(1\,1\,0|0)_{+}$ & A & 4\\[1mm]
1 & $(1)$ & $(1|0)_{-}$ & C & 1 &
                  5 & $(1\,0\,0)$ & $(1\,0\,0|0)_{-}$ & C & 5 \\[1mm]
2 & $(2)$ & $(2|0)_{-}$ & C & 2 &
                  5 & $(5)$ & $(5|0)_{-}$ & C & 5 \\[1mm]
3 & $(0\,0\,0|0\,0)$ & $(0\,0\,0|0)_{+}$ & A & 2 &
                  6 & $(1\,1\,1|0\,0)$ & $(0\,0\,0|1)_{-}$ & B & 5 \\[1mm]
3 & $(3)$ & $(3|0)_{-}$ & C & 3 &
                  6 & $(3\,0\,0|0\,0)$ & $(3\,0\,0|0)_{+}$ & A & 5 \\[1mm]
4 & $(1\,0\,0|0\,0)$ & $(1\,0\,0|0)_{+}$ & A & 3 &
                  6 & $(2\,1\,0|0\,0) $ & $(2\,1\,0|0)_{+}$ & A & 5 \\[1mm]
4 & $(0\,0\,0)$ & $(0\,0\,0|0)_{-}$ & C & 4 &
                  6 & $(6)$ & $(6|0)_{-}$ & C & 6 \\[1mm]
4 & $(4)$ & $(4|0)_{-}$ & C & 4 &
                  6 & $(2\,0\,0)$ & $(2\,0\,0|0)_{-}$ & C & 6\\[1mm]
5 & $(2\,0\,0|0\,0)$ & $(2\,0\,0|0)_{+}$ & A & 4 &
                  6 & $(1\,1\,0)$ & $(1\,1\,0|0)_{-}$ & C & 6 \\[1mm]
\hline
\end{tabular}
\end{center}
\end{table}

\subsubsection{RG mapping between finitized characters}
In this section we show that the three mechanisms A,B,C are compatible with the
counting of states, given by the finitized
characters \cite{FinChar}, at the two conformal endpoints of the flow.

 From the definitions, we observe that for each
IR state there is precisely one applicable mechanism
so that the counting of states is complete.
Moreover, using (\ref{recourrence}), the IR finitized character
naturally splits into three terms precisely associated with the three 
mechanisms
\begin{eqnarray}
\chi_{2,1}^{(N)}(q) & = &
q^{-\frac{c}{24}+\Delta_{2,1}-\frac{1}{2}} \sum_{m_1,\,m_2\irt}
q^{\frac{1}{4}{\boldsymbol{m}\irt}C\boldsymbol{m}\irt}
\gausst{m_1+n_1\irt}{m_1} \gausst{m_2\irt+n_2\irt}{m_2\irt} \nonumber \\
& = &
q^{-\frac{c}{24}+\Delta_{2,1}-\frac{1}{2}} \left\{ \sum_{\text{A}}
q^{\frac{1}{4}{\boldsymbol{m}\irt}C\boldsymbol{m}\irt}
\gausst{m_1-1+n_1\irt}{m_1-1}  \gausst{m_2\irt-1+n_2\irt}{m_2\irt-1}
\right. \nonumber  \\
&&\quad\mbox{} +  \sum_{\text{B}}
q^{\frac{1}{4}{\boldsymbol{m}\irt}C\boldsymbol{m}\irt} q^{m_1}
\gausst{m_1+n_1\irt-1}{m_1} \gausst{m_2\irt-1+n_2\irt}{m_2\irt-1}
\nonumber \\
&&\quad\mbox{} +\left. \sum_{\text{C}}
q^{\frac{1}{4}{\boldsymbol{m}\irt}C\boldsymbol{m}\irt} q^{m_2\irt}
\gausst{m_1+n_1\irt}{m_1} \gausst{m_2\irt+n_2\irt-1}{m_2\irt} \right\}
\nonumber \\
& = & q^{-\frac{c}{24}+\Delta_{2,1}-\frac{1}{2}} \left\{ \sum_{\text{A}}
q^{\frac{1}{4}{\boldsymbol{m}\irt}C\boldsymbol{m}\irt}
\gausst{(N+m_2\irt)/2-1}{m_1-1}  \gausst{(m_1-1)/2}{m_2\irt-1}
\right. \nonumber \\
&&\quad\mbox{} + \sum_{\text{B}}
q^{\frac{1}{4}{\boldsymbol{m}\irt}C\boldsymbol{m}\irt} q^{m_1}
\gausst{(N+m_2\irt)/2-1}{m_1} \gausst{(m_1-1)/2}{m_2\irt-1}
\nonumber \\
&&\quad\mbox{} +\left. \sum_{\text{C}}
q^{\frac{1}{4}{\boldsymbol{m}\irt}C\boldsymbol{m}\irt} q^{m_2\irt}
\gausst{(N+m_2\irt)/2}{m_1} \gausst{(m_1-1)/2}{m_2\irt} \right\}
\label{char21}
\end{eqnarray}
Here and below we attach the labels IR, UV to the variables that 
change under the flow. The labels A,B,C on the  sums indicate that 
the sums on $m_1$, $m_2\irt$ are
restricted by the constraints imposed by the corresponding mechanism.
The sum constrained by A can be understood using (\ref{Izero}) for both
the $q$-binomial factors in (\ref{char21}) and corresponds to summing
under the constraint $I^{(1)}_{m_1}={I^{(2)}_{m_2\ir}}\ir=0$.
Similarly, the sum on B becomes apparent using (\ref{Ipos}) for the first strip
$q$-binomial and (\ref{Izero}) for the second strip $q$-binomial
and likewise the sum on C uses (\ref{Ipos}) for the second strip only.
The relations in Table~\ref{new_critTBA} were used to rewrite the
$q$-binomials as functions of $m_1,\,m_2\ir,\,N$ only.

We now use the properties of the three mechanisms to map this
expression for the finitized IR character (\ref{char21}) into the 
finitized UV character.
An IR energy level at the base of a tower of states with string 
content fixed by
$(m_1,m_2\ir)$ maps to a UV energy level according to the energy
expression (\ref{energy}) that holds at the two conformal endpoints of the flow
\begin{equation} \label{energyABC}
\begin{array}{l@{\hspace{7mm}}l@{~~\mapsto~~}l}
\mbox{A:} & q^{\Delta_{2,1}-\frac{1}{2}} \,
q^{\frac{1}{4}\boldsymbol{m}\irt C\boldsymbol{m}\irt}
& q^{\Delta_{1,3}-\frac{1}{2}} \,
q^{\frac{1}{4}\boldsymbol{m}\uvt C\boldsymbol{m}\uvt} \,
q^{-\frac{1}{2}(m_1-m_2\uvt)}
\\[2mm]
\mbox{B:} & q^{\Delta_{2,1}-\frac{1}{2}}  \,
q^{\frac{1}{4}\boldsymbol{m}\irt C\boldsymbol{m}\irt} \, q^{m_1}
& q^{\Delta_{1,3}-\frac{1}{2}} \,
q^{\frac{1}{4}\boldsymbol{m}\uvt C\boldsymbol{m}\uvt} \,
q^{+\frac{1}{2}(m_1-m_2\uvt)} \,q^{m_2\uvt} \\[2mm]
\mbox{C:} & q^{\Delta_{2,1}-\frac{1}{2}}  \,
q^{\frac{1}{4}\boldsymbol{m}\irt C\boldsymbol{m}\irt} \, q^{m_2\irt}
& q^{\Delta_{1,3}-\frac{1}{2}} \,
q^{\frac{1}{4}\boldsymbol{m}\uvt C\boldsymbol{m}\uvt} \,
q^{+\frac{1}{2}(m_1-m_2\uvt)}.
\end{array}
\end{equation}
Using (\ref{mech2A}) to (\ref{mech2C}) we rewrite the $q$-binomials
appearing in the last three lines of (\ref{char21}). Also taking into
account the mapping of the energies (\ref{energyABC}) we obtain
\begin{eqnarray} \chi_{2,1}^{(N)}(q)
& \mapsto & q^{-\frac{c}{24}+\Delta_{1,3}-\frac{1}{2}} \left\{ \sum_{\text{A}}
q^{\frac{1}{4}{\boldsymbol{m}\uvt}C\boldsymbol{m}\uvt} \,
q^{-\frac{1}{2}(m_1-m_2\uvt)}
\gausst{(N+m_2\uvt-1)/2}{m_1-1}  \gausst{(m_1-1)/2}{m_2\uvt}
\right. \nonumber \\
&&\quad\mbox{} +  \sum_{\text{B}}
q^{\frac{1}{4}{\boldsymbol{m}\uvt}C\boldsymbol{m}\uvt}\,
q^{\frac{1}{2}(m_1-m_2\uvt)} q^{m_2\uvt}
\gausst{(N+m_2\uvt-1)/2}{m_1} \gausst{(m_1-1)/2}{m_2\uvt}
\nonumber \\
&&\quad\mbox{}+ \left. \sum_{\text{C}}
q^{\frac{1}{4}{\boldsymbol{m}\uvt}C\boldsymbol{m}\uvt} \,
q^{\frac{1}{2}(m_1-m_2\uvt)}
\gausst{(N+m_2\uvt-1)/2}{m_1} \gausst{(m_1-1)/2}{m_2\uvt-1} \right\}
\nonumber\\
& = & q^{-\frac{c}{24}+\Delta_{1,3}-\frac{1}{2}} \left\{ \sum_{\text{A}}
q^{\frac{1}{4}{\boldsymbol{m}\uvt}C\boldsymbol{m}\uvt} \,
q^{-\frac{1}{2}(m_1-m_2\uvt)}
\gausst{m_1-1+n_1\uvt}{m_1-1}  \gausst{m_2\uvt+n_2\uvt}{m_2\uvt}
\right. \nonumber \\
&&\quad\mbox{}+ \sum_{\text{B}}
q^{\frac{1}{4}{\boldsymbol{m}\uvt}C\boldsymbol{m}\uvt}\,
q^{\frac{1}{2}(m_1-m_2\uvt)} q^{m_2\uvt}
\gausst{m_1+n_1\uvt}{m_1} \gausst{m_2\uvt+n_2\uvt-1}{m_2\uvt}
\nonumber \\
&&\quad\mbox{}+ \left. \sum_{\text{C}}
q^{\frac{1}{4}{\boldsymbol{m}\uvt}C\boldsymbol{m}\uvt} \,
q^{\frac{1}{2}(m_1-m_2\uvt)}
\gausst{m_1+n_1\uvt}{m_1} \gausst{m_2\uvt-1+n_2\uvt}{m_2\uvt-1} \right\}
\nonumber \\
& = & \chi_{1,3}^{(N)}(q)
\label{char13}
\end{eqnarray}
The last equality in (\ref{char13}) follows from the
$q$-binomial identities in Appendix A.
Again this argument shows the consistency of the IR and UV counting of states
with the three mechanisms A,B,C.

\subsubsection{Order 1 analyticity and TBA equations}
The steps required to solve the functional equation for the scaling 
functions $t_j(x,\xi)$
described in Sections~\ref{sss_func} to \ref{sss_qn} also apply for 
the flow $\chi_{1,3} \mapsto \chi_{2,1}$, the only differences
being related to the boundary terms $g_j$ and the order~1 analyticity
contained in the factors $S_1(u)$, $S(u)$, $S_3(u, \xilatt)$ of (\ref{normal}).
$S_1(u)$ cancels with terms from $B^{1,1}$ and disappears.
Because of the factor $S(u)$, $t_1(x)$ and $t_2(x)$ both have a 
double zero at $x=0$
and two simple poles at $x=\pm i\pi$, independent of $\xi$.
We now have
\begin{equation} \label{s3}
S_{3}(u,\xilatt)=\frac{\sin \!\lambda \, \sin(u+\lambda+\xilatt)\,
\sin(u+3\lambda-\xilatt)\,\cosh 2\im(\xilatt)}{\sin(u+\xilatt)\,
\sin(u+2\lambda-\xilatt)\,\sin(u-\lambda-\xilatt)\,\sin(u+2\lambda+\xilatt)}
\end{equation}
where $\xilatt=-3\lambda/2+i\xi/5$, so we find simple zeros at
$u=\lambda/2\pm i\xi/5$, that are in the first strip at $x=\pm \xi$, and
simple poles at $u=-\lambda/2\pm i\xi/5,~3\lambda/2\pm i\xi/5 $
that are in the first strip at $x=\pm i \pi \pm \xi$ (all four combinations of
signs). Remarkably, the second strip is free of zeros and poles 
dependent on $\xi$.
This leads to the same plot as in Figure~\ref{order1anal} with the exchange
of the two strips.
Observe that the indicated poles are on the edges of the strips
and the zeros in the middle so, using (\ref{t1system}), (\ref{t2system})
we conclude that $1+t_1(x)$ and $1+t_2(x)$ are free of poles and
are nonzero inside $|\im(x)|<\pi/2 $ (\ref{interval}).
This implies that the order~1 factors can be removed from the functional
system (\ref{t1system}), (\ref{t2system}) using
\begin{eqnarray}
\label{g1eq} g_1(x+i\frac{\pi}{2}) \: g_1(x-i\frac{\pi}{2}) &= &1 \\
\label{g2eq} g_2(x+i\frac{\pi}{2})\: g_2(x-i\frac{\pi}{2}) &= &1
\end{eqnarray}
with the indicated analytic properties.
This system has the same form as for the constant $r$ flows (\ref{g1equation}),
(\ref{g2equation}), the difference being in the analytic content.
The solution is the previous one (\ref{g1}), (\ref{g2}) but with the 
exchange of
the two strips
\begin{eqnarray}
\label{g1v} g_1(x,\xi) &= &\tanh ^2 \frac{x}{2} \tanh \frac{x-\xi}{2}
\tanh \frac{x+\xi}{2} \\
\label{g2v} g_2(x,\xi) &= &\tanh ^2 \frac{x}{2}
\end{eqnarray}
so the scaling limit yields
\begin{eqnarray}
\label{hatg1v} \hat{g}_1(x,\xi) &= & \tanh \frac{x+\xi}{2} \\
\label{hatg2v} \hat{g}_2(x,\xi) &= & 1.
\end{eqnarray}
We are led to the same TBA equations (\ref{tba1}), (\ref{tba2}), energy
expression (\ref{scalingenergy}) and quantization conditions
(\ref{quant1}) to (\ref{psi2}) as in the constant $r$ flows.
Similarly, the expressions for the quantum numbers (\ref{qn1}), 
(\ref{qn2}) are not modified
from the IR expressions so that
\begin{equation}
m_1 \mbox{~odd}, \qquad m_2 \mbox{~even}, \qquad s_1=-1,\qquad s_2=1.
\end{equation}

As a complement to the given description of the mechanisms and to the TBA
equations, we summarize the main
results concerning this flow starting from the IR point:
\\[3mm]\indent
\begin{tabular}{l}
{\bf mechanism A and B, mechanism C before the collapse:} \\[2mm]
\hspace*{10mm} $ \begin{array}{l}
m_1 \mbox{ ~odd}, ~m_2 \mbox{ ~even}, \qquad
n_2=\frac{m_1+1}{2}-m_2 \geqslant 0 ,\qquad
s_1=-1, ~s_2=1,  \\[2mm]
n_k^{(1)} = 2(I^{(1)}_k+m_1-k)+1-m_2, \\[2mm]
n_k^{(2)} = 2(I^{(2)}_k+m_2-k)+1-m_1;  \end{array} $ \\[12mm]
{\bf mechanism C after the collapse:} \\[2mm]
\hspace*{10mm} $ \begin{array}{l}
m_2^A=m_2+2, \qquad n^{(2)}_{m_2^A-1}=n^{(2)}_{m_2^A}=1-m_1 , \\[2mm]
\mbox{the values of $~m_1, ~n_1, ~n_k^{(1)}, ~n_k^{(2)}$ with $~k<m_2^{A}-1$,
$~s_j$, ~remain unchanged.}
\end{array} $
\end{tabular}

\subsection{RG flow $\chi_{1,2}\mapsto\chi_{2,1}$\label{ss_1221}}

The preliminarly treatment of this case was presented in the letter~\cite{FPR}
but we now have a more
complete understanding of the mechanisms of movement of the zeros
and of the role played by $\re(\xilatt)$ along the flow so we need to correct
and clarify some statements.

\begin{table}[htb]
\caption{\label{t_s1221}
\small
Flow $\chi_{2,1} \mapsto \chi_{1,2}$ (reverse of the physical flow).
We present the explicit mapping of states from IR to UV up to the UV 
level 6. Here
$n\ir,\:n\uv$ are the excitation levels
above the ground states, respectively $h=7/16$ and $h=1/10$.}
\begin{center}
\begin{tabular}{|c|r@{$\:\:\mapsto\:\:$}l|c|c||c|r@{$\:\:\mapsto\:\:$}l|c|c|}
\hline
$n\ir$\rule[-4mm]{0mm}{10mm}& \multicolumn{3}{c|}{Mapping of states --
mechanism} & $n\uv$ &
   $n\ir$ & \multicolumn{3}{c|}{Mapping of states -- mechanism} & $n\uv$\\
\hline
0\rule[-1mm]{0mm}{6mm} & $(0|0)$ & $(0)_{+}$ & A & 0 &
                  5 & $(5|0)$ & $(4)_{-}$ & B & 5\\[1mm]
1 & $(1|0)$ & $(0)_{-}$ & B & 1 &
                  5 & $(1\,1\,0|0)$ & $(1\,1\,0)_{+}$ & A & 5 \\[1mm]
2 & $(2|0)$ & $(1)_{-}$ & B & 2 &
                  5 & $(2\,0\,0|0)$ & $(2\,0\,0)_{+}$ & A & 5 \\[1mm]
3 & $(3|0)$ & $(2)_{-}$ & B & 3 &
                  6 & $(1\,1\,0|1)$ & $(1\,1\,0|0\,0)_{-}$ & C & 6 \\[1mm]
3 & $(0\,0\,0|0)$ & $(0\,0\,0)_{+}$ & A & 3 &
                  6 & $(2\,0\,0|1)$ & $(2\,0\,0|0\,0)_{-}$ & C & 6 \\[1mm]
4 & $(0\,0\,0|1)$ & $(0\,0\,0|0\,0)_{-}$ & C & 4 &
                  6 & $(6|0) $ & $(5)_{-}$ & B & 6 \\[1mm]
4 & $(4|0)$ & $(3)_{-}$ & B & 4 &
                  6 & $(1\,1\,1|0)$ & $(0\,0\,0)_{-}$ & B & 6 \\[1mm]
4 & $(1\,0\,0|0)$ & $(1\,0\,0)_{+}$ & A & 4 &
                  6 & $(2\,1\,0|0)$ & $(2\,1\,0)_{+}$ & A & 6\\[1mm]
5 & $(1\,0\,0|1)$ & $(1\,0\,0|0\,0)_{-}$ & C & 5 &
                  6 & $(3\,0\,0|0)$ & $(3\,0\,0)_{+}$ & A & 6 \\[1mm]
\hline
\end{tabular}
\end{center}
\end{table}

The analysis done in the previous section for the flow
$\chi_{1,3}\mapsto\chi_{2,1}$ can be repeated for this flow with
$\mathbf{D}(u,\xi) \equiv \mathbf{D}_{1,1|2,1}^N (u,\xilatt)$
and $\xilatt=\frac{3}{2}\lambda+i\frac{\xi}{5}$. According to (\ref{rtype}) and
(\ref{stype}), this choice of the boundary field gives an $r=2$ type
boundary for $\xi=0$ and an $s=2$ type boundary for
$\xi \rightarrow \pm \infty$.
Moreover, this is the only value of $\re(\xilatt)$ that leads to
real analytic transfer matrices (\ref{reality}) and, consequently, to
real positive energies.
We need to analyse the order~1 analyticity.
The relevant term is
\begin{equation}
S_{2}(u,\xilatt)=\frac{\sin \!\lambda \, \sin(\xilatt-u-\lambda)\,
\sin(\xilatt+u+3\lambda)\,\cosh 2\im(\xilatt)}
{\sin(\xilatt-u)\,\sin(\xilatt-u-2\lambda)
\,\sin(\xilatt+u-\lambda)\,\sin(\xilatt+u+2\lambda)}
\end{equation}
that has exacly the same zeros and poles obtained for the
previous flow (\ref{s3}): in the first strip there are two single zeros at
$u=\lambda/2 \pm i\xi/5$ ($x=\pm \xi$) and four single
poles at $u=-\lambda/2 \pm i\xi/5$, $u=3\lambda/2 \pm i\xi/5$
($x=\pm i\pi \pm \xi$), with no zeros or poles in the second strip.
This leads to the same TBA equations and the same boundary term
\begin{eqnarray}
\hat{g}_1(x,\xi) &= & \tanh \frac{x+\xi}{2} \\
\hat{g}_2(x,\xi) &= & 1
\end{eqnarray}
obtained in Section~\ref{ss_1321}, the only difference being
in the following parameters that we take from \cite{FPR}:
$m_1$,  $m_2$ are both odd, $s_1=s_2=-1$ as at the IR starting point.

The mechanisms were explained in~\cite{FPR} and the mapping between
characters (see also Table~\ref{t_s1221})
\begin{equation}
\chi_{2,1}^{(N)}(q)  \mapsto  \chi_{1,2}^{(N)}(q)
\end{equation}
was explicitly computed showing the consistency of the mechanisms with the
IR and UV counting.
Notice that $\hat{t}_j(x)$ actually has no trace of the ``pole''
discussed erroneously there, instead the zero of $\hat{g}_1$ at $x=-\xi$ gives
a zero to $\hat{t}_1(-\xi)$ and, correspondingly, a zero to $L_1(-\xi)$.

We summarize the main results concerning this flow starting from
the IR point:
\\[3mm] \indent
\begin{tabular}{l}
{\bf mechanism A and B, mechanism C before the collapse:}  \\[2mm]
\hspace*{10mm} $ \begin{array}{l}
m_1, ~m_2 \mbox{ ~odd}, \qquad n_2=\frac{m_1+1}{2}-m_2 \geqslant 0 ,\qquad
s_1=s_2=-1,  \\[2mm]
n_k^{(1)} = 2(I^{(1)}_k+m_1-k)+1-m_2, \\[2mm]
n_k^{(2)} = 2(I^{(2)}_k+m_2-k)+1-m_1;  \end{array} $ \\[12mm]
{\bf mechanism C after the collapse:} \\[2mm]
\hspace*{10mm} $ \begin{array}{l}
m_2^A=m_2+2, \qquad n^{(2)}_{m_2^A-1}=n^{(2)}_{m_2^A}=1-m_1 , \\[2mm]
\mbox{the values of $~m_1, ~n_1, ~n_k^{(1)}, ~n_k^{(2)}$ with $~k<m_2^{A}-1$,
$~s_j$, ~remain unchanged.}
\end{array} $
\end{tabular}

\goodbreak
\section{Numerical solution of TBA equations\label{s_numerical}}

The TBA equations (\ref{tba1}), (\ref{tba2}) can be solved analytically at
the two conformal fixed points, which occur at the endpoints $\xi=\pm 
\infty$ of the flow,
leading to the results in \cite{OPW}. In particular, the energy
expression (\ref{energy}) and the finitized characters can be 
obtained analytically.
It appears that the TBA equations cannot be solved analytically along 
the flow, however, with numerical computations we can obtain in 
detail the behaviour of the
flow for intermediate values of $\xi$ to interpolate between the 
known conformal fixed points.

Our numerical algorithm to solve the TBA equations and auxiliary 
equations for the zeros is iterative. We make suitable initial 
guesses for the pseudo-energies and the locations of the zeros close 
to one of the conformal fixed points.
We then iteratively update in turn the pseudo-energies and locations 
of the zeros
to find new values until the iteration scheme converges to the 
required accuracy. We then increment or decrement the value of the 
field $\xi$ and repeat the process until we scan the full range in 
$\xi$.
This procedure is natural for
the pseudo-energies in the TBA equations but iteration of the 
quantization conditions for the location of the zeros requires 
inversion of some complicated expressions associated with phases that 
wind along the flow.
 From the expression for $\psi_2$ (\ref{quant1}), (\ref{psi2}), we
can extract the zero $y_\ell^{(1)}$ by one of the following:\\[6pt]
\noindent {\bf first strip algorithms:} \\[3pt]
\hspace*{4mm} \begin{tabular}{l@{\hspace{5mm}}l}
{\em exp}: & inversion of the exponential term $e^{-x}$; \\
$g_2$: & inversion of the boundary term, for the constant $r$ flows only;\\
$\psi_2$: & scan the domain of the function
$\psi_2(x)-\pi n_\ell^{(1)}$ to find its zero; \\
{\em phase}: & inversion of one of the phases
$i\log \tanh (\frac{y_k^{(2)}-y_\ell^{(1)}}{2}-\frac{i\pi}{4})$
(not used here).
\end{tabular}

\noindent
Similarly, from the expression for $\psi_1$ (\ref{quant2}), (\ref{psi1}) we
can extract the zero $y_\ell^{(2)}$ by one of the following:\\[6pt]
\noindent {\bf second strip algorithms:} \\[3pt]
\hspace*{4mm} \begin{tabular}{l@{\hspace{5mm}}l}
$g_1$: & inversion of the boundary term, for the variable $r$ flows only;\\
$\psi_1$: & scan the domain of the function
$\psi_1(x)-\pi n_\ell^{(2)}$ to find its zero; \\
{\em phase}: & inversion of one of the phases
$i\log \tanh (\frac{y_k^{(1)}-y_\ell^{(2)}}{2}-\frac{i\pi}{4})$.
\end{tabular}

\medskip
\noindent
We start with the simplest choice, algorithm {\em exp, phase} in strips 1, 2
respectively.
The computations at the critical points in \cite{OPW}
can all be done with this choice and do not require the introduction of other
algorithms.
We emphasize that these different algorithms are not equivalent under iteration
and it can happen that some schemes fail to converge in some
intervals. In every case, however, the validity of our use of 
different algorithms in different intervals is confirmed by the fact 
that the plots of the scaling energies, and the locations of all of 
their associated zeros, join smoothly in some typically small 
overlapping interval of convergence.
         
The {\em contraction mapping theorem}, says that if a 
mapping $f: M \rightarrow M$ on a
complete metric space $M$ is a contraction
then there exists a unique fixed point $x_0=f(x_0)$ and
all the sequences obtained under iteration starting from an arbitrary 
initial point $x\in M$
converge to the fixed point.
For our purposes, we take $M \subseteq \mathbb{R}$ to be a closed interval
and a contraction means that the first derivative is $|f'(x)|\leqslant k<1$
inside $M$, for a fixed $k$. In particular,
$|f'(x_0)|$ is a measure of the rate of convergence. We will see that, in
some cases, $|f'(x_0)| \gg 1$ and although the fixed point exists it cannot be
reached by iterating the mapping $f$.

\subsection{Flow $\chi_{1,2}\mapsto\chi_{1,1}$\label{ss_n1211}}
\begin{figure}\begin{center}
\includegraphics[width=0.85\linewidth ]{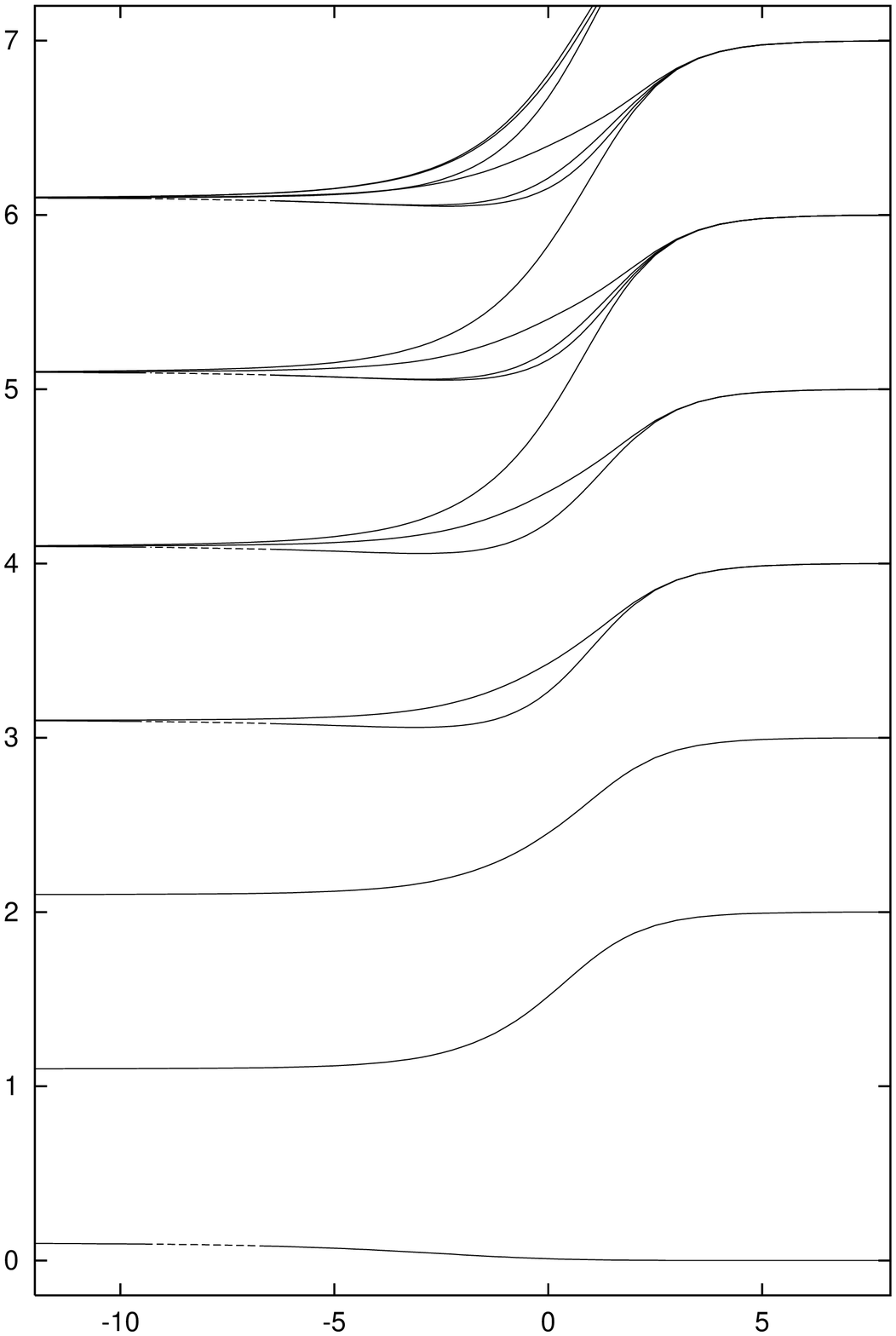}
\caption{\small Scaling energies for the flow $\chi_{1,2} \mapsto \chi_{1,1}$.
The list of states is given in Table~\ref{t_s1211}. The intermediate region 
of the mechanism~A levels (shown dashed) are schematic and have not 
been obtained from the solution of the TBA equations.
\label{energy1211}}
\end{center}
\end{figure}

The scaling energies for the states in Table~\ref{t_s1211}
are presented in Figure~\ref{energy1211}.
The second strip is a spectator and is not effected by mechanisms A and B in
(\ref{mechanismA}), (\ref{mechanismB}), so it always has $m_2$ even and is not
relevant in the following discussion.
As usual, for numerical convenience, we describe the reverse of the
physical flow.

\begin{figure}[ht] \begin{center}
\includegraphics[scale=0.9 ]{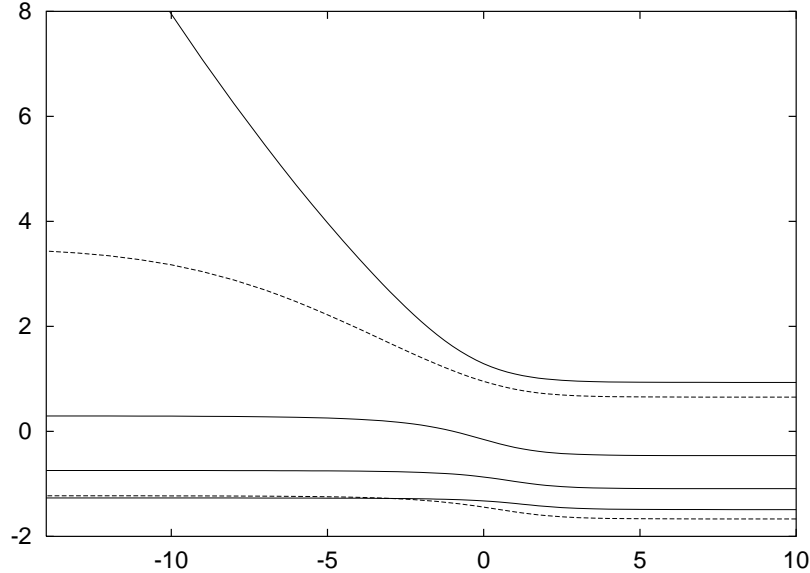}
\caption{\small Movement of the zeros versus $\xi$ for the state
$(0\,0\,0|0\,0)_{-} \mapsto (0\,0\,0\,0|0\,0)$, described by mechanism B,
in the flow $\chi_{1,2} \mapsto \chi_{1,1}$.
Solid lines from bottom to top are $y_1^{(1)},\ldots,y_{4}^{(1)}$,
dashed lines from bottom to top are $y_1^{(2)},~y_2^{(2)}$.
This picture is consistent with
the lattice predictions in part B of Fig.~\ref{ABmech}.\label{zeros1211-}}
\end{center}
\end{figure}

 From Table~\ref{t_s1211}, we begin by following the simpler
{\bf mechanism B} that first occurs in the excitation
$(0\,0)\mapsto (0)_{-}$ and has only one strip of zeros.
We start decreasing $\xi$ from the IR point $\xi \rightarrow +\infty$ with
algorithm {\em exp} and make the following observations:
\begin{itemize}
\item the numerical iteration converges in the interval
$\xi \gtrsim -2$ but fails for smaller values;
\item the zero $y_{2}^{(1)}$ moves towards
$+\infty$, consistent with the lattice description of mechanism B while
$y_{1}^{(1)}$ moves slowly in the finite region, as shown in
Figure~\ref{zeros1211-}.
\end{itemize}
If we denote by $f(x)$ the mapping obtained
by inverting the exponential term $e^{-x}$ in (\ref{psi2}) to solve for $x$
and follow the value of $f'(y_2^{(1)})$ as we vary $\xi$ we see that
$f'(y_2^{(1)})\lesssim 0.1$ for large $\xi \gg 0$ but 
$f'(y_2^{(1)})\to -\infty$
for $\xi \rightarrow -\infty$ while it
remains bounded for the other zero, $|f'(y_1^{(1)})| \lesssim 0.1$, for all
values of $\xi$.
This is a clear indication that the zero $y_{2}^{(1)}=y_{m_1}^{(1)}$
cannot be reached by iteration. We certainly know that it exists because
in the whole interval $\xi \gtrsim -2$ we can track it and at
$\xi=-2$ it is at a finite position, $y_{2}^{(1)} \approx 1.49$.
An analytic estimation of $f'(y_2^{(1)})$ shows that the dominant behaviour
is $f'(y_2^{(1)}) \sim - e^{-y_2^{(1)}}$ if $y_2^{(1)} $ is sufficiently
large, as we expect in the region $\xi\rightarrow -\infty$,
see also Figure~\ref{zeros1211-}.
These considerations show that we need a different algorithm to determine
the location of this zero.

If we try with algorithm $g_2$ we observe the opposite behaviour for
the mapping $f(x)$ obtained by inverting the boundary term $\log g_2$ in
(\ref{psi2}):  $f'(y_2^{(1)}) \rightarrow -\infty$
for large $\xi \gg 0$ and $f'(y_2^{(1)}) $ is small for $\xi \lesssim 
-1.5 $ so we can start
from the UV point and combine the results
from this algorithm with the previous ones, thanks to the existence 
of a small overlapping
region of applicability.

Actually, we find that the algorithm $\psi_2$ is more general in the sense
that it can be applied everywhere, being intrinsically non-iterative.
We always use this choice for the largest first strip zero $y_{m_1}^{(1)}$
(the other first strip zeros can be fixed with the slightly faster algorithm
{\em exp}).
A similar problem shows up even in the second strip for higher
excited states so we use the non-iterative algorithm
$\psi_1$ for all the second strip zeros $y_{k}^{(2)}$, $k=1,\ldots,m_2$.

Once the appropriate algorithms are applied, the numerical analysis 
of the states
proceeds smoothly from IR to UV and the predictions of mechanism B 
are completely
confirmed. In particular,
the IR description of zeros and quantum numbers can be used all
along the flow and $y_{m_1}^{(1)}$ only reaches $+\infty$ at the UV 
critical point.

\setlength{\unitlength}{1mm}
\begin{figure}[ht] \begin{center} \setlength{\unitlength}{1mm}
\begin{picture}(100,122)(12,3)
\put(0,83){\includegraphics[]{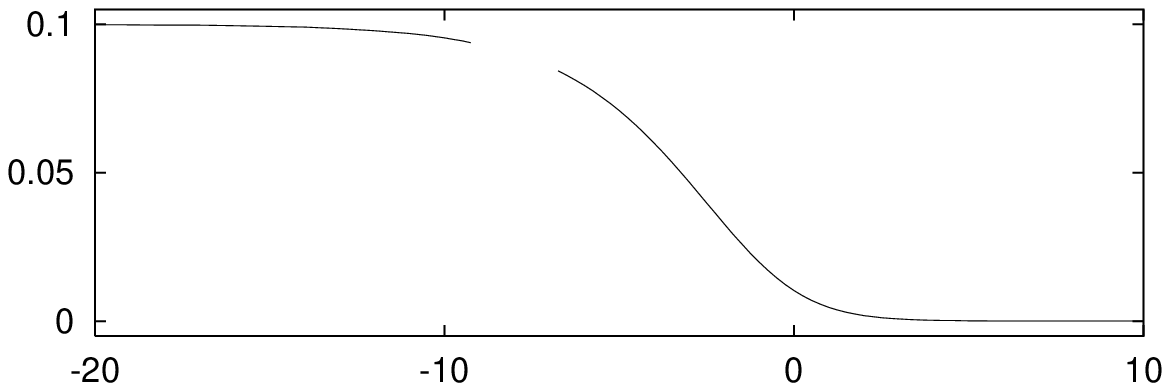}}
\put(4.3,5){\includegraphics[scale=0.996]{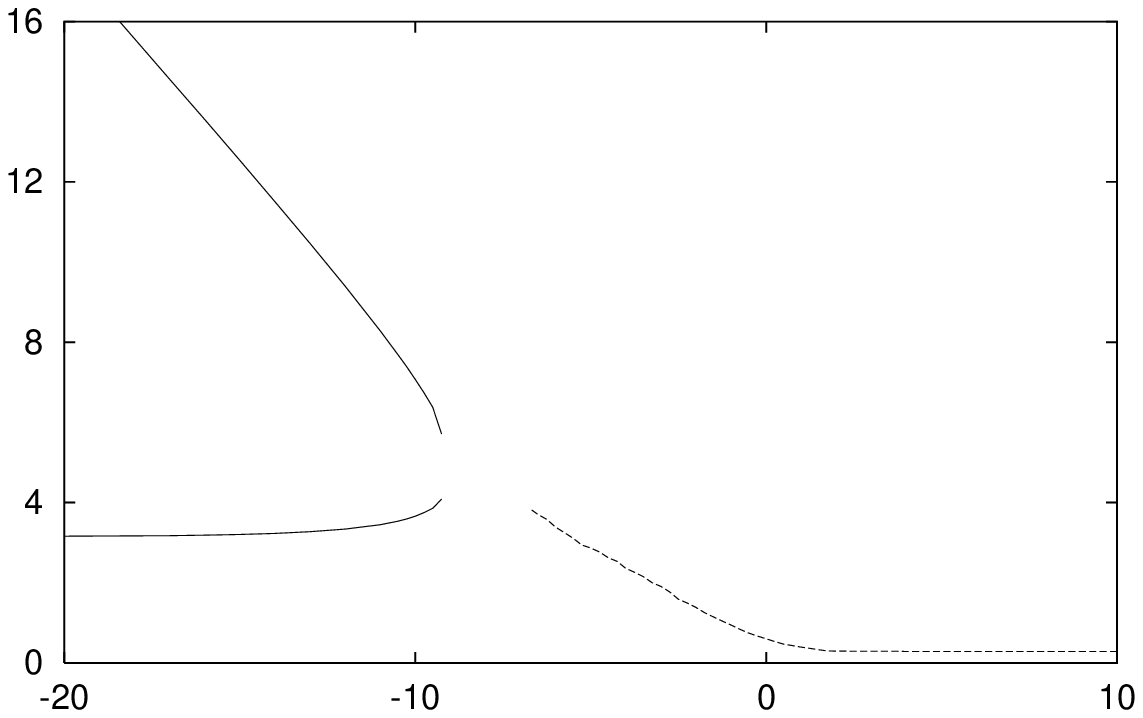}}
\multiput(68,3)(0,78){2}{$\xi$}
\put(-3,70){\makebox(0,0)[l]{zeros}}
\put(-3,115){\makebox(0,0)[l]{$E(\xi)$}}
\end{picture}
\caption{\small Scaling energy and movement of the zeros for the state
$(0)_{+} \mapsto (\,)$, involving mechanism A, in the flow
$\chi_{1,2} \mapsto \chi_{1,1}$.
In the lower plot, the solid line corresponds to the position of
the correlated 1-strings and the dashed one  to the top 2-string in strip 1.
\label{zeros1211}}
\end{center}
\end{figure}

The {\bf mechanism A} requires some care. We examine the ground state $(~)$,
as in Table~\ref{t_s1211}, starting in
the IR where there are no zeros and decrease $\xi$. This procedure 
must fail somewhere
because we expect the top 2-string in strip~1 to transform into a pair of
correlated 1-strings, from Figure~\ref{ABmech}. Keeping track of the top
2-string (with algorithm {\em exp}) we see that the TBA equations reproduce
exactly the movement toward infinity, as shown in
Figure~\ref{zeros1211}, up to a value $\xi\gtrsim -6.5$.
We next start from the other end of the flow, the UV point, now 
increasing $\xi$
from $-\infty$. There and only there, one of the correlated  1-strings sits
exactly at $+\infty$ and is essentially decoupled from the system,
restoring the correct odd parity of $m_1$. Away from the UV point we need to
consider both of the correlated 1-strings, so now $m_1=2$.
To get the code working, we need to use the algorithm $g_2$ for the
largest zero $y_2^{(1)}$ ($y_{m_1}^{(1)}$ in general) and use the 
usual algorithm
{\em exp} for the other zero ($k=1,\ldots,m_1-1$, in general).
We reproduce the expected behaviour, as shown in Figure~\ref{zeros1211},
for $\xi \lesssim -9.5$.

In the interval $\xi\in (-9.5,-6.5)$ containing the collapse region
the TBA equations are difficult
to solve numerically. Indeed, the indicated zeros
are not in the standard positions; in terms of $u$ their real part is not
fixed and must be determined as well as the imaginary part.
In solving (\ref{quant1}) to obtain the position of the two 1-strings
for $\xi \lesssim -9.5$ we see they have the same quantum number
$n^{(1)}_1=n^{(1)}_2=1$ as they had when they were a 2-string
($\xi \gtrsim -6.5$).
This explains the name {\em correlated 1-strings} because
they are on a different footing from the ordinary 1 and 2-strings.
At criticality, the furthest among them is at $+\infty$ and the closest
is the so called frozen zero already observed in \cite{OPW}.
The monotonic decreasing behaviour of the function $\psi_2(x)$ assumed
to introduce the quantum numbers is observed to fail in the present case,
in a neighbourhood of the furthest zero among the correlated 1-strings.
In this way the space to allocate two degenerate quantum numbers is created.
We see the same behaviour in all the mechanism A states.

We have a consistency check. At the critical point $\chi_{1,2}$, using
the data in Table~\ref{critical_TBA}, we have $m_1$ odd, $m_2$ even,
$n_k^{(1)}$ even and $n_k^{(2)}$ odd. Comparing with the the limits
(\ref{lim_psi+}), we have agreement only if $\psi_j(+\infty)=-\pi m_j$
and $y_{m_j}^{(j)}<+\infty$. For the first strip, $\psi_2(+\infty)=-m_2 \pi $
is an even multiple of $\pi$ and $n_k^{(1)}$ is even so, in addition to the
known zeros labelled $1,\ldots,m_1$, we have room for another zero
located at $+\infty$. This zero was not counted in \cite{OPW} because it is
``almost'' decoupled from the system but now we know exactly its origin and
its role from the off critical behaviour.
In the mechanism A case, that is the frozen case ($\sigma=+1$),
this zero is the partner of the frozen one in the correlated 1-strings;
in the mechanism B case, corresponding to the unfrozen case ($\sigma=-1$), it
is the zero that left the finite region; in this case it is a normal 1-string.

Repeating the previous argument for the second strip, it seems that we have
room for an additional zero. Actually,
it is only there at the conformal point $\chi_{1,2}$: during the flow 
it cannot exist
either at $+\infty$ or at a finite location so we don't need to consider
its presence at all.

\subsection{Flows $\chi_{3,2}\mapsto\chi_{3,1}$ and
$\chi_{2,2}\mapsto\chi_{2,1}$}

\begin{figure}\begin{center}
\includegraphics[width=0.85\linewidth ]{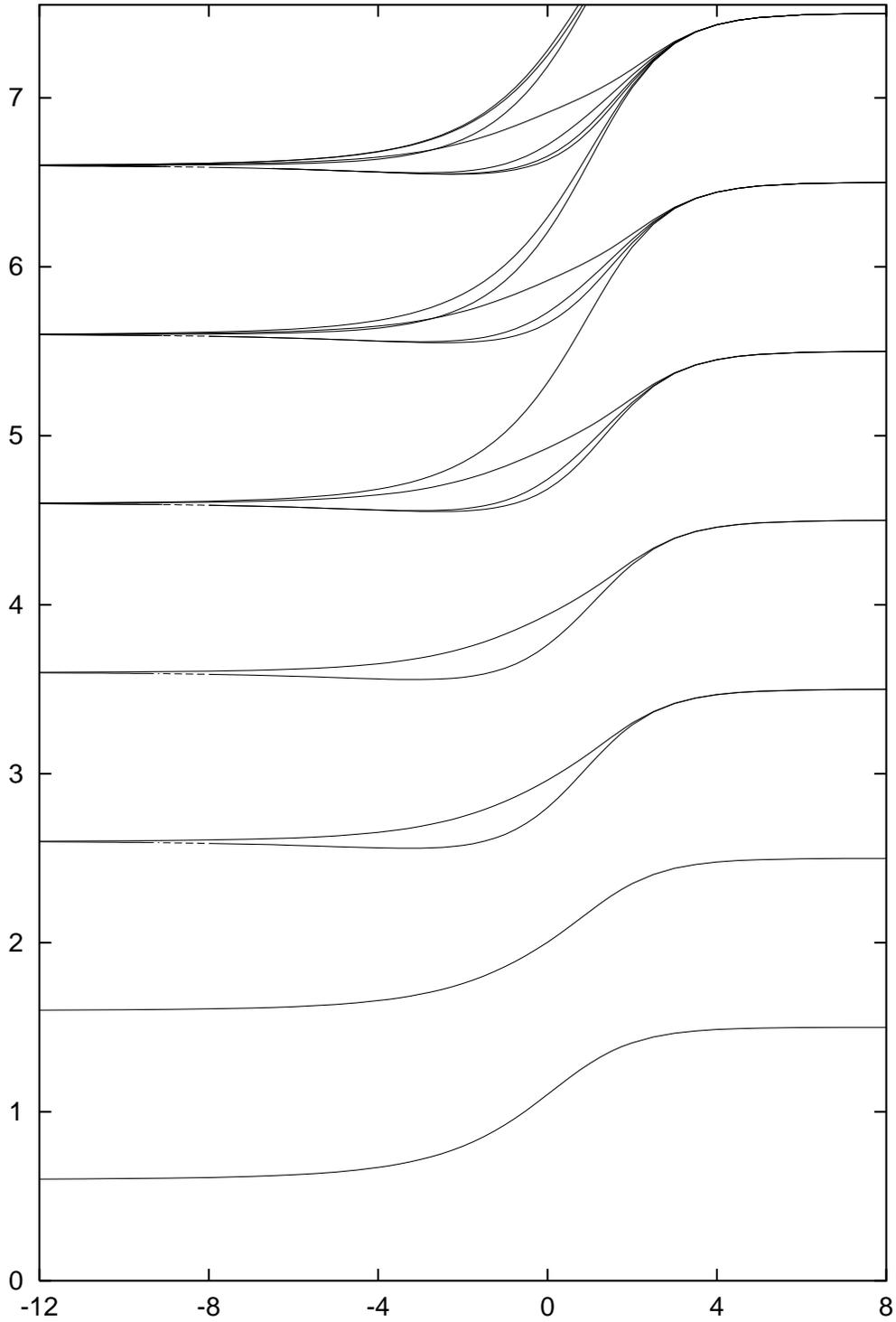}
\caption{\small Scaling energies for the flow $\chi_{3,2} \mapsto \chi_{3,1}$.
The list of states is given in Table~\ref{t_s3231}. The intermediate 
region of the mechanism~A levels (shown dashed) are schematic and 
have not been obtained from the solution of the TBA 
equations.\label{energy3231}}
\end{center}
\end{figure}

\begin{figure} \begin{center}
\includegraphics[width=0.85\linewidth ]{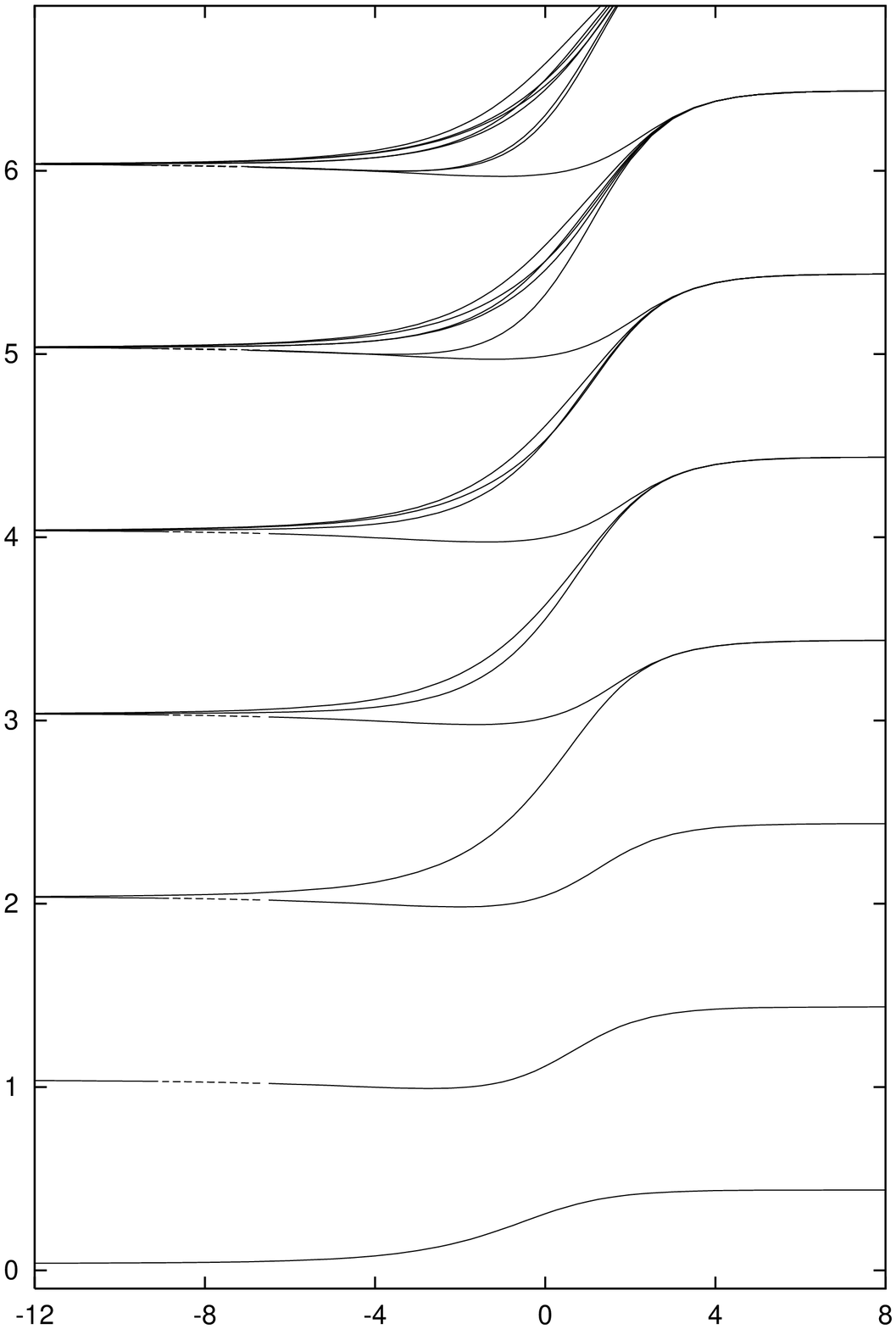}
\caption{\small Scaling energies for the flow $\chi_{2,2} \mapsto \chi_{2,1}$.
The list of states is given in Table~\ref{t_s2221}. The intermediate 
region of the mechanism~A levels (shown dashed) are schematic and 
have not been obtained from the solution of the TBA 
equations.\label{energy2221}}
\end{center}
\end{figure}

These flows behave similarly to $\chi_{1,2}\mapsto\chi_{1,1}$ described
in the previous section and the algorithms to solve the TBA system
are exactly the same.
In Figures~\ref{energy3231} and \ref{energy2221} we plot the scaling energies
for the states in Tables~\ref{t_s3231} and \ref{t_s2221}
respectively. They are complete up to the top level in each plot.

A comment for the second of these flows is useful.
At the IR point and all along the flow except at the UV point, the
value of $n_k^{(2)}$ is even and, in addition, $\psi_1(-\infty)=0$ is even
(\ref{psi1_asymp2}) so that at $x=-\infty$ there can be a zero $y_1^{(2)}$.
Actually, a 1-string occurs whenever its quantum number equals the number
of 2-strings in the same strip, $I_1^{(2)}=n_2$. In this case, the zero
sticks to its position for all values of $\xi$ and the TBA equations can be
slightly simplified taking the appropriate limit.

\subsection{Flows $\chi_{1,3}\mapsto\chi_{2,1}$ and
$\chi_{1,2}\mapsto\chi_{2,1}$}
\begin{figure}\begin{center}
\includegraphics[width=0.85\linewidth ]{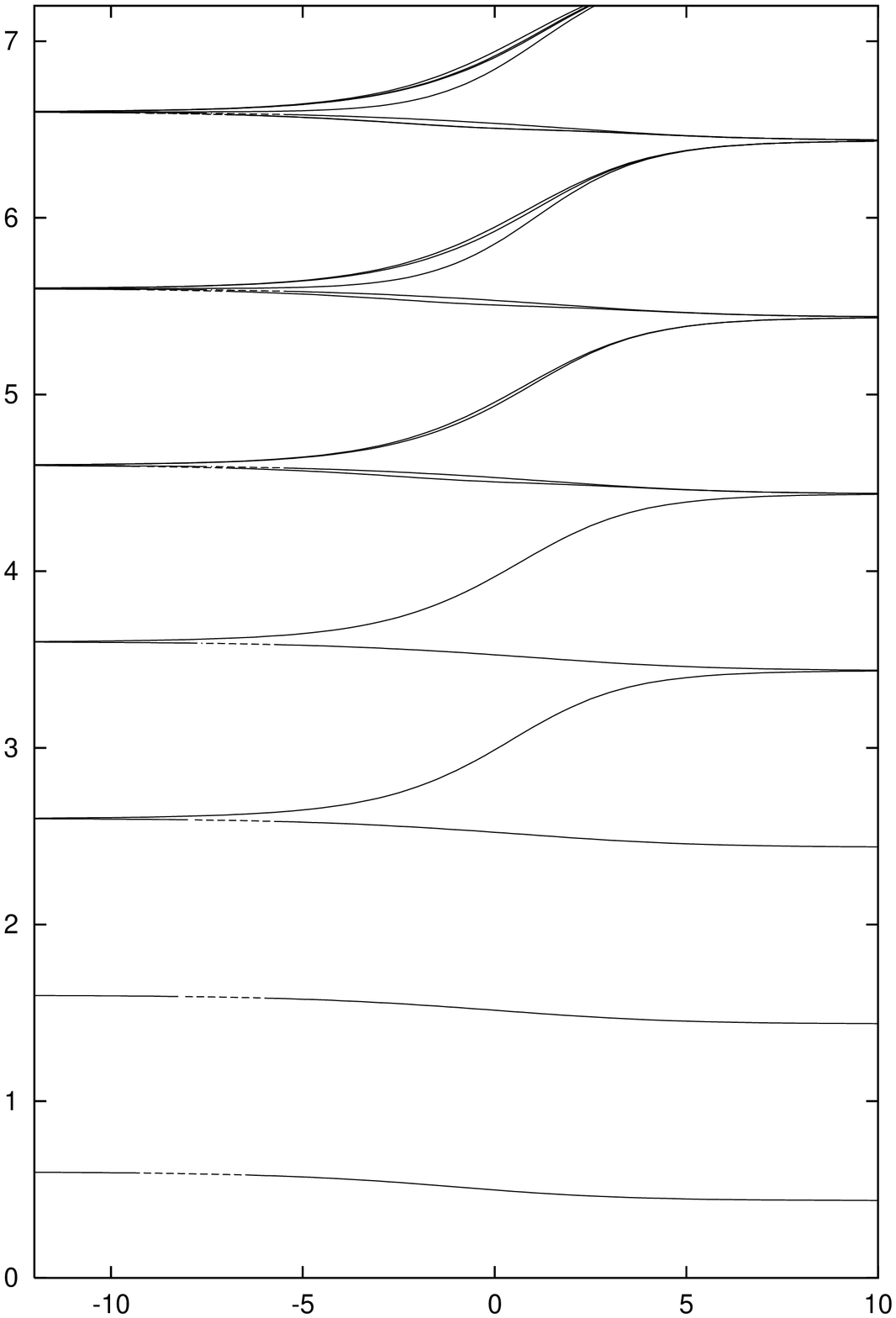}
\caption{\small Scaling energies for the flow $\chi_{1,3} \mapsto \chi_{2,1}$.
The list of states is given in Table~\ref{t_s1321}. The intermediate 
region of the mechanism~C levels (shown dashed) are schematic and 
have not been obtained from the solution of the TBA 
equations.\label{energy1321}}
\end{center}
\end{figure}
\begin{figure}\begin{center}
\includegraphics[width=0.85\linewidth ]{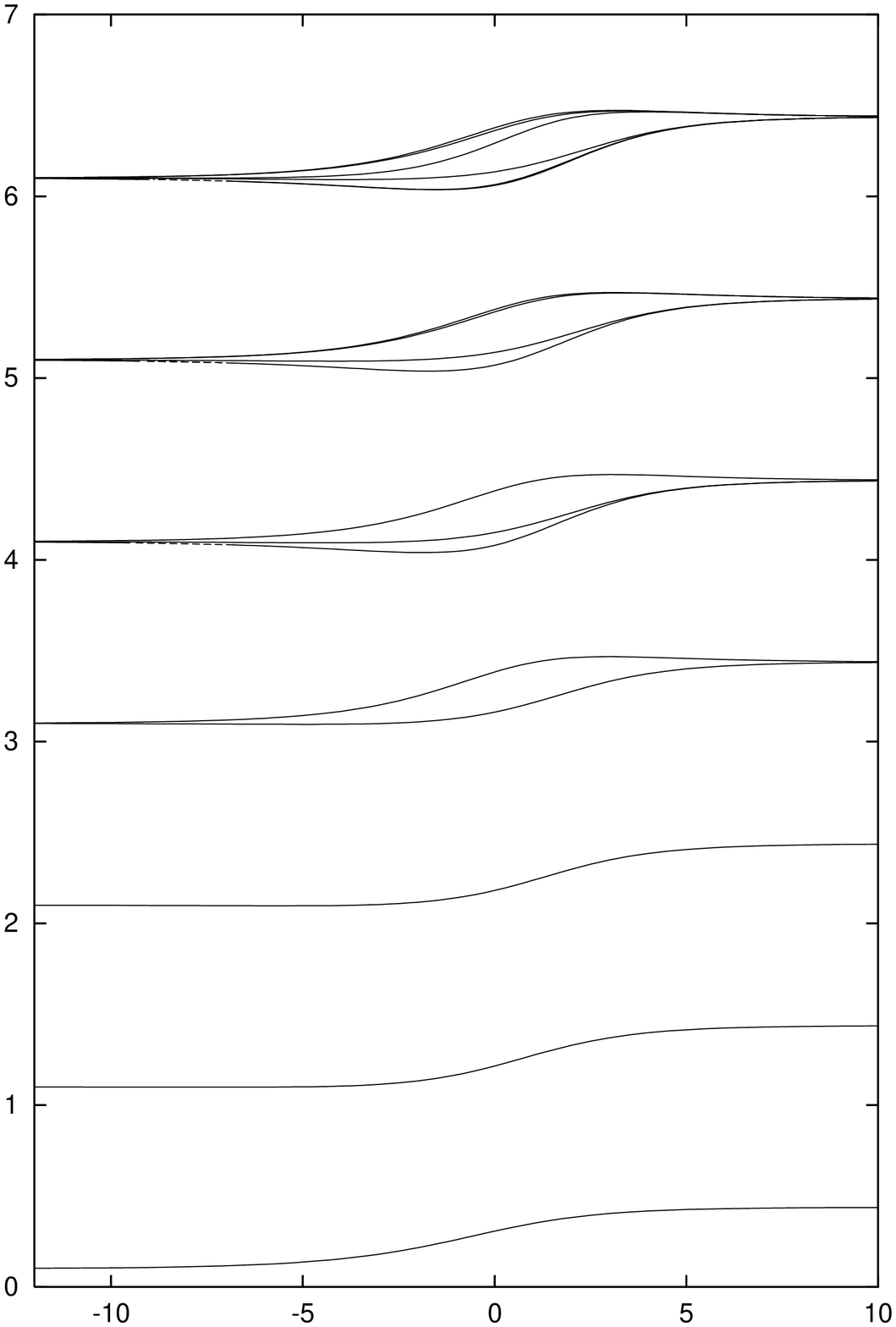}
\caption{\small Scaling energies for the flow
$\chi_{1,2} \mapsto \chi_{2,1}$.
The list of states is given in Table~\ref{t_s1221}. The intermediate 
region of the mechanism~C levels (shown dashed) are schematic and 
have not been obtained from the solution of the TBA 
equations.\label{energy1221}}
\end{center}
\end{figure}

We consider first $\chi_{1,3}\mapsto\chi_{2,1}$, displayed in
Figure~\ref{energy1321}. It doesn't present any new difficulties
with respect to the description in Section~\ref{ss_n1211}.
In this case, we have three algorithms for each strip to get the position of
the zeros and we need to choose those that converge in the appropriate
regimes and depending on the mechanism.

In general the first strip zeros labelled by $1,\ldots,m_1-1$ do not require
special care and the simplest algorithm {\em exp} can be applied to them.
So, in the following discussion, we refer just to the remaining
first strip zero $m_1$ and the second strip zeros. In general, they
can be determined by the algorithms $\psi_j, ~j=1,2$ except for a few 
exceptions.

Considering \textbf{mechanism A}, we just have a zero escaping
to $+\infty$ which causes no problems for the numerical solution.
The flow of these states can be followed without interruptions from
IR to UV.

Considering \textbf{mechanism B}, we start from the IR and
observe that the movement of the 2-strings depicted in Figure~\ref{ABCmech}
doesn't influence the TBA equations, as expected because
they are just spectators; in fact the scaling energies are determined
by the 1-string dynamics only.
The zero moving toward infinity, $y_{m_2\ir}^{(2)}$, is most conveniently
determined by inverting the boundary term $\hat{g}_1$.

Considering \textbf{mechanism C}, we start from the IR and we expect 
the iteration
to fail because new objects must appear in the region
where the transformation of the zeros depicted in Figure~\ref{ABCmech}C
occurs, namely $-9.5 \lesssim \xi \lesssim -6.5 $.
Data for smaller values $\xi \lesssim -9.5$ can be obtained by starting from
the UV point and including the zero at infinity,
$y_{m_2\ir+2}^{(2)}=+\infty$. Again, this particular zero is most conveniently
determined by inverting the boundary term $\hat{g}_1$.

A preliminary discussion of the flow $\chi_{1,2} \mapsto \chi_{2,1}$ 
was given in \cite{FPR}. It is very similar to the flow 
$\chi_{1,3}\mapsto \chi_{2,1}$.
For completeness, we present our current, more accurate numerical 
results in Fig. 11.

\section{Conclusions}

In this paper we have used a lattice approach to derive exact TBA equations
for all excitations of the 5 $\varphi_{1,3}$ integrable boundary 
flows of the tricritical Ising model. We have shown that, along these 
boundary flows, the patterns of zeros classifying the states can 
change by one of  2 or 3 mechanisms which have been explicitly 
identified. These mechanisms produce  precise mappings between the 
relevant finitized characters describing the finitized energy spectra 
at the conformal UV and IR fixed points. The TBA equations were also 
solved numerically to determine the interpolating boundary flows for 
the leading excitations.

Even for the TIM there remain other 
questions of interest. First, it is desirable to have a comparison of 
our results with the results of the TCSA. The methods used here could 
usefully be generalized to study other flows, such as $\chi_{2,2}(q) 
\mapsto \chi_{1,1}(q)+\chi_{3,1}(q)$, which involve linear 
combinations of Virasoro characters. It would also be of interest to 
study the missing 2 $\varphi_{1,2}$ integrable boundary flows, but 
presumably, this would involve a study of the relevant dilute $A$ 
lattice model. Likewise, our approach should be applied to the 
boundary flows of the TIM  from the superconformal perspective, 
especially since the integrable boundary conditions on the lattice 
that correspond to the  superconformal boundary conditions in the 
continuum scaling limit are already known~\cite{RichardP}. Similarly, 
it remains to incorporate the flow of boundary entropies into our 
approach. Ultimately, of course, the considerations of this paper 
should be generalized to all the minimal models.

\section*{Acknowledgments}
This research was supported by the Australian Research Council and in 
part by the European Network EUCLID (Grant no. HPRN-CT-2002-00325). 
Some of the work was carried out while GF and PAP visited IPAM, UCLA. 
We thank Kevin Graham and Gerard Watts for discussions.

\appendix
\renewcommand{\theequation}{A.\arabic{equation}}
\setcounter{equation}{0}

\section*{Appendix A. Gaussian polynomials}
The expressions for the finitized characters are based on the following
relations for the $q$-binomials:
\begin{eqnarray}\label{gauss}
\gauss{m+n}{m} &=&
\sum_{I_1=0}^n \sum_{I_2=0}^{I_1}\cdots \sum_{I_m=0}^{I_{m-1}}
q^{I_1+\ldots+I_m} \\[4mm]
\label{recourrence}
\gauss{n}{m} &=&\gauss{n-1}{m-1} + q^m \gauss{n-1}{m}
\end{eqnarray}
The following expressions are useful to understand the contribution of
each mechanism to the full character. Indeed, the restriction $I_m>0$
leads to
\begin{equation} \label{Ipos}
\sum_{I_1=1}^{n} \sum_{I_2=1}^{I_1}
\ldots \sum_{I_{m}=1}^{I_{m-1}} q^{I_1+\ldots+I_{m}}  =
q^{m} \sum_{I'_1=0}^{n-1} \sum_{I'_2=0}^{I'_1}
\ldots \sum_{I'_{m}=0}^{I'_{m-1}}
q^{I'_1+\ldots+I'_m} =  q^{m} \gausst{m+n-1}{m}
\end{equation}
whereas the restriction $I_m=0$ leads to
\begin{equation} \label{Izero}
\sum_{I_1=0}^{n} \sum_{I_2=0}^{I_1}
\ldots \sum_{I_{m-1}=0}^{I_{m-2}} q^{I_1+\ldots+I_{m-1}}
= \gausst{m-1+n}{m-1}.
\end{equation}

\section*{Appendix B. Braid limit}
\renewcommand{\theequation}{B.\arabic{equation}}
\setcounter{equation}{0}
In this section we consider the braid limit of the normalized transfer
matrix and its eigenvalues.
By direct computation, we obtain the following limits:
\setlength{\unitlength}{10mm}
\begin{eqnarray}
\lim_{\mbox{\scriptsize Im}(u) \rightarrow \pm \infty} e^{\pm 4iu} \sum_{g}
\raisebox{-1.3\unitlength}[1.3\unitlength][1.1\unitlength]
{\begin{picture}(3,2.4)(-0.7,0.1)
\put(0,0.5){\line(1,1){1}}\put(0,2.5){\line(1,-1){1}}
\multiput(-0.5,0.5)(0,2){2}{\line(1,0){0.5}}
\multiput(-0.5,0.5)(0,0.3){7}{\line(0,1){0.2}}
\multiput(1,0.5)(1,0){2}{\line(0,1){2}}
\multiput(1,0.5)(0,1){3}{\line(1,0){1}}
\put(0,0.4){\spos{t}{r}}\put(1,0.4){\spos{t}{r}}
\put(2.05,0.4){\spos{t}{b}}
\put(0,2.6){\spos{b}{r}}\put(1,2.6){\spos{b}{r}}
\put(2.05,2.6){\spos{b}{d}}
\put(1.05,1.45){\spos{tl}{g}}\put(2.07,1.55){\spos{tl}{c}}
\put(1.5,1){\spos{}{u}}
\put(1.5,2){\spos{}{\lambda\!-\!u}}
\put(0.12,1.5){\makebox(0,0){\scriptsize $ \begin{matrix}r ,\, 1 \\
\lambda \!-\!u ,\,\xi \end{matrix} $}}
\multiput(0.0,0.5)(0,2){2}{\makebox(1,0){\dotfill}}
\end{picture}} & = &
\lim_{\mbox{\scriptsize Im}(u) \rightarrow \pm \infty}
\raisebox{-1.3\unitlength}[1.3\unitlength][1.1\unitlength]
{\begin{picture}(2,2.4)(-0.7,0.1)
\put(0,0.5){\line(1,1){1}}\put(0,2.5){\line(1,-1){1}}
\multiput(-0.5,0.5)(0,2){2}{\line(1,0){0.5}}
\multiput(-0.5,0.5)(0,0.3){7}{\line(0,1){0.2}}
\put(0,0.4){\spos{t}{b}}
\put(0,2.6){\spos{b}{b}}
\put(1.07,1.55){\spos{tl}{c}}
\put(0.12,1.5){\makebox(0,0){\scriptsize $ \begin{matrix} b,\, 1 \\
\lambda \!-\!u ,\,\xi \end{matrix} $}}
\end{picture}} \;
\delta_{bd} \: \frac{e^{\pm 2iu} e^{\pm i\lambda}}{4\sin^2 \lambda}, \\[4mm]
\lim_{\mbox{\scriptsize Im}(u) \rightarrow \pm \infty} e^{\pm 4iu} \sum_{g}
\raisebox{-1.3\unitlength}[1.3\unitlength][1.1\unitlength]
{\begin{picture}(3.5,2.4)(-0.7,0.1)
\put(0,0.5){\line(1,1){1}}\put(0,2.5){\line(1,-1){1}}
\multiput(-0.5,0.5)(0,2){2}{\line(1,0){0.5}}
\multiput(-0.5,0.5)(0,0.3){7}{\line(0,1){0.2}}
\put(1,1.5){\line(1,1){1}}\put(1,1.5){\line(1,-1){1}}
\multiput(2,0.5)(0,2){2}{\line(1,0){0.5}}
\multiput(2.5,0.5)(0,0.3){7}{\line(0,1){0.2}}
\put(1,0.4){\spos{t}{r}}
\put(1,2.6){\spos{b}{r}}
\put(1,1.3){\spos{}{g}}
\put(0.12,1.5){\makebox(0,0){\scriptsize $ \begin{matrix}r,\, 1 \\
\lambda \!-\!u ,\,\xi_1 \end{matrix} $}}
\put(1.9,1.5){\makebox(0,0){\scriptsize $ \begin{matrix}r,\, 1 \\
u ,\,\xi_2 \end{matrix} $}}
\multiput(0.0,0.5)(0,2){2}{\makebox(2,0){\dotfill}}
\end{picture}} &=&
\frac{e^{\pm 2 i \lambda}\, \cos \lambda}
{8 \sin^2 \lambda \,\sin 2 \xi_1 \,\sin 2 \xi_2}.
\end{eqnarray}
We can recursively apply the first result to the normalized transfer matrix
to remove all the columns then use the second expression to take care of
the boundary terms. This leads to the conclusion that
$\mathbf{t}(u,\xi_1,\xi_2)$ is diagonal and
the actual value of the limit is the same for all eigenvalues
and all integrable boundary conditions,
that is to say that it is proportional to
the identity. For a generic eigenvalue we thus have
\begin{equation} \label{tlimit}
\lim_{\mbox{\scriptsize Im}(u) \rightarrow \pm \infty}
t(u,\xi_1,\xi_2)=2 \cos \lambda =\frac{1+\sqrt{5}}{2}.
\end{equation}
In \cite{OPW} the same expression was boundary dependent
because of different normalizations in the right boundary term.


\end{document}